\documentclass[preprint]{revtex4-2}
\usepackage{nopageno,times,epsf,amsmath,amssymb}
\usepackage[T1]{fontenc} %Especial para español (for spanish)
\usepackage[spanish, mexico]{babel}
\usepackage[]{caption2}
\usepackage{graphicx}
\usepackage{blindtext}
\usepackage{color}
\usepackage{hyperref}
\usepackage{gensymb}
\usepackage{cancel}
\usepackage{array}
\usepackage{esint}
\begin{document}
%\markboth{ Curso Introductorio de Cristales L\'{\i}quidos}{ Fases y Propiedades Estructurales }

%%%%%
%
% Please provide the following information
%
%%%%%
\title{Curso introductorio de cristales l\'{\i}quidos I: fases y propiedades estructurales \\
\vspace{ 2pt}
\textit{Introductory Course on Liquid Crystals I: Phases and Structural Properties}}
\author{Humberto H\'{\i}jar}
\affiliation{Research Center, La Salle University Mexico, Benjamin Franklin 45, 06140, Mexico City, Mexico}
%%%%%
%
% To be filled by the Editorial Team of RMF, RMF-E 
% and SRMF
%
%%%%%
%\recibido{}{Enero 2024
%\vspace{-12pt}}
\begin{abstract}
\vspace{1em} 
%%%%%%%%%%%%%%%%%%%%%%%%%%%%%%%%%%%%%%%%%%%%%%%%%%%%%%%%%%%%%%%%%%%%%%%%%%%%%%%%%%%%%%%%%%%%%%%%%%%%%%%%%%%%%%%%%%%%%%%
Los cristales l\'{\i}quidos son el prototipo de la llamada materia condensada suave. En t\'erminos simples, son 
``l\'{\i}quidos con estructura'' que hist\'oricamente han recibido mucho inter\'es tanto por ser fuente de nuevos 
conceptos y conocimientos en F\'isica, como por tener aplicaciones electro\'opticas importantes, \textit{e. g.} en las 
pantallas de computadoras port\'atiles y tel\'efonos celulares. M\'as recientemente, se ha descubierto que los 
cristales l\'{\i}quidos podr\'{\i}an aplicarse en la fabricaci\'on de metamateriales y en biomedicina, donde son 
potencialmente \'utiles en la identificaci\'on de tejidos, la reparticipaci\'on controlada de f\'armacos y la 
detecci\'on de bacterias y virus. No obstante, en M\'exico, los cursos sobre cristales l\'{\i}quidos se incluyen en muy 
pocos programas universitarios de F\'isica y en la mayor\'ia de los casos son optativos. Por ello, es escasa la 
literatura en espa\~nol para la ense\~nanza sobre cristales l\'iquidos. En este art\'{\i}culo discutiremos sobre la 
F\'{\i}sica elemental de los cristales l\'{\i}quidos y las herramientas matem\'aticas que permiten analizarlos. 
Principalmente, explicaremos c\'omo caracterizar anal\'iticamente la simetr\'ia y el orden de los cristales l\'iquidos 
nem\'aticos uniaxiales est\'aticos, la fase de cristal l\'iquido m\'as sencilla de todas. Conduciremos esta 
explicaci\'on en t\'erminos b\'asicos, apropiados para estudiantes de ciencias o ingenier\'ias a un nivel intermedio de 
licenciatura, con conocimientos de \'algebra lineal y c\'alculo vectorial, que deseen acercarse por primera vez a este 
tema. Nuestro objetivo es objetivo es apoyar a la comunidad cient\'ifica mexicana en la formaci\'on de recursos humanos 
en este campo. El art\'{\i}culo est\'a escrito en espa\~nol para contribuir al establecimiento y la comunicaci\'on 
de los conceptos que en la literatura especializada s\'olo existen en otros lenguajes. 
%%%%%%%%%%%%%%%%%%%%%%%%%%%%%%%%%%%%%%%%%%%%%%%%%%%%%%%%%%%%%%%%%%%%%%%%%%%%%%%%%%%%%%%%%%%%%%%%%%%%%%%%%%%%%%%%%%%%%%%
\vspace{1em}
%%%%%%%%%%%%%%%%%%%%%%%%%%%%%%%%%%%%%%%%%%%%%%%%%%%%%%%%%%%%%%%%%%%%%%%%%%%%%%%%%%%%%%%%%%%%%%%%%%%%%%%%%%%%%%%%%%%%%%%

\textit{Liquid crystals are the prototype of the so-called soft condensed matter. In simple terms they are ''structured
liquids'' that historically have received a lot of interest because they help to generate new concepts and knowledge in Physics,
and possess important electro--optical applications, e. g., in displays of mobile computers and telephones. More recently,
it has been discovered that liquid crystals could be applied in fabrication of metamaterials and in biomedicine where they
are potentially useful for tissue identification, controlled drug delivery and detection of bacteria and viruses. However,
in Mexico, liquid crystals courses are included only in a few undergraduate programs on Physics being elective in most
of the cases. Therefore, literature for teaching about liquid crystals in Spanish is scarce. In this paper we will discuss
about elementary Physics of liquid crystals and the mathematical tools that permit to analyze them. We will explain, mainly,
how to analytically characterize the symmetry and order of static uniaxial nematic liquid crystals, the most simple of
all liquid crystal phases. We will conduct this explanation in basic terms, suitable for science and engineering students
at intermediate undergraduate level, with knowledge about linear algebra and vector calculus, willing to approach for the first
time to this subject. Our goal is to support the Latin American scientific community in preparing human resources in this field. 
The article is written in Spanish to contribute to the establishment and communication of concepts that only exist in other
languages in the specialized literature.
}
%%%%%%%%%%%%%%%%%%%%%%%%%%%%%%%%%%%%%%%%%%%%%%%%%%%%%%%%%%%%%%%%%%%%%%%%%%%%%%%%%%%%%%%%%%%%%%%%%%%%%%%%%%%%%%%%%%%%%%%

\vspace{1em}
\end{abstract}
%\keys{ \bf{\textit{Materia suave, Anisotrop\'ia, Autoensamblaje, Transiciones de fase, Par\'ametros de orden, \'Algebra tensorial}} \vspace{-8pt}}
%\begin{multicols}{2}

\maketitle
% ============================================================================================================================================
\section{Introducci\'on \label{introduccion_seccion}}
% ============================================================================================================================================

Los cristales l\'iquidos (CL) son fases de la materia que ocurren entre los cristales s\'olidos y los l\'iquidos ordinarios~\cite{hirst_2012}. 
Estas fases combinan propiedades de cristalinidad y de fluidez y, sin embargo, exhiben fen\'omenos exclusivos que no se presentan ni en los 
cristales, ni en los l\'iquidos simples. Lo anterior genera un enorme inter\'es por estudiarlas y hace factible su aplicaci\'on en numerosas 
tecnolog\'ias~\cite{palffy_phys_today_2007}. No obstante, el car\'acter \'unico de los CL requiere de la formulaci\'on de variables nuevas y 
modelos que resultan m\'as complejos que la simple combinaci\'on de las teor\'ias del estado s\'olido y la mec\'anica de 
fluidos~\cite{de_gennes_1993,blinov_2011}.

Aunque a primera vista nos pueda parecer contradictorio, una manera breve de explicar qu\'e son los CL, ser\'ia decir que son ``l\'iquidos con 
estructura''~\cite{hirst_2012}. La ciencia de estos ``l\'iquidos estructurados'' es multifac\'etica, multidisciplinaria, compleja matem\'aticamente 
y enorme en cantidad de fen\'omenos. As\'i que resulta imposible abarcarla totalmente s\'olo en una monograf\'ia. Ese no es el prop\'osito de este 
manuscrito. M\'as bien, se toma un enfoque pedag\'ogico para presentar algunos de los conceptos b\'asicos de la f\'isica y las matem\'aticas de los
CL relacionados con la descripci\'on de su estructura. En la secci\'on~\ref{breve_historia_seccion} se incluye un resumen hist\'orico de c\'omo los 
CL se establecieron como un campo de estudio relevante en la ciencia. En la secci\'on~\ref{caracteristicas_fundamentales_seccion} se discute sobre 
los conceptos de autoensamblaje y anisotrop\'ia. En las secciones~\ref{fases_liquidocristalinas_seccion} y \ref{cristales_liquidos_liotropicos_seccion}, 
respectivamente, se describen de las fases l\'iquido-cristalinas termotr\'opicas y liotr\'opicas. En la secci\'on~\ref{descripcion_matematica_seccion} 
se da una explicaci\'on pormenorizada de las variables que describen la estructura de fases de CL con la simetr\'ia m\'as simple (llamadas fases 
nem\'aticas). En la secci\'on~\ref{energias_seccion} se explica la manera en la que se cuantifican dos de las contribuciones energ\'eticas caracter\'isticas 
de las fases nem\'aticas: la energ\'ia de Landau--de Gennes y la energ\'ia el\'astica. Todos estos temas se abordan de manera autocontenida y con detalle 
a\'un en los c\'alculos b\'asicos, para que puedan estudiarse autodid\'acticamente o bien para que el material puedan servir como gu\'ia para su ense\~nanza. 
Finalmente, en la secci\'on~\ref{reflexiones_finales_seccion} se realizan algunas reflexiones finales y se plantean las pr\'oximas extensiones de este curso.

% ================================================================================================================================================================
\section{El establecimiento de la ciencia de los CL \label{breve_historia_seccion}}
% ================================================================================================================================================================

%-----------------------------------------------------------------------------------------------------------------------------------------------------------------
\subsection{El descubrimiento \label{descubrimiento_seccion}}
%-----------------------------------------------------------------------------------------------------------------------------------------------------------------

El descubrimiento de los CL se atribuye a Friedrich Reinitzer (1857--1927)~\cite{slukin_2004,lagerwall_liq_cryst_2013,dunmur_2014,mitov_chemphyschem_2014}. Como 
muchos descubrimientos importantes en la ciencia, el de los CL fue casual. Reitnitzer trabajaba como bot\'anico en la secci\'on alemana de la Universidad de Praga 
y realizaba experimentos para determinar si el colesterol extra\'{\i}do de fuentes vegetales y animales era un compuesto \'unico o un conjunto de compuestos 
emparentados. En 1888, al analizar el comportamiento f\'{\i}sico de un compuesto en particular, el \emph{benzoato de colesterilo}, descubri\'o para su sorpresa 
que \'{e}ste no se fund\'{\i}a ni se solidificaba como lo hac\'{\i}an las substancias estudiadas hasta entonces. Las substancias ordinarias puras 
son s\'olidas a temperaturas bajas y se funden a una temperatura precisa y repetible cuando se calientan. Si el l\'{\i}quido resultante se enfr\'{\i}a, se 
solidifcar\'a a la misma temperatura a la cual se fundi\'{o} previamente. 

El benzoato de colesterilo difer\'{\i}a en el hecho de que aparentaba tener dos temperaturas o puntos de fusi\'on. A $145.5\degree\text{C}$ su fase s\'olida 
se fund\'{\i}a en un l\'{\i}quido turbio. Si este l\'{\i}quido turbio se calentaba m\'as, se volv\'{\i}a transparente a los $178.5\degree\text{C}$. Si el 
l\'{\i}quido transparente se enfriaba, el proceso se revert\'{\i}a a las mismas temperaturas~\cite{reinitzer_monatshefte_1888}.~\footnote{La mayor\'ia de los
primeros art\'iculos sobre CL est\'an escritos en alem\'an y franc\'es. Algunos de ellos pueden encontrarse traducidos en la referencia~\cite{slukin_2004},
que es una compilaci\'on comentada de art\'iculos hist\'oricamente importantes sobre este tema.} Cerca de las temperaturas previas, Reinitzer observ\'o una 
notoria coloraci\'on azul--violeta del benzoato de colesterilo. Sin embargo, dado que su formaci\'on acad\'emica era en bot\'anica, no pudo dar una explicaci\'on 
m\'as profunda de este fen\'omeno fisicoqu\'imico. Por ello, decidi\'o buscar tal explicaci\'on con la ayuda de alguien m\'as y en marzo de 1888 comparti\'o su 
descubrimiento con un profesor experto en microscop\'ia y cristalograf\'ia,
%asistente del polit\'ecnico de \textit{Aachen} (Aquisgr\'an), 
llamado  Otto Lehmann (1855--1922). No es muy claro por qu\'e Reinitzer opt\'o precisamente por Lehmann, pero es probable que al buscar en la literatura y preguntar, haya 
identificado que la coloraci\'on llamativa en sus muestras ten\'ia que ver con un fen\'omeno cristalino. %por ser un experto en microscop\'ia y cristalograf\'ia.

Para analizar los cambios del benzoato de colesterilo, Lehmann utiliz\'o el ``microscopio de cristalizaci\'on'', un microscopio inventado por \'el mismo, equipado
con polarizadores y con una platina que pod\'ia calentar o enfriar controladamente las muestras. En particular, Lehmann estudi\'o el paso de la luz a trav\'es del 
benzoato de colesterilo y concluy\'o que, en el rango de temperaturas donde tiene el aspecto de l\'iquido turbio, exhibe la propiedad de \textit{birrefringencia}. 
En los materiales birrefringentes, el \'indice de refracci\'on depende de la direcci\'on de propagaci\'on de la luz, lo que se manifiesta en que un haz de luz 
incidente puede refractarse en dos haces diferentes. Hasta entonces, la birrefringencia se hab\'ia observado exclusivamente en los cristales. 

Lehmann estaba ante un fen\'omeno nuevo. En este estado el benzoato de colesterilo era birrefringente, pero tambi\'en pod\'{\i}a fluir. Si, \textit{e. g.}, levantaba 
o presionaba la l\'amina superior del portaobjetos, la substancia era succionada hacia arriba o se desparramaba hacia los lados, respectivamente. El estudio
sistem\'atico de este y otros compuestos relacionados, le condujo a concluir que estos ``l\'iquidos con birrefringencia'' eran una nueva fase de la materia,
que combinaban todas las propiedades de fluidez y cristalinidad. En 1900, tras haber utilizado varios nombres tentativos para ellas como \textit{fliessende Kristalle} 
(cristales que fluyen) o \emph{kristallinischer Fl\"ussigkeiten} (l\'{\i}quidos cristalinos)~\cite{lehmann_annalen_1890,lehmann_zeitschrift_1890}, las
nombr\'o como las conocemos hoy, \textit{fl\"ussige Kristalle}, en espa\~nol, CL~\cite{lehmann_1904}.

%-----------------------------------------------------------------------------------------------------------------------------------------------------------------
\subsection{Primeras investigaciones: entre el asombro y el escepticismo \label{primeras_investigaciones_seccion}}
%-----------------------------------------------------------------------------------------------------------------------------------------------------------------

Los estudios de Lehmann abrieron un campo nuevo en el que participaron inicialmente muchos f\'isicos y qu\'imicos alemanes y franceses. Un ejemplo es el de Ludwig 
Gattermann (1860--1920) y A. Ritschke, quienes en 1890 sintetizaron los primeros cristales l\'iquidos artificiales~\cite{gattermann_berichte_1890}. Uno de \'estos, el 
\textit{para}-azoxianisol (PAA), mostr\'o cambios con la temperatura y propiedades \'opticas y de fluidez similares a las de los compuestos de Lehmann y se convirti\'o en 
uno de los materiales est\'andar para el an\'alisis de los CL. Tambi\'en puede mencionarse a Daniel Vorl\"ander (1867--1941), quien clasific\'o diferentes compuestos de acuerdo
con su capacidad de dar lugar o no a CL, siendo su contribuci\'on m\'as importante el notar que aquellos que s\'i lo hacen est\'an formadas por mol\'eculas 
alargadas~\cite{vorlaender_berichte_1906,vorlaender_1908}. Esto hecho ser\'a muy importante, tal como lo veremos en la secci\'on~\ref{caracteristicas_fundamentales_seccion}.

Sin embargo, la idea de que pod\'ian existir cristales que fluyen no 
fue aceptada de manera generalizada. Resultaba f\'acil desestimar a los CL como l\'iquidos ordinarios que albergaban impurezas en la forma de peque\~nos cristales 
o bien, como emulsiones, en las que gotas de un l\'iquido est\'an suspendidas en otro.~\footnote{Algunas emulsiones nos son muy familiares, como la leche (gotas de aceite en 
agua) o la vinagreta (gotas de aceite en vinagre). Ambos ejemplos son l\'iquidos turbios y es por ello que superficialmente se parecen al estado de l\'iquido
turbio identificado por Reinitzer.}
Algunos cr\'iticos suger\'ian con muy poco tacto que el concepto de CL se basaba en experimentos con errores, descuidos o interpretaciones incorrectas. 
Tal fue el caso de Gustav Tammann (1861--1938) quien en los primeros a\~nos del siglo XX public\'o diversos art\'iculos, algunos con t\'itulos sarc\'asticos como ``Sobre los as\'i 
llamados cristales l\'iquidos''~\cite{tammann_annalen_1901,tammann_annalen_1902}, en los que defend\'ia fuertemente la hip\'otesis de la emulsi\'on. Seg\'un Tammann,
las substancias de Lehmann eran mezclas en las que se observaban diferentes puntos de fusi\'on porque las diferentes componentes cambiaban de fase a temperaturas
distintas. Otro ejemplo es el de Georges Friedel (1865--1933) y Fran\c{c}ois Grandjean (1882--1975), quienes llegaron a reportar en 1910 que si los ``cristales l\'iquidos de Lehmann'' 
se analizaban correctamente, en realidad no eran birrefringentes~\cite{friedel_bulletin_1910}. 

Estas cr\'iticas fueron desmentidas por el propio Lehmann y otros de sus contempor\'aneos, como Rudolf Schenck (1870--1965), quien demostr\'o mediante el uso de diferentes t\'ecnicas 
y sin dejar lugar a dudas, que los CL eran substancias puras~\cite{schenck_1904}. Por otra parte, el mismo a\~no en que Fridel y Grandjean negaron
la birrefringencia de los CL, Charles Mauguin (1878--1958) report\'o observaciones, mediante un m\'etodo independiente al de Lehmann, que arrojaban una birrefringencia de una
magnitud similar~\cite{mauguin_comptes_1910}. A \'estas le sigui\'o una serie de experimentos~\cite{mauguin_bulletin_1911,mauguin_comptes_1911} con una calidad cient\'ifica 
tal, que no s\'olo result\'o in\'util volver a cuestionar la birrefringencia de los CL, sino que la investigaci\'on sobre ellos pas\'o de una etapa m\'as bien cualitativa a 
otra de an\'alisis sistem\'atico que permitir\'ia comprenderlos a un nivel m\'as profundo. 

%-----------------------------------------------------------------------------------------------------------------------------------------------------------------
\subsection{Decaimiento y retorno \label{decaimiento_retorno_seccion}}
%-----------------------------------------------------------------------------------------------------------------------------------------------------------------

Para finales de los a\~nos 1920, los CL eran ya completamente aceptados como estados genuinos de la materia. Parad\'ojicamente, la investigaci\'on sobre ellos declin\'o 
y se volvi\'o m\'as lenta, si bien nunca se detuvo por completo. Alrededor de ese periodo y en las d\'ecadas subsecuentes (tiempo en el que debe tomarse en cuenta la ocurrencia
de las dos guerras mundiales), son notables las contribuciones de Carl Wilhelm Oseen (1879--1944), Hans Zocher (1893--1969) y Frederick Charles Frank (1911--1998) a la teor\'ia
de la elasticidad de los CL~\cite{oseen_1929,zocher_discussions_1933,frank_discussions_1958}, la cual presentaremos m\'as adelante en la secci\'on~\ref{energia_elastica_seccion}. 
Asimismo, puede destacarse el trabajo de Wilhelm Maier (1913--1964) y Alfred Saupe (1925--2008) sobre una teor\'ia estad\'istica que permite predecir la transici\'on hacia fases 
l\'iquido-cristalinas~\cite{maier_z_naturforschg_1959}. 

Industrias como RCA, Bell Laboratories e IBM renovaron la investigaci\'on sobre CL en los a\~nos 1960. Sus actividades fueron acompa\~nadas intensamente por grupos acad\'emicos
entre los que destacaron los de las universidades de Orsay (hoy Paris-Saclay) en Francia, as\'i como Harvard y Kent State en Estados Unidos. Por encima de todos sobresale el 
nombre de Pierre-Gilles de Gennes (1932--2007), fundador del grupo de Orsay, quien fue pionero en notar similitudes profundas de los CL con sistemas aparentemente ajenos como 
los superconductores, los pol\'imeros y los materiales magn\'eticos~\cite{joanny_biogr_mem_fellows_r_soc_2019}. Esto le permiti\'o, entre otras cosas, unificar modelos 
matem\'aticos para los CL y as\'i interpretar correctamente su naturaleza f\'isica, explicar la dispersi\'on de luz que les da su apariencia turbia y describir transiciones 
entre algunas de sus diversas fases (el modelo de de Gennes para la transici\'on m\'as sencilla ser\'a analizado en la secci\'on~\ref{energia_formacion_fase_seccion}).
Por el descubrimiento de estos m\'etodos le fue otorgado el Premio Nobel de f\'isica a de Gennes en 1991, un poco m\'as de cien a\~nos despu\'es de los hallazgos originales de 
Reitnizer y Lehmann.~\footnote{Lehmann fue nominado, sin \'exito, para recibir el Premio Nobel de 1913 a 1922.} 

En gran medida, el resurgimiento de la ciencia de los CL en la segunda mitad del siglo XX se 
produjo porque fue hasta entonces que se destac\'o su enorme potencial tecnol\'ogico, siendo su aplicaci\'on m\'as popular la de las pantallas planas de bajo 
consumo de energ\'ia, que evolucionaron desde las car\'atulas de siete segmentos de los relojes digitales y calculadoras, hasta los LCDs (por las siglas del ingl\'es 
\textit{Liquid crystal displays}) que utilizamos a diario en nuestros tel\'efonos celulares, computadoras y pantallas de televisi\'on~\cite{geelhaar_angew_chem_int_ed_2013}.
Tambi\'en son notables los sistemas de distribuci\'on de f\'armacos basados en la habilidad de los CL para encapsular y liberar 
substancias~\cite{kim_j_pharm_investig_2015,mo_liq_cryst_rev_2017}, al igual que los sistemas \'opticos y optoelectr\'onicos como ventanas inteligentes, l\'aseres sintonizables, 
lentes ajustables y moduladores \'opticos en fibras \'opticas y sistemas de 
comunicaci\'on~\cite{algorri_crystals_2019,zola_adv_mater_2019,bisoyi_chem_rev_2021,mysliwiec_nanophotonics_2021}. Todos estos avances han requerido de continuar la s\'intesis y 
producci\'on de nuevos CL con propiedades aptas para las aplicaciones y desarrollar un conocimiento profundo de su naturaleza e interacci\'on con otras fases de la materia y campos 
externos. 

Hoy en d\'ia, la sensibilidad, respuesta y estructura de los CL pueden dise\~narse a tal grado, que se vuelve factible su uso para la fabricaci\'on de 
sistemas de separaci\'on de iones y mol\'eculas~\cite{salikolimi_langmuir_2020,kloos_j_membr_sci_2021}, sistemas de control de bacterias~\cite{peng_science_2016,chi_commun_phys_2020},
detectores de bacterias y virus~\cite{vallooran_adv_funct_mater_2016,wang_biosensors_2022}, m\'usculos artificiales~\cite{mehta_appl_phys_lett_2020}  
y metamateriales (materiales artificiales ideados para presentar propiedades que no existen en los materiales naturales)~\cite{musevic_liquid_crystal_colloids_2017}.
Por todo esto, es posible afirmar que la investigaci\'on sobre CL no se agotar\'a en el futuro cercano, sino que estas fases de la materia seguir\'an siendo
de inter\'es en las ciencias fundamentales y aplicadas.
%Aunado a esto, recientemente se ha extendido la  los CL para actuar como materia activa, .

% ===================================================================================================
\section{Caracter\'{\i}sticas fundamentales cristales l\'{\i}quidos \label{caracteristicas_fundamentales_seccion}}
% ===================================================================================================

%----------------------------------------------------------------------------------------------------------------------
\subsection{Anisotrop\'{\i}a y autoensamblaje \label{anisotropia_autoensamblaje_seccion}}
%----------------------------------------------------------------------------------------------------------------------

El rasgo esencial que distingue a los CL de los l\'{\i}quidos simples es la \emph{anisotrop\'{\i}a}. Este t\'ermino se 
refiere al hecho de que en los CL existen una o m\'as direcciones espaciales en las que las propiedades materiales toman 
valores diferentes. En contraste, los l\'{\i}quidos ordinarios son \emph{isotr\'opicos}, es decir, sus propiedades son 
id\'enticas en todas la direcciones. La anisotrop\'{\i}a se manifiesta en muchos par\'ametros f\'{\i}sicos relevantes de 
los CL. Por ejemplo, en un l\'{\i}quido isotr\'opico el calor se propaga al mismo ritmo en todas las direcciones, mientras 
que en un CL la transferencia de energ\'{\i}a en forma de calor depende de la direcci\'on en la que esta ocurre, de tal 
manera que un CL puede tener al menos dos coeficientes de conducci\'on del calor. De manera similar, los CL tienen por lo 
menos dos \'indices de refracci\'on (de all\'i que sean birrefringentes), dos susceptibilidades diel\'ectricas, dos coeficientes 
de difusi\'on, etc. La anisotrop\'{i}a de los CL es consecuencia de dos efectos fundamentales: la anisotrop\'{\i}a molecular y el 
\emph{autoensamblaje}, ambos conceptos se explican a continuaci\'on.

%.......................................................................................................................
\subsubsection{Anisotrop\'{\i}a molecular \label{anisotropia_molecular_section}}
%.......................................................................................................................

\begin{figure}[h]
\centering
\includegraphics[scale = 0.60]{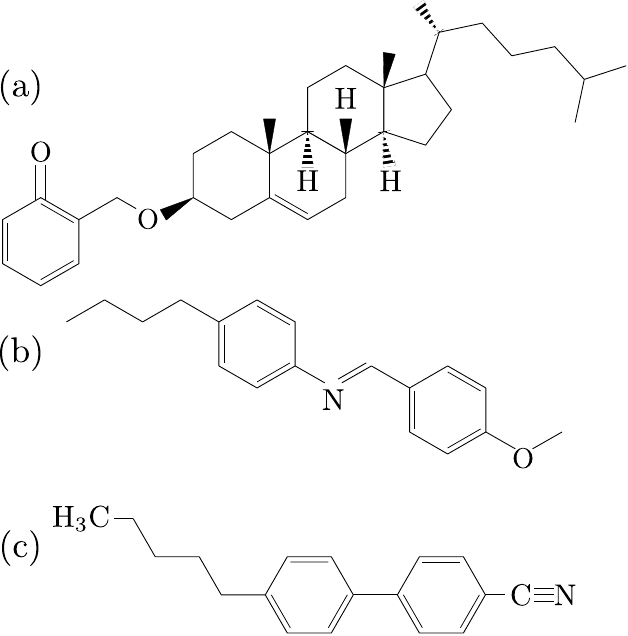}
\caption{Mol\'eculas que forman fases de CL: (a)~benzoato de colesteril, (b) MBBA y (c) 5CB.}
\label{figura_001}
\end{figure}

La anisotrop\'{\i}a a nivel molecular significa que las mol\'eculas que forman a los CL tienen una estructura marcadamente 
alejada de la simetr\'{\i}a esf\'erica y en la que se distinguen direcciones preferentes. Por ejemplo, las mol\'eculas que 
constituyen a los CL pueden tener una forma alargada. Tal es el caso de las mol\'eculas que se muestran en la 
figura~\ref{figura_001}, las cuales correponden al benzoato de colesteril estudiado por Reinitzer y a los compuestos 
org\'anicos %N--(4--Butylphenyl)--1--(4--methoxyphenyl)methanimine y %4--Cyano--4'--pentylbiphenyl, mejor 
conocidos como MBBA y 5CB. Tanto el MBBA como el 5CB fueron sintetizados a principios de los a\~nos 1970 siendo los primeros 
compuestos en exhibir fases de CL a temperatura ambiente, concretamente entre los $21.0\degree\text{C}$ y 
los $48.0\degree\text{C}$, en el caso del MBBA, y entre los $22.5\degree\text{C}$ y los $35.0\degree\text{C}$, en el caso del 
5CB. La figura~\ref{figura_001} illustra que la masa en estas mol\'eculas se distribuye mayoritariamente a lo largo de una
direcci\'on espacial que puede considerarse el eje de simetr\'{\i}a principal de la estructura. Los CL formados por mol\'eculas
como las que se ilustran en la figura~\ref{figura_001}, cuya forma puede aproximarse burdamente como la de una varilla o una 
barra, son llamados CL \emph{calam\'{i}ticos}.

\begin{figure}[h]
\centering
\includegraphics[scale=0.60]{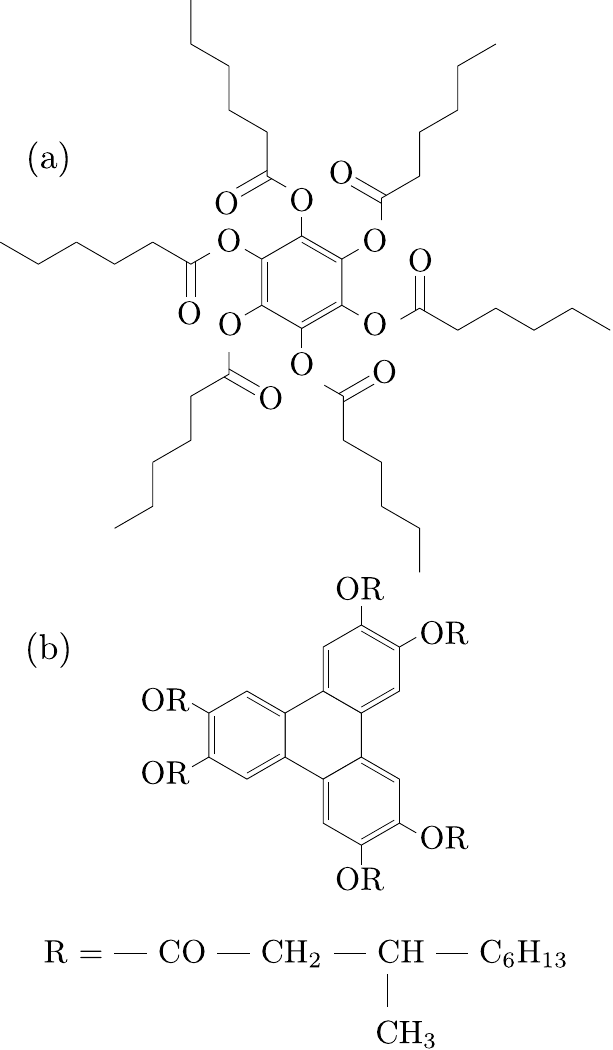}
\caption{Mol\'eculas con forma de disco que forma fases de CL: (a) benceno-hexa-$n$-hexanoato y (b) mol\'ecula con un trifenileno en el centro.
En el caso (b), R representa la estructura molecular que se especifica en la parte inferior.}
\label{figura_002}
\end{figure}

Otro grupo de mol\'eculas que dan lugar a CL lo forman estructuras ``aplastadas'' que semejan discos debido a que sus \'atomos est\'an 
distribuidos principalmente en un plano y de manera sim\'etrica. Estas estructuras est\'an basadas en anillos de benceno, los cuales tienen
una forma hexagonal regular en la que cada v\'ertice est\'a ocupado por un \'atomo de carbono unido a uno de hidr\'ogeno. El benceno es muy 
estable y le proporciona una alta rigidez a la estructura molecular. Los anillos de benceno tambi\'en pueden identificarse en las mol\'eculas 
de los CL calam\'{\i}ticos de la figura~\ref{figura_001}. Ejemplos de mol\'eculas con forma de disco que originan CL son los llamados 
benceno-hexa-$n$-alcanoatos y diversas mol\'eculas con cuatro anillos de benceno unidos en el centro (trifenilenos)~\cite{woehrle_chem_rev_2016}.
%El benceno es  altamente estable, de forma hexagonal, en el que cada v\'ertice es ocupado por un \'atomo de carbono unido a uno de
%hidr\'ogeno~\cite{chandrasekhar_pramana_1977}. Por otra parte, los trifenilenos consisten de cuatro anillos de benceno unidos, lo que le
%brinda a la estructura central de la mol\'ecula una alta estabilidad. 
La figura~\ref{figura_002} ilustra la estructura particular del benceno-hexa-$n$-hexanoato. Este fue analizado experimentalmente en 1977 por 
el f\'{\i}sico indio Sivaramakrishna Chandrasekhar (1930--2004) siendo uno de los primeros compuestos con forma de disco en presentar una 
fase de CL. Tambi\'en se muestra en la figura~\ref{figura_002} la estructura general de las mol\'eculas con trifenilenos centrales que
originan CL. Los CL constituidos por mol\'eculas tipo disco se denominan \emph{disc\'oticos}.

% ......................................................................................................................
\subsubsection{Autoensamblaje \label{autoensamblaje_seccion}}
% ......................................................................................................................

El autoensamblaje puede definirse como la aparici\'on espont\'anea de ordenamiento molecular en un material como resultado del 
balance entre fuerzas intermoleculares y efectos t\'ermicos~\cite{hirst_2012}. Un sistema autoensamblado no es forzado por ning\'un 
agente externo a adorptar ninguna configuraci\'on particular. En vez de ello, las fuerzas entre sus mol\'eculas son quienes promueven 
una estructura colectiva ordenada en la que el sistema se encuentra en equilibrio. Tal equilibrio es termodin\'amicamente riguroso.
En \'el, el potencial termodin\'amico que describe al sistema adopta un valor extremo, es decir, m\'aximo o m\'{\i}nimo. Considermos 
el caso de un sistema con temperatura fija $T$. La funci\'on caracter\'{\i}stica es la llamada energ\'{\i}a libre de Helmholtz,
\begin{equation}
F = U - T S,
\label{autoensamblaje_001}
\end{equation}
en donde $U$ es la energ\'{\i}a interna y $S$ la entrop\'{\i}a~\cite{zemansky_1997}. En equilibrio, $F$ es m\'{\i}nima. 

En la ecuaci\'on~(\ref{autoensamblaje_001}), $U$ consiste de la suma de la energ\'{\i}a potencial 
almacenada en las interacciones moleculares m\'as la energ\'{\i}a cin\'etica de las mol\'eculas. Por otra parte, $S$ indica el n\'umero 
de posibles configuraciones que puede adoptar el ensamble molecular. La entrop\'{\i}a crece con la cantidad de informaci\'on que se 
requiere para especificar la posici\'on y velocidad de todas mol\'eculas del sistema. El t\'ermino $T S$ en la 
ecuaci\'on~(\ref{autoensamblaje_001}) representa la cantidad de energ\'{\i}a disipada en los grados de libertad moleculares y que no es 
utilizable como trabajo. Siguiendo estas ideas, $F$ consiste de la energ\'{\i}a que cuesta formar el material menos la energ\'{\i}a de 
los movimientos t\'ermicos. 

Por una parte, las mol\'eculas son atra\'idas unas con otras por las fuerzas intermoleculares. Por otra, la temperatura las mantiene 
movi\'endose en una constante agitaci\'on err\'atica. El delicado balance entre estos dos efectos determina la estructura que adquirir\'a 
el material a nivel molecular a la temperatura $T$. Si $T$ es alta, la agitaci\'on t\'ermica impedir\'a que las mol\'eculas formen una
estructura regular. Por el contrario, si $T$ es baja, la interacci\'on molecular podr\'a sobreponerse al ruido t\'ermico. En este caso, 
dependiendo del potencial intermolecular, el material podr\'{\i}a condensarse en una fase suave en donde a\'un persista alguna regularidad 
a nivel molecular. Dada la enorme diversidad de estructuras moleculares naturales y artificiales, los arreglos resultantes pueden ser 
tambi\'en muy diversos y muchas de estas fases suaves estructuradas son, de hecho, CL. 

Los CL son m\'as ordenados que los l\'{\i}quidos isotr\'opicos. La formaci\'on espont\'anea de estas fases, bajo condiciones apropiadas,
puede entenderse al interpretar la ecuaci\'on~(\ref{autoensamblaje_001}) desde otro punto de vista. Si bien la aparici\'on de ordenamiento
molecular implica una disminuci\'on de la entrop\'{\i}a y un consecuente aumento de $F$, el proceso est\'a acompa\~nado de una disminuci\'on
m\'as pronunciada de la energ\'ia interna que logra llevar la energ\'{\i}a libre a un m\'{\i}nimo.

El concepto de autoensamblaje se ha vuelto muy importante no s\'olo por el estudio de los CL sino en el \'ambito m\'as amplio de la materia 
condensada suave, dentro de la cual se encuentran materiales tan diversos como los pl\'asticos, las gomas, los geles, las pinturas y la 
mayor\'{\i}a de los alimentos y tejidos biol\'ogicos. Todos estos son de enorme importancia en los niveles indrustrial y de la ciencia 
b\'asica.

% =====================================================================================================================
\section{Fases l\'{\i}quido-cristalinas \label{fases_liquidocristalinas_seccion}}
% =====================================================================================================================

Como se ha discutido en la secci\'on~\ref{autoensamblaje_seccion}, las mol\'eculas de los CL se encuentran m\'as ordenadas que las de los
l\'{\i}quidos isotr\'opicos. Sin embargo, este orden no alcanza a ser tan alto como el que presentan las mol\'eculas en un cristal s\'olido.
Entre el orden perfecto de un cristal y el desorden total de un l\'{\i}quido isotr\'opico, pueden tenerse diferentes niveles de orden 
asociados con simetr\'{\i}as que dan lugar a muchas fases distintas de CL. Estas fases de la materia se denominan en general, \emph{mesofases},
al encontrarse a medio camino entre los s\'olidos y los l\'{\i}quidos ordinarios. Algunas de \'estas se analizar\'an en esta secci\'on,
iniciando con aquella que tiene la estructura m\'as sencilla.

% ---------------------------------------------------------------------------------------------------------------------
\subsection{Fase nem\'atica \label{fase_nematica_seccion}}
% ---------------------------------------------------------------------------------------------------------------------

\begin{figure}[h]
\includegraphics[width=\linewidth]{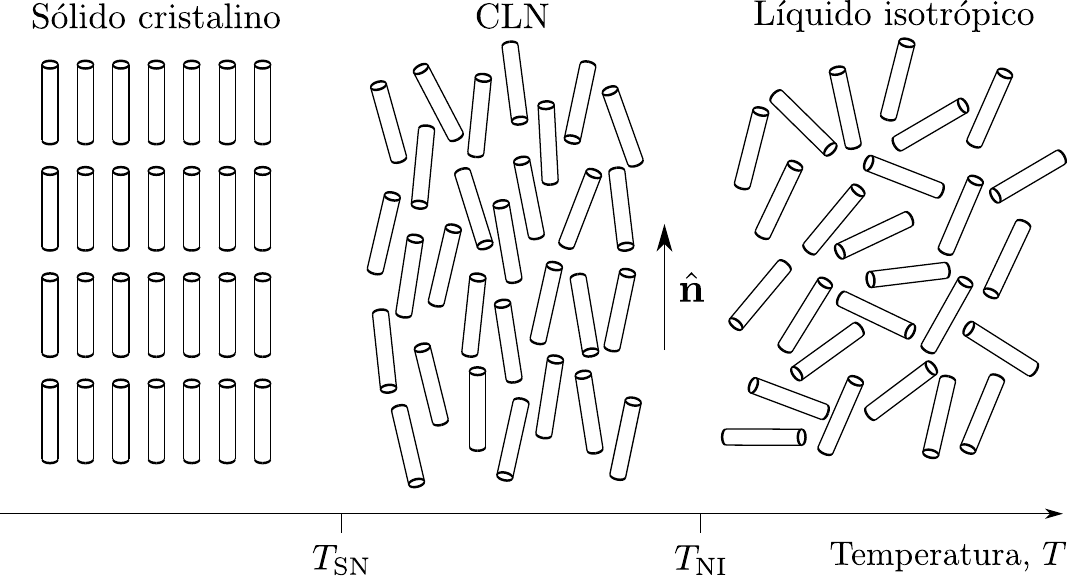}
\caption{Esquema de los arreglos moleculares en un s\'olido cristalino (izquierda), un CLN (centro) y un l\'{\i}quido isotr\'opico (derecha).
El material ilustrado transita entre los tres estados como funci\'on de la temperatura.}
\label{figura_003}
\end{figure}

La mesofase con el ordenamiento molecular m\'as simple es la mesofase \emph{nem\'atica}.~\footnote{El t\'ermino ``nem\'atico'', proviene del
griego antiguo \textit{n\'ematos}, $\nu\acute{\eta}\mu\alpha\tau o\varsigma$, que significa \textit{cabello}. La raz\'on por la cual la fase 
nem\'atica recibe este nombre es que al mirarse en un microscopio con polarizadores cruzados, se aprecian curvas delgadas obscuras que asemejan 
cabellos. El nombre fue sugerido por Marie Friedel, hija de Georges Friedel~\cite{mitov_chemphyschem_2014}. \'Este \'ultimo, despu\'es de afirmar que los CL no 
eran birrefringentes, cambi\'o de opini\'on y acept\'o la existencia de fases que combinan propiedades cristalinas y de fluidez.} 
Esta se ilustra esquem\'aticamente en la 
figura~\ref{figura_003}, en la cual se muestran tambi\'en los arreglos moleculares del s\'olido cristalino y el l\'{\i}quido isotr\'opico. 
En la figura~\ref{figura_003} se ha adoptado una escala en la cual los detalles at\'omicos de las mol\'eculas no son relevantes y su forma
se ha simplificado a la de barras r\'{\i}gidas debido a su simetr\'{\i}a. La figura~\ref{figura_003} corresponde, entonces, a un material 
calam\'{\i}tico. El eje horizontal en la figura~\ref{figura_003} indica la temperatura y el CL nem\'atico (CLN) se muestra en el centro.

A temperaturas bajas, menores a $T_{\text{SN}}$, el material forma un cristal perfecto en el que las mol\'eculas ocupan los sitios de una 
red regular y todas apuntan en la misma direcci\'on. En el cristal, el conjunto molecular tiene tanto orden en la posici\'on como en la 
orientaci\'on. En el extremo opuesto de temperaturas, cuando $T>T_{\text{NI}}$, la agitaci\'on t\'ermica hace que ambos tipos de orden se 
pierdan. En esa situaci\'on, las mol\'eculas pueden desplazarse por todo el volumen de la muestra y giran arbitrariamente, pudiendo apuntar 
con la misma probabilidad en cualquier direcci\'on. Lo m\'as notable es que para muchos materiales existe un rango 
de temperaturas, $T_{\text{SN}}<T<T_{\text{NI}}$, en el cual las mol\'eculas tienen un orden intermedio: no presentan orden en 
la posici\'on pero el autoensamblaje les permite preservar orden en la orientaci\'on. 

De manera m\'as detallada, en un CLN los centros de masa moleculares se mueven de manera similar a como lo hacen en un l\'{\i}quido 
isotr\'opico, teniendo la libertad de desplazarse por toda la muestra. Sin embargo, mientras las mol\'eculas se trasladan arbitrariamente, 
su eje de simetr\'{\i}a principal permanece orientado cerca de una direcci\'on que es com\'un al ensamble. En la figura~\ref{figura_003},
tal direcci\'on se ha representado mediante el vector unitario $\hat{\mathbf{n}}$, al cual se le llama \emph{vector director} o, simplemente,
\emph{el director}.

El preservar \emph{orden orientacional} y carecer de \emph{orden posicional} es la caracter\'istica distintiva de los CL. Esta dualidad es 
tambi\'en el rasgo m\'as valioso de los CL por lo que respecta a sus aplicaciones tecnol\'ogicas. Esto se debe a que el desorden posicional 
implica que los CL son suaves y capaces de fluir como los l\'{\i}quidos com\'unes, mientras que el orden orientacional
los hace comportarse \'opticamente como un cristal. As\'{\i}, los CL son materiales con propiedades \'opticas cristalinas que pueden 
manipularse con energ\'{\i}as muy bajas debido a que son extremadamente suaves.

La figura~\ref{figura_003} permite entender que la anisotrop\'{\i}a de las mesofases se origina por el orden orientacional que poseen. En la fase 
isotr\'opica se carece de ese orden y el material luce id\'entico (igualmente desordenado), en cualquier direcci\'on que 
se explore. Por el contrario, en un CLN, $\hat{\mathbf{n}}$ representa una direcci\'on en la que podemos esperar que las propiedades materiales 
tomen un valor especial, distinto al que se observa en la direcci\'on perpendicular. Por ejemplo, la luz se propagar\'a con velocidades 
distintas si viaja paralela o perpendicularmente al director y, consecuentemente, el CLN tendr\'a dos \'indices de refracci\'on. La luz que
viaja en la direcci\'on perpendicular a $\hat{\mathbf{n}}$ ser\'a afectada por un \'indice de refracci\'on llamado ordinario, $n_{\text{o}}$,
mientras que aquella que se propaga paralelamente a $\hat{\mathbf{n}}$ lo har\'a a una velocidad dictada por un \'indice de refracci\'on
llamado extraordinario, $n_{\text{e}}$.~\footnote{No se deben confundir los s\'imbolos para los \'indices de refracci\'on ordinario y 
extraordinario con las componenetes de $\hat{\mathbf{n}}$ para las cuales se reservar\'a la notaci\'on $n_{\alpha}$, con $\alpha = 1,2,3$.}
De manera similar, al aplicar un campo magn\'etico a un CLN, la magnetizaci\'on 
resultante es distinta si dicho campo apunta en la direcci\'on de $\hat{\mathbf{n}}$ o en la direcci\'on perpendicular.

Otra caracter\'{\i}stica importante de la fase nem\'atica es que es invariante ante rotaciones alrededor de $\hat{\mathbf{n}}$. El director 
representa el \'unico eje preferencial de esta mesofase, a la cual tambi\'en se le asigna el nombre m\'as preciso de \emph{fase nem\'atica 
uniaxial}. En la secci\'on~\ref{descripcion_matematica_seccion} se describir\'an otros aspectos y variables que son relevantes en el an\'alisis 
de los CLN. 

% =====================================================================================================================
\subsection{Fases esm\'ecticas \label{fases_esmecticas_seccion}}
% =====================================================================================================================

\begin{figure}[h]
\includegraphics[width=\linewidth]{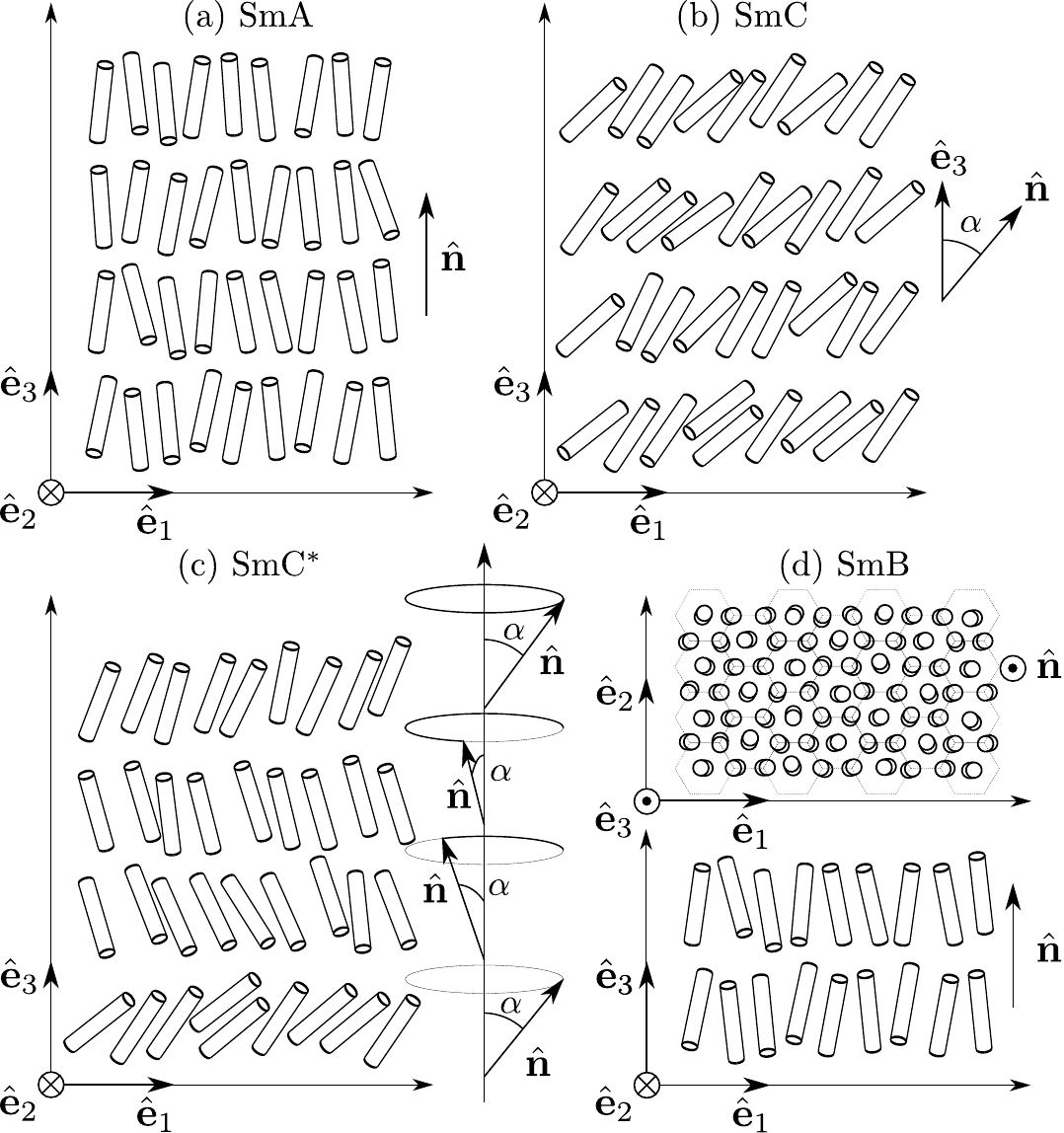}
\caption{Estructura de cuatro tipos comunes de CLE: (a)~Esm\'ectico A (SmA), (b)~Esm\'ectico C (SmC), (c)~Esm\'ectico C$^{*}$ (SmC$^{*}$) y 
(d)~Esm\'ectico B (SmB). \'Este \'ultimo se muestra en dos perspectivas que permiten apreciar mejor el acomodo molecular en las capas que forman la
fase. }
\label{figura_004}
\end{figure}

Los CLN carecen de orden posicional. Otras mesofases pueden formarse al incluir cierta cantidad de orden en la posici\'on molecular, sin que 
este llegue a ser tan alto como lo es en los cristales. La familia de los CL esm\'ecticos (CLE) se distingue porque las mol\'eculas exhiben 
orden orientacional similar al que tienen en la fase nem\'atica y, adem\'as, orden orientacional en una dimensi\'on.~\footnote{El t\'ermino ``esm\'ectico'',
tiene su origen en la palabra del griego antiguo %\textit{sm}$\bar{\text{\textit{e}}}$\textit{ktik\'os}, $\sigma\mu\eta\kappa\tau\iota\kappa\text{\textit{\'o}}\varsigma$, 
%que a su vez deriva de 
\textit{sm$\acute{\bar{\text{\textit{e}}}}$khein}, $\sigma\mu\grave{\eta}\chi\varepsilon\iota\nu$, que significa \textit{limpiar}. La causa
de esta designaci\'on es que una muestra de CLE tiene capas apiladas que pueden deslizarse unas sobre otras comport\'andose de manera similar 
a un jab\'on, el cual es, de hecho, un CLE~\cite{mitov_chemphyschem_2014}.} La figura~\ref{figura_004} 
ilustra la estructura de algunas fases esm\'ecticas representativas. A simple vista resalta el hecho de que los CLE forman capas 
igualmente espaciadas de mol\'eculas. Las mol\'eculas tienen libertad de moverse dentro de cada capa, pero la probabilidad de que transiten de 
una capa a otra es despreciable. Mientras las mol\'eculas se desplazan, permanecen orientadas a lo largo de una direcci\'on que es com\'un a la 
capa. As\'{\i}, cada capa molecular puede considerarse un CLN bidimensional mientras que el conjunto completo tiene orden posicional en una 
dimensi\'on debido a la periodicidad con la que las capas est\'an distribuidas.

Las fases esm\'ecticas se presentan a temperaturas m\'as bajas que las nem\'aticas, debido a que poseen m\'as orden. La estructura estratificada 
de los CLE fue verificada por primera vez mediante experimentos de difracci\'on de rayos X realizados por Edmont Friedel y Maurice de Broglie 
(hermano de Louis de Broglie quien es famoso por formular la hip\'otesis de la dualidad onda--part\'{\i}cula)~\cite{de_broglie_compt_rend_1923}. 
T\'{\i}picamente, la separaci\'on entre las capas moleculares en los CLE es de $2$ a $4~\text{nm}$~\cite{lagerwall_chem_phys_chem_2006}.   

La categor\'{\i}a esm\'ectica abarca m\'ultiples fases distintas que se distinguen unas de otras por el comportamiento de la orientaci\'on molecular 
promedio en las diferentes capas, as\'i como por poder presentar un tipo especial de orden posicional dentro de ellas. Las fases esm\'ecticas se 
representan mediante la abreviatura Sm (por esm\'ectico en ingl\'es), seguida de una letra asignada de acuerdo a un sistema est\'andar que refleja 
las caracter\'isticas del orden posicional y orientacional de las mol\'eculas en las capas. Las estructuras en la figura~\ref{figura_004} representan 
esquem\'aticamente a cuatro integrantes comunes de la familia esm\'ectica. 

En un esm\'ectico A, SmA, el director es perpendicular a las capas y uniforme, \textit{i. e.}, no cambia de capa a capa. Esta fase se presenta en la 
figura~\ref{figura_004}~(a) donde el sistema cartesiano expandido por la base ortonormal de vectores $\{\hat{\mathbf{e}}_{1},\hat{\mathbf{e}}_{2},\hat{\mathbf{e}}_{3}\}$ 
se ha introducido como referencia,
siendo $\hat{\mathbf{n}} \parallel \hat{\mathbf{e}}_{3}$. En un esm\'ectico C (SmC) figura~\ref{figura_004}~(b), $\hat{\mathbf{n}}$ tambi\'en es uniforme 
pero forma un \'angulo de inclinaci\'on $\alpha$ con repecto a $\hat{\mathbf{e}}_{3}$. En la fase esm\'ectica C$^{*}$, SmC$^{*}$, tambi\'en llamada \emph{fase 
esm\'ectica quiral}, la proyecci\'on de $\hat{\mathbf{n}}$ a lo largo del eje $\hat{\mathbf{e}}_{3}$ es constante, al igual que en la fase SmC, pero
$\hat{\mathbf{n}}$ rota de capa a capa, dibujando una h\'elice alrededor de $\hat{\mathbf{e}}_{3}$, tal como se aprecia en la figura~\ref{figura_004}~(c).
La figura~\ref{figura_004}~(d) ilustra a un esm\'ectico B, SmB, cuya estructura es muy similar a la de la fase SmA, s\'olo que en la primera las mol\'eculas
dentro de cada capa se acomodan de acuerdo con un patr\'on hexagonal.

Dado que la estructura de los CLE es peri\'odica en la direcci\'on de $\hat{\mathbf{e}}_{3}$, la densidad de masa, $\rho$, puede expandirse en
una serie de Fourier
\begin{equation}
\rho(x_{3}) = \rho_{0} + \sum_{m = 1}^{\infty} \rho_{m} \cos\left(2\pi m  \frac{x_{3}}{d}\right),
\label{fases_esmecticas_001}
\end{equation}
donde $\rho_{0}$ representa la densidad promedio, mientras que 
\begin{equation}
\rho_{m} = \frac{1}{d} \int_{-d/2}^{d/2}dx_{3}\, \rho(x_{3}) \cos\left(2 m \pi\frac{x_{3}}{d}\right),
\label{fases_esmecticas_002}
\end{equation}
es el $m$--\'esimo coefficiente de Fourier y $d$ es la distancia de separaci\'on entre las capas esm\'ecticas. La expansi\'on dada por las 
ecuaciones~(\ref{fases_esmecticas_001}) y (\ref{fases_esmecticas_002}) es exacta dada la periodicidad de la densidad. El orden translacional de 
los CLE se mide en funci\'on de los coeficientes $\rho_{m}$. Usualmente, el primer harm\'onico domina la modulaci\'on de la densidad y la expansi\'on
en la ecuaci\'on~(\ref{fases_esmecticas_001}) puede reducirse a
\begin{equation}
\rho(x_{3}) = \rho_{0} + \rho_{1} \cos\left(2\pi \frac{x_{3}}{d}\right).
\label{fases_esmecticas_003}
\end{equation}

El coeficiente $\rho_{1}$
%al primero de ellos
%\begin{equation}
%\sigma_{1} = \frac{1}{d} \int_{-d/2}^{d/2}dz\, \rho(z) \cos\left(2\pi\frac{z}{d}\right),
%\label{fases_esmecticas_002}
%\end{equation}
tiene la caracter\'{\i}stica de anularse cuando no hay orden posicional. Esto puede verificarse al reemplzar $\rho(x_{3})$ por el valor
constante $\rho_{0}$ y $m=1$ en la ecuaci\'on~(\ref{fases_esmecticas_002}) e integrar, 
\begin{equation}
\rho_{1} = \frac{1}{d} \int_{-d/2}^{d/2}dx_{3}\, \rho_{0} \cos\left(2 \pi\frac{x_{3}}{d}\right) = 0.
\nonumber
\end{equation}

Adem\'as, cuando la distancia entre capas es constante, $\rho(x_{3})$ tiene m\'aximos pronunciados en $x_{3} = j\,d$, para $j \in \mathbb{Z}$. 
Lo anterior implica valores grandes de $\rho_{1}$, quien sin embargo, no puede ser mayor que $\rho_{0}$, pues ello implicar\'{\i}a densidades negativas.
En otras palabras, para un sistema muy estructurado en capas, $\rho_{1}$ se aproxima a $\rho_{0}$, sin sobrepasarlo.

Para medir el grado de orden posicional unidimensional puede definirse el llamado \emph{par\'ametro de orden de espaciamiento entre capas} 
como
\begin{equation}
S_{\text{LS}} = \frac{\rho_{1}}{\rho_{0}}.
\label{fases_esmecticas_004}
\end{equation}

Por lo discutido anteriormente se tiene $S_{\text{LS}} \in [0,1]$. $S_{\text{LS}}$ se anular\'a cuando no existen capas y ser\'a mayor que cero
en la fases esm\'ecticas. Los valores t\'ipicos de $S_{\text{LS}}$ para algunas fases SmA, SmC y SmC$^{*}$, se encuentran entre $0.2$ y 
$0.8$~\cite{mcmillan_phys_rev_a_1972,mukherjee_liq_cryst_2019,blinov_2011}.
Valores de $S_{\text{LS}}$ cercanos a $1$ indican un alto grado de ordenamiento molecular en capas.

% ---------------------------------------------------------------------------------------------------------------------
\subsection{Fase colest\'erica\label{fase_colesterica_seccion}}
% ---------------------------------------------------------------------------------------------------------------------

\begin{figure}[h]
\includegraphics[width=\linewidth]{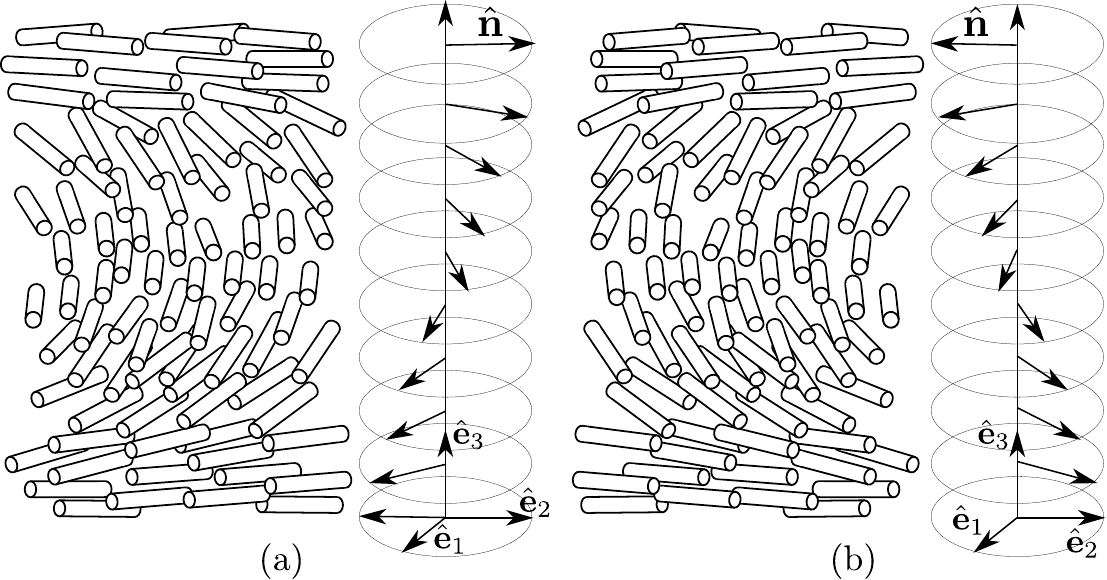}
\caption{Estructura de la fase colest\'erica. El director gira de manera continua describiendo una h\'elice alrededor de un eje fijo que en este caso
coincide con el vector $\hat{\mathbf{e}}_{3}$. (a) Corresponde a una fase colest\'erica derecha y (b) a una fase colest\'erica izquierda.}
\label{figura_005}
\end{figure}

Otra mesofase importante es la llamada fase \emph{colest\'erica} o \emph{quiral}.~\footnote{Las fases ``colest\'ericas'' se denominan as\'i por 
tener carcater\'isticas similares a las del benzoato de colesterilo descubierto por Reinitzer. El t\'ermino ``quiral'' es un nombre alternativo 
que deriva de la palabra griega para \textit{mano}, \textit{kheir}, $\chi\varepsilon\acute{\iota}\rho$.} La estructura de esta fase, la cual se ilustra en la
figura~\ref{figura_005}, se caracteriza porque los ejes moleculares se alinean alrededor de una orientaci\'on com\'un cuando se observan
en planos con alguna coordenada constante. Adem\'as, conforme se recorren diferentes planos, el director gira peri\'odicamente y dibuja una 
h\'elice alrededor del eje normal a los planos. En la figura~\ref{figura_005}, los planos referidos est\'an descritos por la ecuaci\'on 
$x_{3} = \text{cte}$.

Es usual encontrar esquemas en internet y la literatura especializada en los que los CL colest\'ericos (CLC) se 
representan mediante un conjunto de planos apilados dentro de los cuales las mol\'eculas est\'an casi perfectamente 
alineadas~\cite{mallia_chem_soc_rev_2004,oswald_2009,matsui_2011,soni_2013}. En tales esquemas, la separaci\'on 
entre planos se considera constante y al pasar de un plano al siguiente tambi\'en se supone un giro constante de la orientaci\'on promedio. 
Es importante resaltar que este tipo de representaciones, si bien sirven para visualizar la estructura helicoidal de los CLC, no pueden 
considerarse f\'isicamente v\'alidas ya que las mol\'eculas no forman capas en la fase colest\'erica. En vez de ello, est\'an distribu\'idas 
uniformemente en el espacio. Por tanto, el director no cambia de manera abrupta de un plano a otro, sino que lo hace de manera continua.

\begin{figure}[h]
\centering
\includegraphics[scale=0.60]{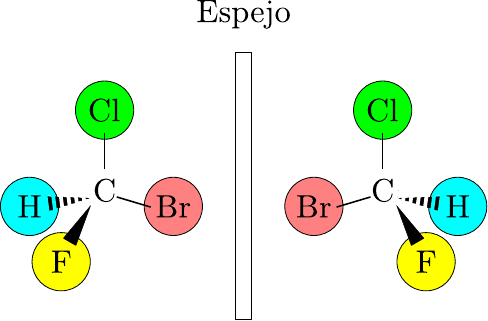}
\caption{Ejemplo de una mol\'ecula quiral, el bromoclorofluorometano.}
\label{figura_006}
\end{figure}

Las mol\'eculas que forman las fases colest\'ericas son anisotr\'opicas y \textit{quirales}. La quiralidad es una propiedad asociada con 
estructuras asim\'etricas que no pueden hacerse coincidir, mediante ninguna rotaci\'on, con su imagen reflejada en un espejo plano.
Una forma sencilla de entender la quiralidad es mediante la laterilidad de nuestras manos. La mano izquierda es id\'entica a la imagen 
en el espejo de la mano derecha y viceversa. Aunque las manos derecha e izquierda tienen las mismas caracter\'isticas, no podemos hacer 
que todas ellas coincidan simult\'aneamente, no importa qu\'e tanto las giremos. Por la misma raz\'on, no podemos calzarnos un guante
derecho en la mano izquierda y nos resulta inc\'omodo estrachar con la mano izquierda, la mano derecha de otra persona al saludarla. El 
benzoato de colesterilo, del que ya hemos hablado por ser la primera substancia en la que se indetificaron los CL, tiene mol\'eculas 
quirales. Un grupo importante de mol\'eculas quirales pueden formarse a partir de un \'atomo de carbono enlazado con cuatro elementos
o sustituyentes diferentes. Como ejemplo se muestra en la figura~\ref{figura_006} el caso particular del bromoclorofluorometano en
el que el carbono se enlaza a \'atomos de hidr\'ogeno, fl\'ur, bromo y cloro. Los enlaces pueden disponerse en dos variedades distintas
que resultan en dos mol\'eculas que forman la imagen especular una de la otra y que no pueden hacerse coincidir por ninguna rotaci\'on.
Las dos variaciones con esta caracter\'istica se conocen como en qu\'imica como \textit{enanti\'omeros}.

Uno de los par\'ametros que caracterizan la estructura colest\'erica es el sentido del giro del director, el cual puede ser derecho o izquierdo.
Para distinguir entre estos casos, puede mirarse el sistema a lo largo del eje de la h\'elice. En un CLC derecho, la rotaci\'on del director 
ocurre en el sentido en el que giran las manecillas del reloj. En un CLC izquierdo, esta rotaci\'on ocurren en el sentido contrario. %ido de las
%manecillas del reloj, mientras que en un CLC izquierdo los dedos de la mano izquierda girar\'an en el sentido contrario al de las manecillas del reloj.
Los adjetivos \textit{dextr\'ogiro} y \textit{lev\'ogiro} se utilizan con frecuencia para nombrar, de manera respectiva, a los dos casos previos.
Puedes verificar que el CLC ilustrado en la figura~\ref{figura_005}~(a) es derecho, mientras que el de la figura~\ref{figura_005}~(b) es izquierdo.

Otro par\'ametro importante de las fases colest\'ericas es la distancia a lo largo del eje de la h\'elice en la cual el director realiza un giro de 
$2 \pi$ radianes. A esta se le llama en ingl\'es \textit{pitch}, siendo ``tono'' o ``paso'' dos traducciones comunes al espa\~nol. El \textit{pitch}, 
$p$, es una medida de qu\'e tan cercanos est\'an los giros del director entre s\'{\i}. Sus valores espec\'ificos dependen de la estructura qu\'imica 
de las mol\'eculas que forman la fase y  abarcan un rango muy amplio que va desde algunas decenas de nan\'ometros hasta varios micr\'ometros. 
Notablemente, este intervalo cubre la escala correspondiente a las longitudes de onda del espectro visible (de los $380~\text{nm}$ a los 
$750~\text{nm}$), lo que resulta enormemente \'util en las aplicaciones.

Los CLC son \'opticamente multifuncionales debido a su estructura peri\'odica. La propiedad \'optica m\'as popular y utilizada de los
CLC es la reflexi\'on selectiva de la luz, siendo esta, la capacidad de reflejar la luz en colores espec\'ificos que dependen de $p$. Este efecto 
se genera por la interacci\'on de luz con la estructura helicoidal del CLC. Un haz de luz no polarizada consiste de un
campo el\'ectrico y un campo magn\'etico que oscilan en varias direcciones. Cuando este haz incide en el CLC, los campos experimentar\'an un \'indice
de refracci\'on que var\'ia peri\'odicamente. Para algunas longitudes de las ondas electromagn\'eticas la interacci\'on ser\'a destructiva y para otras 
constructiva. M\'as espec\'ificamente, \'unicamente aquellas ondas cuya longitud se ajuste con $p$ interferir\'an constructivamente y ser\'an reflejadas, 
mientras que el resto interferir\'an destructivamente y s\'olo podr\'an seguir transmiti\'endose. As\'i, la luz reflejada por el CLC ser\'a de un color
particular determinado por $p$. Es posible demostrar que la longitud de onda para la cual ocurre la m\'axima reflexi\'on es, de hecho,
\begin{equation}
\lambda_{\text{max}} = n\,p ,
\label{fase_colesterica_000}
\end{equation}
donde $n = \left( n_{\text{o}} + n_{\text{e}} \right)/2$ es el \'indice de refracci\'on promedio.

El valor del \textit{pitch} se modifica al someter al CLC a cambios en la temperatura, esfuerzos mec\'anicos y otros factores. A su vez, se modifica 
$\lambda_{\text{max}}$, haciendo que el CLC luzca de diferentes colores para diferentes condiciones externas. Este fen\'omeno permite aplicar las 
mesofases colest\'ericas en sensores de temperatura, humedad o flujo basados en cambios de 
colores~\cite{smith_exp_in_fluids_2001,stasiek_opt_laser_technol_2006,saha_chem_commun_2012}.

La estructura helicoidal del director tambi\'en les da a los CLC la habilidad de rotar el plano de polarizaci\'on de la luz
que viaja a trav\'es de ellos. Dicho plano contiene las oscilaciones del campo el\'ectrico del haz de luz. La habilidad para rotar el plano 
de polarizaci\'on es conocida como \textit{actividad \'optica} y no es excluisva de los CLC pues tambi\'en lo
presentan cristales como el quarzo y suspensiones de mol\'eculas quirales como az\'ucar en agua. Sin embargo, la rotaci\'on \'optica,
definida como el \'angulo que gira el plano de polarizaci\'on por unidad de longitud que la luz atraviesa, es excepcionalmente m\'as alta 
en los CLC que en otros materiales. Para la luz visible, la rotaci\'on \'optica en los CLC puede ir de $1000^{\circ}$ a 
$100000^{\circ}~\text{mm}^{-1}$, mientras que los valores t\'ipicos para otras substancias van de los $0.01^{\circ}$ a los 
$100^{\circ}~\text{mm}^{-1}$. 

Las propiedades \'opticas excepcionales de los CLC encuentran m\'ultiples aplicaciones en l\'aseres de color 
ajustable~\cite{coles_nat_photonics_2010}, ventanas inteligentes~\cite{mitov_adv_mater_2012} y filtros de rayos UV~\cite{oh_rsc_adv_2021}, s\'olo por 
nombrar algunas. A pesar de todo esto, la principal importancia de los CLC radica quiz\'a
en otro aspecto, a saber: que est\'an universalmente presentes en los seres vivos. Por lo tanto, las fases colest\'ericas
juegan un papel fundamental en muchos procesos biol\'ogicos~\cite{mitov_soft_matter_2017}. S\'olo por mencionar algunos ejemplos, 
el ADN, la celulosa, el col\'ageno y muchos virus pueden exhibir organizaci\'on colest\'erica. Muchas de las estructuras que le 
dan forma y estructura a los sistemas biol\'ogicos resultan del arreglo helicoidal de bloques anisotr\'opicos como mol\'eculas, 
macromol\'eculas o microfibrillas que, sin embargo, no guardan orden en la posici\'on. 

% =====================================================================================================================
\section{Cristales l\'{\i}quidos liotr\'opicos \label{cristales_liquidos_liotropicos_seccion}}
% =====================================================================================================================

Es com\'un clasificar a los cristales l\'{\i}quidos como termotr\'opicos o liotr\'opicos. En los primeros, el orden molecular
se determina por la temperatura. Todos los ejemplos de fases l\'{\i}quido-cristalinas que hemos presentado en las 
secciones~\ref{fase_nematica_seccion} a \ref{fase_colesterica_seccion} han sido termotr\'opicos. Sin embargo, los CL que m\'as
abundan en la naturaleza son liotr\'opicos~\cite{hirst_2012}. Estos se forman al cambiar la conecntraci\'on de cierto tipo de
mol\'eculas en un solvente.

La gran mayor\'ia de las fases liotr\'opicas se producen en soluciones acuosas y el mecanismo principal que promueve su existencia
es una combinaci\'on de dos efectos conocidos como \textit{efecto hidrof\'ilico} y \textit{efecto hidrof\'obico}, los cuales
explicaremos a continuaci\'on.

\subsection{Efectos hidrof\'ilico e hidrof\'obico \label{efectos_hidro_seccion}}

Los efectos hidrof\'ilico e hidrof\'obico deben su existencia a %principales responsables de su existencia 
%son los llamados puentes de hidr\'ogeno. Estos, a su vez, se originan por 
la naturaleza polar de las mol\'eculas de agua, $\text{H}_{2}\text{O}$. En ellas, 
los \'atomos de hidr\'ogeno quedan parcialmente desprovistos de sus electrones y exhiben carga positiva, mientras el \'atomo de 
ox\'igeno, al recibir estos electrones, muestra una carga efectiva negativa. Esto ocasiona que cada mol\'ecula de agua sea un 
peque\~no dipolo el\'ectrico.~\footnote{Recordemos que un dipolo el\'ectrico consiste esencialmente de dos cargas el\'enctricas
de la misma magnitud y signo opuesto, separadas por una distancia fija.}

Cuando varias mol\'eculas de agua se encuentran cerca, tienden a generar estructuras ordenadas debido a las interacciones el\'ectricas 
entre los \'atomos con carga efectiva: b\'asicamente un hidr\'ogeno de una mol\'ecula es atra\'ido por el ox\'igeno de otra. Estas 
fuerzas de atracci\'on se conocen como \textit{puentes de hidr\'ogeno}. En la fase l\'iquida del agua, los puentes de hidr\'ogeno 
son muy d\'ebiles en comparaci\'on con otro tipo de interacciones y la agitaci\'on t\'ermica. Debido a esto, las estructuras ordenadas
que se forman a partir de ellos se destruyen muy f\'acilmente y tienen un periodo de existencia muy corto. No obstante, los puentes de 
hidr\'ogeno juegan un papel crucial cuando otras mol\'eculas se sumergen en el agua. 

Si se trata de mol\'eculas polares, estas podr\'an formar tambi\'en puentes de hidr\'ogeno con las de agua.
Estas mol\'eculas son solubles en agua y se les llama \textit{hidrof\'ilicas}.
Por el contrario, al introducir mol\'eculas no polares en agua, no se formar\'an puentes de hidr\'ogeno. En vez de ello, las mol\'eculas 
de agua deber\'an formar una estructura ordenada alrededor de cada mol\'ecula intrusa, lo que causa una disminuci\'on en la entrop\'ia del 
sistema y un costo energ\'etico por haberlas introducido. Debido a que existe una barrera de energ\'ia asociada con el proceso de introducir 
mol\'eculas no polares en agua, se dice que estas son \textit{hidrof\'obicas}. Las substancias formadas por este tipo de mol\'eculas no son solubles 
en agua. Un ejemplo son los aceites o hidrocarburos, formados por \'atomos de hidr\'ogeno y carbono unidos por enlaces covalentes en una cadena  
que es no polar. Debido a esto, mezclar aceites y agua no es sencillo.

\begin{figure}[h]
\centering
\includegraphics[scale=1.40]{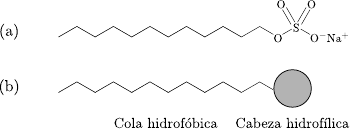}
\caption{(a) La estructura molecular del laurilsulfato s\'odico, que es la base de muchos productos de higiene personal, combina una cadena
hidrocarbonada no polar (parte hidrof\'ofica), con una secci\'on polar (parte hidrof\'ilica). (b) Representaci\'on simplificada de una
mol\'ecula anfif\'ilica.}
\label{figura_007}
\end{figure}

Cierto tipo de mol\'eculas naturales y artificiales combinan en su estrutura una parte hidrof\'ilica y otra hidrof\'obica. A estas mol\'eculas se les 
llama \textit{anfif\'ilicas}. Como ejemplo consideremos el laurilsulfato s\'odico, tambi\'en llamado dodecilsulfato s\'odico, con el que interactuamos
todos los d\'ias por ser un compuesto com\'un en los productos de higiene personal como los champ\'us, los jabones de ba\~no y las pastas de dientes. 
La parte hidrof\'ilica de esta mol\'ecula, que se ilustra en la figura~\ref{figura_007}, es una ``cabeza polar'' formada por un \'atomo de azufre unido 
a cuatro de ox\'igeno y uno de sodio. La parte hidrof\'obica, es una ``cola hidrocarbonada'' de doce \'atomos de carbono. 

%----------------------------------------------------------------------------------------------------------------------
\subsection{Algunas fases liotr\'opicas \label{algunas_fases_liotropicas_seccion}}
%----------------------------------------------------------------------------------------------------------------------

Al sumergir mol\'eculas anfif\'ilicas en agua a concentraciones bajas, la configuraci\'on energ\'eticamente m\'as favorable es aquella en la que sus cabezas 
est\'an en contacto con el agua pero sus colas no. As\'i, las mol\'eculas anfif\'ilicas tienden a dirigirse hacia la superficie y se orientan de tal manera 
que sus cabezas quedan dentro del agua y sus colas apuntando hacia afuera. Cabe mencionar que poseer una secci\'on hidrof\'ilica y una hidrof\'obica es 
propiedad esencial de los materiales tensoactivos o surfactantes, que son substancias que, precisamente, se localizan en la superficie que separa
a dos fluidos inmiscibles, \textit{e. g.}, agua y aire o agua y aceite, y act\'uan sobre esa superficie modificando la tensi\'on superficial.
As\'i, la mayor\'ia de las mol\'eculas anfif\'ilicas se localizar\'an en la superficie del agua,
mientras que el resto se encontrar\'an dilu\'idas en el volumen realizando movimientos aleatorios permitidos por las fluctuaciones en la energ\'ia y la 
agitaci\'on t\'ermica. A este estado se le llama una \textit{fase dispersa}.

Cuando el n\'umero de mol\'eculas anfif\'ilicas aumenta, la superficie del agua se satura de ellas y comienza a aumentar el n\'umero de las que est\'an disueltas 
en el volumen. Entonces, \'estas podr\'an actuar colectivamente para adoptar configuraciones que tambi\'en les sean m\'as fovorables energ\'eticamente. Este es un
mecanismo de autoensamblaje que da lugar a una gran variedad de arreglos conocidos como fases liotr\'otipcas debido a que su ocurrencia est\'a dictada por la
concentraci\'on de los solutos anfif\'ilicos, $c$.~\footnote{El t\'ermino ``liotr\'opico'' proviene del griego \textit{lyo} ($\lambda\acute{\upsilon\omega}$) que
significa \textit{disolver} y \textit{tr\'opos} ($\tau\rho\acute{o}\pi o \varsigma$) que significa \textit{direcci\'on} o \textit{sentido}.} 
Las fases liotr\'opicas suelen clasificarse como CL ya que sus mol\'eculas t\'ipicamente presentan orden de corto alcance, \textit{i. e.}, un orden que abarca distancias
comparables a las de un conjunto de unidades moleculares y que no se extiende a distancias macrosc\'opicas.

\begin{figure}[h]
\centering
\includegraphics[scale=1.00]{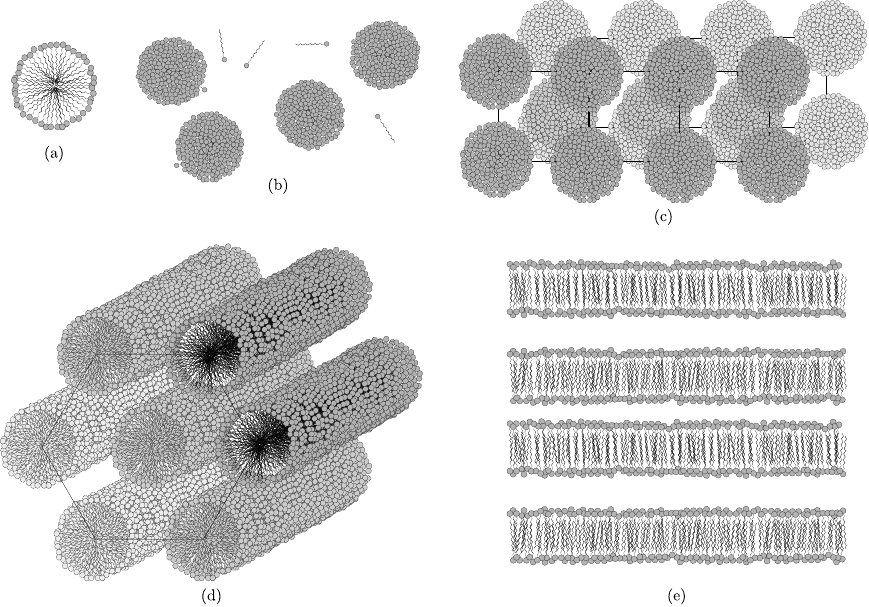}
\caption{(a) Corte transversal de una micela, esfera formada por mol\'eculas anfif\'ilicas donde las cabezas hidrof\'ilicas
protegen a las colas hidrof\'obicas del contacto con el agua. (b) Fase micelar. (c) Micelas en un arreglo c\'ubico centrado en
el cuerpo. Aqu\'i se han representado micelas en diferentes tonos de gris para distinguir su acomodo espacial. (d) Micelas cil\'indricas 
en un arreglo hexagonal. Al igual que en el caso (c), diferentes tono s de gris se han utilizado para distinguir las difrentes micelas. 
(e) Conte transversal de una fase lamelar. }
\label{figura_008}
\end{figure}

En particular, cuando aumenta $c$, se alcanza un punto en el que el n\'umero de mol\'eculas anfif\'ilicas es suficiente para que estas formen estructuras esf\'ericas llamadas
\textit{micelas}. En cada micela, las colas hidrof\'obicas se agrupan en el interior y son protegidas del agua por un cascar\'on de cabezas hidrof\'ilicas, como se ilustra 
esquem\'aticamente en la figura~\ref{figura_008}~(a). El estado en el que se forman micelas a trav\'es de la soluci\'on se conoce como \textit{fase micelar}. En ella, las micelas 
siguen trayectorias aleatorias y son capaces de absorber o liberar mol\'eculas anfif\'ilicas, las cuales pueden encontrarse a\'un, en un n\'umero reducido, vagando en el 
volumen de agua. La fase micelar se ilustra esquem\'aticamente en la figura~\ref{figura_008}~(b).

%\end{multicols}

%\medline
%\begin{multicols}{2}
%\noindent

La transici\'on de la fase diluida a la fase micelar ocurre a un valor de $c$ conocido como \textit{concentraci\'on micelar cr\'itica}~(CMC). \'Este depende de la estructura espec\'ifica 
de las mol\'eculas anfif\'ilicas. En general, para un grupo polar fijo, la reducci\'on de la energ\'ia es m\'as notoria y las micelas son m\'as estables si la longitud de las cadenas 
hidrof\'obicas es mayor, por lo que CMC disminuye para solutos anfif\'ilicos m\'as largos~\cite{dierking_crystals_2020}. 
Si los grupos polares ocupan la superficie esf\'erica de manera muy escasa, permitir\'an a 
las mol\'eculas de agua penetrar y desestabilizar la micela. La formaci\'on de las micelas obedece a un balance muy delicado entre la estructura de los 
solutos anfif\'ilicos, su interacci\'on y la reducci\'on de la energ\'ia ocasionada por su empaquetamiento. Para cada geometr\'ia molecular, la micela se forma con un radio \'optimo 
lo suficientemente grande como para que las cabezas cubran su superficie sin dejar huecos, pero no tan grande como para que su interior no pueda ser llenado por las colas hidrof\'obicas. 

Cuando aumenta $c$, crece el n\'umero de micelas y la interacci\'on entre ellas se vuelve relevante. Al valor de $c$ m\'as alto correspondiente a la fase micelar,
las micelas pueden empaquetarse para ocupar el espacio disponible de manera \'optima. Entonces, se acomodan en las posiciones de una red c\'ubica centrada en el cuerpo, 
como la que se ilustra en la figura~\ref{figura_008}~(c), que es similar al arreglo cristalino del mismo nombre~\cite{kittel_2015}. Este acomodo le permite a las micelas ocupar el $68$\% del volumen del sistema.

Adem\'as de las micelas, pueden formarse otras estructuras que minimicen la energ\'ia y permitan una ocupaci\'on \'optima del espacio a concentraciones a\'un m\'as altas. 
Por ejemplo, las mol\'eculas pueden proteger sus colas hidrof\'obicas concentr\'andose tambi\'en en estructuras cil\'indricas muy alargadas como las que se ilustran en la 
figura~\ref{figura_008}~(d). A su vez, estas \textit{micelas cil\'indricas} se agrupan en un arreglo hexagonal que les permite ocupar hasta un $91$\% del volumen del sistema. Tal
arreglo recibe el nombre de \textit{fase hexagonal}. A concentraciones todav\'ia m\'as altas se da lugar a otra fase conocida como \textit{fase lamelar}, en la cual las
mol\'eculas anfif\'ilicas se agrupan en diversas capas dobles que exponen a los grupos polares al agua y protegen a las colas hidrof\'ilicas, tal como se muestra en la 
figura~\ref{figura_008}~(e). Un ejemplo muy conocido e importante de estructuras lamelares son las bicapas de las membranas celulares en la cuales el rol 
anfif\'ilico lo juegan fosfol\'ipidos con un grupo hidrof\'ilico fosfato y dos colas hidrof\'obicas de \'acido graso~\cite{raicu_2008}.

Si $c$ aumenta m\'as, se da lugar a las llamadas fases invertidas. Por ejemplo, la fase hexagonal invertida, que consiste en cilindros de agua rodeados de 
tensoactivo, y la fase micelar inversa, en la que gotas de agua se encuentran rodeadas por tensoactivo. 

Por brevedad, no discutiremos m\'as detalles acerca de las fases liotr\'opicas y a partir de este punto nos enfocaremos en profundizar sobre las caracter\'isticas de
la fase termotr\'opica nemt\'atica que es la m\'as sencilla de todas las fases l\'iquido-cristalinas.

% =====================================================================================================================
\section{Descripci\'on matem\'atica del orden orientacional \label{descripcion_matematica_seccion}}
% =====================================================================================================================

Con el prop\'osito de estudiar a los CL a un nivel f\'{\i}sico y matem\'atico que permita predecir acertadamenete su comportamiento, 
se introducen nuevos conceptos y teor\'{\i}as que describen simult\'aneamente sus caracter\'{\i}sticas de fluidez y estructurales, 
haciendo \'enfasis en su propiedad distintiva: el orden orientacional de las mol\'eculas. A continuaci\'on discutiremos sobre las
variables que se utilizan para describir dicho orden, restringi\'endonos al caso de los CLN por ser aquellos con la estructura m\'as 
simple.

% --------------------------------------------------------------------------------------------------------------------
\subsection{Director \label{director_seccion}}
% --------------------------------------------------------------------------------------------------------------------

Una manifestaci\'on del orden orientacional es la existencia del director, cuya definici\'on  preliminar ha sido la de representar la 
orientaci\'on com\'un a lo largo de la cual tienden a dirigirse las mol\'eculas. Usualmente, se describe a $\hat{\mathbf{n}}$ como la 
orientaci\'on molecular promedio. Sin embargo, esta descripci\'on requiere de muchas precisiones. 

Primero, es correcto que para establecer el valor de $\hat{\mathbf{n}}$ se requiere un n\'umero grande de mol\'eculas sobre 
el cual las orientaciones deben promediarse de alguna manera. En este sentido, 
$\hat{\mathbf{n}}$ est\'a en el mismo nivel de descripci\'on que los campos de las teor\'{\i}as de la materia continua. Para 
definir $\hat{\mathbf{n}}$ como un campo vectorial, es decir, como una variable que puede tomar valores distintos de un punto 
a otro, se establece un elemento de volumen, $\Delta V$, alrededor de una posici\'on $\mathbf{r}$. Posteriormente, se calcula 
el promedio de las orientaciones moleculares dentro de $\Delta V$, en la forma que se explicar\'a m\'as adelante, y se toma el 
l\'{\i}mite $\Delta V \rightarrow 0$.
%
%El cociente 
%de la masa en el interior del elemento, $M$, entre $\Delta V$ aproxima la densidad, $\rho$. \'Esta se define como un campo vectorial, es decir, como una 
%variable que puede tomar valores distintos de un punto a otro, al hacer tender el volumen a cero,
%\begin{equation}
%\rho(\mathbf{r}) = \lim_{\Delta V\rightarrow 0} \frac{M}{\Delta V},
%\nonumber
%\end{equation}
%ya que b
Este proceso l\'imite tiene el mismo significado que en otras \'areas como la mec\'anica de fluidos o la electrodin\'amica,
donde se definen los campos de densidad de masa y carga, flujo, magnetizaci\'on, etc. El l\'{\i}mite $\Delta V \rightarrow 0$, implica 
considerar elementos de volumen muy peque\~nos como para poder asignar un valor local del director, $\hat{\mathbf{n}}(\mathbf{r})$, pero 
lo suficientemente grandes como para contener un n\'umero importante de mol\'eculas y dar sustento estad\'{\i}stico al promedio. El 
director es, por lo tanto, una variable macrosc\'opica.

Por otra parte, es muy importante enfatizar que $\hat{\mathbf{n}}$ no es el promedio algebr\'aico directo de las orientaciones moleculares. 
La raz\'on es que, con una excelente aproximaci\'on, \'estas pueden considerarse no polares, de tal manera que si $\hat{\mathbf{u}}_{i}$ denota la
orientaci\'on de la $i$--\'esima mol\'ecula, las probabilidades asociadas con los vectores $\hat{\mathbf{u}}_{i}$ y $-\hat{\mathbf{u}}_{i}$ 
son id\'enticas. Esto hace que en cada elemento de volumen haya esencialemente el mismo n\'umero de mol\'eculas apuntando en una direcci\'on y 
en la direcci\'on contraria y que el promedio de los vectores $\hat{\mathbf{u}}_{i}$ se cancele id\'enticamente. En la 
secci\'on~\ref{parametro_de_orden_tensorial_seccion} discutiremos cual es la manera matem\'aticamente correcta de calcular $\hat{\mathbf{n}}$.

El hecho de poder intercambiar los ejes moleculares $\hat{\mathbf{u}}_{i}$ por $-\hat{\mathbf{u}}_{i}$ sin producir ninguna alteraci\'on en la 
fase, da lugar a la llamada \textit{simetr\'ia nem\'atica}. De acuerdo con \'esta, tampoco existe ninguna consecuencia f\'isica de invertir el 
director por su vector antiparalelo. 

En ausencia de fuerzas, $\hat{\mathbf{n}}$ puede apuntar con igual probabilidad en cualquier direcci\'on. Sin embargo, en la pr\'actica, 
los CLN est\'an sometidos a campos de flujo o electromagn\'eticos, as\'{\i} como a restricciones impuestas por superficies. Todos estos 
efectos pueden obligar al director a adoptar una direcci\'on espec\'ifica. Al escogerse esa
direcci\'on, se rompe la simetr\'ia del espacio tridimensional pues puede identificarse un eje preferente. Sin embargo, se preservan
simetr\'ias parciales. Cualquier rotaci\'on alrededor del director deja al sistema inalterado. Como los estados $\hat{\mathbf{n}}$ y
$-\hat{\mathbf{n}}$ son equivalentes, el CLN es invariante ante cualquier reflexi\'on en un plano perpendicular a $\hat{\mathbf{n}}$.
Adem\'as, la reflexi\'on en la direcci\'on de cualquier eje perpendicular a $\hat{\mathbf{n}}$ tampoco altera al sistema. As\'i, adem\'as
de la simetr\'ia $\hat{\mathbf{n}}\rightarrow -\hat{\mathbf{n}}$, tambi\'en se tiene la simetr\'ia de reflexi\'on
$\mathbf{r}\rightarrow -\mathbf{r}$. Las ecuaciones que describen el comportamiento  de los CLN deden cumplir siempre con  
%invariantes ante el intercambio entre $\hat{\mathbf{n}}$ y $-\hat{\mathbf{n}}$.
%En la expansi\'on para la densidad de energ\'ia el\'astica s\'olo estar\'an permitidos aquellos
%t\'ermino que cumplan con  
estas dos simetr\'ias.

% --------------------------------------------------------------------------------------------------------------------
\subsection{Par\'ametro de orden escalar \label{parametro_de_orden_seccion}}
% --------------------------------------------------------------------------------------------------------------------

\begin{figure}[h]
\includegraphics[width=\linewidth]{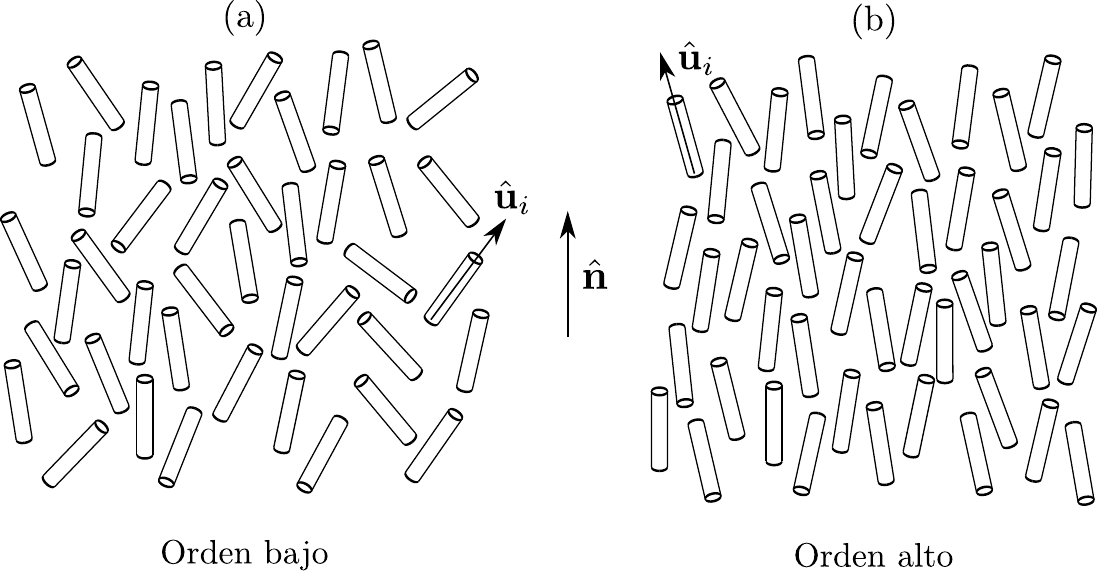}
\caption{Dos CLN con el mismo director pero diferente cantidad de orden. El vector $\hat{\mathbf{u}}_{i}$ indica la orientaci\'on
de la $i$-\'esima mol\'ecula del ensamble. }
\label{figura_009}
\end{figure}

Un aspecto importante de resaltar, es que $\hat{\mathbf{n}}$ no brinda una representaci\'on completa del estado de ordenamiento de un CLN. 
Para ilustrar esto,
pueden considerarse las dos fases nem\'aticas en las figuras~\ref{figura_009}~(a) y (b), las cuales tienen el mismo director y, sin embargo,
exhiben un orden diferente. En la fase de la figura~\ref{figura_009}~(a), las mol\'eculas pueden tener orientaciones m\'as alejadas de 
$\hat{\mathbf{n}}$. Esta fase luce mucho m\'as desordenada que la del caso (b), donde la probabilidad de observar orientaciones
$\hat{\mathbf{u}}_{i}$ cercanas a $\hat{\mathbf{n}}$ es m\'as alta. Para cuantificar esta caracter\'{\i}stica, se introduce otro campo
conocido como el \emph{par\'ametro de orden escalar}, o simplemente, el \emph{par\'ametro de orden}, $S$. Esta es una cantidad adimensional 
definida de tal manera que valdr\'a cero para un sistema completamente desordenado y cuyo valor m\'aximo, igual a la unidad, se alcanzar\'a
cuando todas las mol\'eculas est\'en perfectamente orientadas a lo largo de $\hat{\mathbf{n}}$. $S$ puede calcularse mediante la expresi\'on
%\begin{eqnarray}
%S & = & \frac{1}{2} \langle 3 \left(\hat{\mathbf{u}}_{i} \cdot \hat{\mathbf{n}} \right)^2 - 1 \rangle , \nonumber \\
%  & = & \frac{1}{2} \langle 3 \cos^{2}\theta_{i} - 1 \rangle , \label{descripcion_matematica_001}
%\end{eqnarray}
\begin{equation}
S  =  \frac{1}{2} \langle 3 \left(\hat{\mathbf{u}}_{i} \cdot \hat{\mathbf{n}} \right)^2 - 1 \rangle 
   =  \frac{1}{2} \langle 3 \cos^{2}\theta_{i} - 1 \rangle , \label{descripcion_matematica_001}
\end{equation}
en donde los \textit{brakets} $\langle\,\dots\rangle$ indican el promedio sobre el ensamble molecular y $\theta_{i}$ es el \'angulo entre 
$\hat{\mathbf{u}}_{i}$ y $\hat{\mathbf{n}}$. 

La ecuaci\'on~(\ref{descripcion_matematica_001}) indica que $S$ aumenta conforme la proyecci\'on de los ejes moleculares sobre $\hat{\mathbf{n}}$ 
es mayor. Adem\'as, conduce a los l\'imites esperados, $S = 0$, cuando no hay orden orientacional, y $S = 1$, para el alineamiento perfecto de
las mol\'eculas. La demostraci\'on de las igualdades anteriores se puede encontrar en el ap\'endice~\ref{apendice_001}. $S$ depende de la temperatura
siendo mayor a temperaturas m\'as bajas. Para los CLN en situaciones pr\'acticas, los valores t\'{\i}picos de $S$ se encuentran entre $0.3$ y 
$0.7$~\cite{shen_phys_rev_lett_1973,emsley_chem_phys_lett_1984,denniston_adv_physx_2020}. Por ejemplo, para el 5CB alrededor de los $26^{\circ}\text{C}$, 
$S$ tiene un valor cercano a $S = 0.68$~\cite{mandal_phys_rev_e_2019}.

% --------------------------------------------------------------------------------------------------------------------
\subsection{Par\'ametro de orden tensorial \label{parametro_de_orden_tensorial_seccion}}
% --------------------------------------------------------------------------------------------------------------------

% --------------------------------------------------------------------------------------------------------------------
\subsubsection{Definici\'on \label{definicion_parametro_de_orden_tensorial_seccion}}
% --------------------------------------------------------------------------------------------------------------------

En los CLN existe un eje preferencial, paralelo a $\hat{\mathbf{n}}$, alrededor del cual se pueden hacer rotaciones sin alterar la estructura. Esto parecer\'ia 
sugerir que los CLN tienen la misma simetr\'ia que los sistemas polares, como los materiales con una polarizaci\'on inducida, $\mathbf{P}$, en los 
que los que los dipolos a nivel molecular se encuentran alineados en torno una direci\'on com\'un~\cite{resnick_2021}. Sin embargo, mientras que dos 
sistemas con polarizaciones $\mathbf{P}$ y $-\mathbf{P}$ son distintos, dos CLN con directores $\hat{\mathbf{n}}$ y $-\hat{\mathbf{n}}$ son id\'enticos 
en todas sus propiedades. 

Los CLN se describen matem\'aticamente bajo la idea de que su estructura consiste de una superposici\'on de contribuciones independientes 
llamadas \textit{momentos multipolares de la orientaci\'on}. Este principio nos es familiar por su aplicaci\'on en otras ramas de la f\'isica, como el 
electromagnetismo. Por ello, es conveniente recordar brevemente su significado para el caso de una distribuci\'on est\'atica de cargas.

Al calcular el potencial el\'ectrico alrededor de una distribuci\'on de cargas, es razonable suponer que conforme m\'as alejados
estamos de la fuente, los detalles de \'esta son menos perceptibles. En puntos lejanos, la distribuci\'on de carga puede aproximarse como una conjunci\'on  
de cantidades promediadas que dependen de algunos de sus detalles y guardan informaci\'on de la simetr\'ia con la que las cargas se encuentran repartidas 
en el espacio~\cite{wangsness_1986}. Estas cantidades son, precisamente, los momentos multipolares de la distribuci\'on de carga. La carga neta
%\begin{equation}
%q^{\text{e}} = \int_{V} d\mathbf{r}^{\prime} \rho^{\text{e}}\left(\mathbf{r}^{\prime}\right) ,
%\nonumber
%\end{equation}
recibe el nombre de momento monopolar y es la principal caracter\'istica que se aprecia de la distribuci\'on a distancias grandes. 
%Aqu\'i, 
%$\rho^{\text{e}}\left(\mathbf{r}^{\prime}\right)$ es la densidad de carga el\'ectrica en la posici\'on $\mathbf{r}^{\prime}$ y $V$ es el volumen ocupado por
%la distribuci\'on. En t\'erminos de las mismas cantidades ei
El momento dipolar es la polarizaci\'on integrada sobre la distribuci\'on de carga~\cite{wangsness_1986} y es la caracter\'istica m\'as importante de la
distribuci\'on cuando la carga neta es cero.
%\begin{equation}
%\mathbf{p}^{\text{e}} = \int_{V} d\mathbf{r}^{\prime} \rho^{\text{e}}\left(\mathbf{r}^{\prime}\right) \mathbf{r}^{\prime} .
%\nonumber
%\end{equation}
Ahora, si tanto el momento monopolar como el momento dipolar se anulan, entonces la caracter\'istica dominante de la distribuci\'on es el
\textit{tensor de momento cuadripolar}. Para un conjunto de $N$ cargas puntuales $q_{1}, q_{2}, \dots , q_{N}$, que ocupan las posiciones
$\mathbf{r}_{1}, \mathbf{r}_{2}, \dots, \mathbf{r}_{N}$, respectivamente, la forma del tensor de momento cuadripolar es una matriz de $3\times 3$ con 
componenentes definidas por la expresi\'on
\begin{equation}
Q^{\text{e}}_{\alpha \beta} 
                 = \sum_{i = 1}^{N} q_{i}
                   \left( 
                         3 r_{\alpha,i} r_{\beta,i} - r^{2}_{i} \delta_{\alpha \beta}
                   \right),
\label{parametro_orden_tensorial_001}
\end{equation}
donde $\alpha,\beta =1,2,3$ y $\delta_{\alpha\beta}$ es la delta de Kronecker que, a su vez, se define por
\begin{equation}
\delta_{\alpha\beta} 
= \begin{cases}
1, & \text{si } \alpha = \beta \\
0, & \text{si } \alpha \ne \beta
\end{cases} .
\nonumber
\end{equation}

Adem\'as, en la ecuaci\'on~(\ref{parametro_orden_tensorial_001}), $r^{2}_{i} = \mathbf{r}_{i} \cdot \mathbf{r}_{i}$ y los vectores de posici\'on se han 
desarrollado en la forma $\mathbf{r}_{i} = \left(r_{1,i},r_{2,i},r_{3,i}\right)$, de tal manera que $r_{\alpha,i}$ es la componente $\alpha$ de $\mathbf{r}_{i}$. 
%y $r^{2}_{i} = \mathbf{r}_{i} \cdot \mathbf{r}_{i}$.

Al regresar al caso de los CLN, notamos que los momentos monopolar y dipolar de la distibuci\'on de orientaciones son nulos. En particular, no hay momento dipolar dado que no se 
altera ninguna propiedad f\'{\i}sica al invertir los vectores de orientaci\'on $\hat{\mathbf{u}}_{i}$. El primer multipolo que no se cancela y resulta 
ser el m\'as importante para describir la estructura promedio, es el cuadripolo. En el campo de la materia condensada suave, a este momento se le llama 
el \textit{par\'ametro de orden tensorial} y, para un ensamble de $N$ mol\'eculas con orientaciones $\hat{\mathbf{u}}_{i}$, est\'a definido mediante la 
ecuaci\'on
\begin{equation}
Q_{\alpha \beta} = \frac{1}{2 N}\sum_{i = 1}^{N}
                   \left(
                         3 u_{\alpha,i} u_{\beta,i} - \delta_{\alpha \beta}
                   \right),
\label{parametro_orden_tensorial_002}
\end{equation}
cuya estructura es esencialemente la misma que la de la ecuaci\'on~(\ref{parametro_orden_tensorial_001}), con $\hat{\mathbf{u}}_{i}$ jugando el
papel de $\mathbf{r}_{i}$.

El factor $1/N$ que puede distinguirse en el lado derecho de la ecuaci\'on~(\ref{parametro_orden_tensorial_002}) permite escribirla tambi\'en en t\'erminos
de un promedio sobre el ensamble molecular
\begin{equation}
Q_{\alpha \beta} = \frac{1}{2} \left\langle
                   3 u_{\alpha} u_{\beta} - \delta_{\alpha \beta}
                   \right\rangle.
\label{parametro_orden_tensorial_003}
\end{equation}

% --------------------------------------------------------------------------------------------------------------------
\subsubsection{Propiedades \label{propiedades_parametro_de_orden_tensorial_seccion}}
% --------------------------------------------------------------------------------------------------------------------

El tensor $Q_{\alpha\beta}$ es la caracter\'istica dominante de la distribuci\'on de orientaciones. Tiene la propiedad de ser sim\'etrico, \textit{i. e.},
$Q_{\alpha\beta} = Q_{\beta\alpha}$, dado que el producto $u_{\alpha} u_{\beta}$ y $\delta_{\alpha \beta}$ tambi\'en lo son. 

Adem\'as, la traza de $Q_{\alpha\beta}$ es nula,
\begin{eqnarray}
\text{Tr}\left(\mathbf{Q}\right) & = & Q_{11} + Q_{22} + Q_{33}  \nonumber \\
				 & = & \frac{1}{2} \left\langle 3 u_{1}^{2} - \delta_{11} \right\rangle
				     + \frac{1}{2} \left\langle 3 u_{2}^{2} - \delta_{22} \right\rangle \nonumber \\
				 &  &+ \frac{1}{2} \left\langle 3 u_{3}^{2} - \delta_{33} \right\rangle \nonumber \\
				 & = & \frac{1}{2} 
				       \left[ 3 \left\langle u_{1}^{2} + u^{2}_{2} + u^{2}_{3} \right\rangle - 1 - 1 - 1
                                       \right] \nonumber \\
                                 & = & 0 . \nonumber
\end{eqnarray}

Estas propiedades se deben a la simetr\'ia de la fase nem\'atica mencionada al inicio de esta secci\'on.

Uno de los aspectos m\'as importantes del par\'ametro de orden tensorial es que, al contener la informaci\'on estructural del CLN, permite obtener los valores
de $S$ y $\hat{\mathbf{n}}$. Es decir, si conocemos $Q_{\alpha \beta}$ podemos obtener de \'el la cantidad de orden y ``la 
orientaci\'on promedio''. Espec\'ificamente, $S$ es el eigenvalor m\'as alto del tensor de par\'ametro de orden y $\hat{\mathbf{n}}$ el eigenvector asociado a $S$.
El ap\'edice~\ref{apendice_002} presenta una demostraci\'on de ello, para el caso de un CLN uniaxial, que es el que hemos discutido en este art\'iculo.

En algunos casos especiales es posible escribir al tensor $Q_{\alpha\beta}$ en representaciones que permiten una interpretaci\'on m\'as directa.
En particular, para un nem\'atico uniaxial, si se conocen $S$ y $\hat{\mathbf{n}}$, se puede escribir
\begin{equation}
Q_{\alpha\beta} = \frac{1}{2} S \left( 3 n_{\alpha} n_{\beta} - \delta_{\alpha\beta}\right) ,
\label{parametro_orden_tensorial_004}
\end{equation}
donde el promedio sobre el ensamble de mol\'eculas ya no aparece expl\'icitamente. La ecuaci\'on~(\ref{parametro_orden_tensorial_004}) muestra que el tensor
$Q_{\alpha\beta}$ combina tanto la informaci\'on del orden orientacional como de la orientaci\'on promedio.

Por supuesto, tambi\'en es posible escribir al par\'ametro de orden tensorial en notaci\'on matricial. En \'esta, la 
ecuaci\'on~(\ref{parametro_orden_tensorial_004}) luce
\begin{equation}
\mathbf{Q} = \frac{1}{2} S \left( 3 \hat{\mathbf{n}} \hat{\mathbf{n}} - \mathbf{I} \right) ,
\label{parametro_orden_tensorial_005}
\end{equation}
donde $\mathbf{I}$ es la matriz identidad de $3\times 3$ y $\hat{\mathbf{n}} \hat{\mathbf{n}}$ es tambi\'en una matriz de $3\times 3$, conocida 
com\'unmente como el \textit{producto di\'adico} o \textit{producto externo} de $\hat{\mathbf{n}}$ consigo mismo. En general, el producto di\'adico entre dos 
vectores $\mathbf{a}$ y $\mathbf{b}$, $\mathbf{a}\mathbf{b}$, se define por las identidades siguientes, que expresan c\'omo se multiplica \'el (como matriz) 
por cualquier vector, $\mathbf{v}$,~\cite{symon_mechanics}
%~\footnote{La forma expl\'icita de la diada $\mathbf{a}\mathbf{b}$ es
%\begin{equation}
%\mathbf{a}\mathbf{b}  = \left( \begin{matrix} 
%                        a_{1}b_{1} & a_{1}b_{2} & a_{1}b_{3} \\
%			a_{2}b_{1} & a_{2}b_{2} & a_{2}b_{3} \\
%			a_{3}b_{1} & a_{3}b_{2} & a_{3}b_{3} \\
%                        \end{matrix} \right) . 
%\nonumber
%\end{equation}} 
\begin{equation}
\left({\mathbf{a}} \mathbf{b}\right)\cdot \mathbf{v} = \mathbf{a} \left( \mathbf{b} \cdot \mathbf{v} \right),
\label{parametro_orden_tensorial_006}
\end{equation}
\begin{equation}
\mathbf{v}^{\text{T}} \cdot \left(\mathbf{a}  \mathbf{b} \right) =  \left( \mathbf{v} \cdot \mathbf{a}  \right)\mathbf{b} .
\label{parametro_orden_tensorial_007}
\end{equation}

%son importantes de mencionarse. 

Como cabr\'ia esperar, todas las ecuaciones (\ref{parametro_orden_tensorial_002}) a (\ref{parametro_orden_tensorial_005}) son 
invariantes ante las transformaciones $\hat{\mathbf{u}}\rightarrow -\hat{\mathbf{u}}$ o $\hat{\mathbf{n}}\rightarrow -\hat{\mathbf{n}}$.

% --------------------------------------------------------------------------------------------------------------------
\subsubsection{Invariantes \label{invariantes_parametro_de_orden_tensorial_seccion}}
% --------------------------------------------------------------------------------------------------------------------

Una caracter\'istica importante de cualquier tensor de segundo rango son sus invariantes. Estos son aquellas cantidades formadas por las 
componentes del tensor cuyo valor no cambia si \'este se representa en diferentes sistemas de coordenadas. Los invariantes son importantes 
para expresar diversas propiedades f\'isicas, cuando se sabe que \'estas no pueden depender del marco de referencia que se utilice para 
describir al sistema f\'isico. Un ejemplo de ello es la energ\'ia, la cual se analizar\'a
en detalle en las secciones~\ref{energia_formacion_fase_seccion} y \ref{energia_elastica_seccion}. 

Un tipo especial de invariantes 
son los llamados \emph{invariantes principales}, que son los coeficientes del polinomio caracter\'istico del tensor, \textit{i. e.},
el polinomio que se utiliza en el problema de eigenvalores $\lambda$. Para un tensor $\mathbf{A}$ esta ecuaci\'on se escribe en la forma 
\begin{equation}
\det \left( \mathbf{A} - \lambda \mathbf{I}\right) = 0,
\label{parametro_orden_tensorial_008}
\end{equation}

En particular, cuando $\mathbf{A}$ tiene dimensi\'on $3$, la ecuaci\'on (\ref{parametro_orden_tensorial_008}) se convierte en un polinomio
c\'ubico de $\lambda$, 
\begin{equation}
-\lambda^{3} + A_{1} \lambda^{2} -A_{2} \lambda + A_{3} = 0.
\label{parametro_orden_tensorial_008a}
\end{equation}

No es dif\'icil demostrar que los coeficientes de este polinomio, o sea, los invariantes principales de $\mathbf{A}$, son
\begin{equation}
A_{1} = \text{Tr}\,\left( \mathbf{A} \right) = A_{11} + A_{22} + A_{33},
\label{parametro_orden_tensorial_009}
\end{equation}
\begin{eqnarray}
A_{2} & = & \frac{1}{2} \left\{ \left[\text{Tr}\left(\mathbf{A}\right)\right]^{2} - \text{Tr}\left(\mathbf{A}^{2}\right) \right\} \nonumber \\ 
      & = &   A_{11} A_{22} + A_{11} A_{33} + A_{22} A_{33} \nonumber \\ 
      &   & - A_{12} A_{21} - A_{13} A_{31} - A_{23} A_{32}
\label{parametro_orden_tensorial_010}
\end{eqnarray}
y
\begin{eqnarray}
A_{3} & = & \det\left( \mathbf{A}\right) \nonumber \\
      & = & A_{11} \left( A_{22} A_{33} - A_{23} A_{32} \right) + A_{12}  \left( A_{23} A_{31} \right. \nonumber \\
      &   & \left. -A_{21} A_{33} \right) + A_{13} \left( A_{32} A_{21} - A_{31} A_{22} \right) . 
\label{parametro_orden_tensorial_011}
\end{eqnarray}

Dado que $\mathbf{Q}$ es sim\'etrico y sin traza, sus invariantes principales se simplifican, \textit{e. g.}, $Q_{1} = 0$ y
$Q_{2} = -\text{Tr} \left( \mathbf{A}^{2} \right)/2$. Cuando, adem\'as, $\mathbf{Q}$ se escribe en la forma dada por las 
ecuaciones~(\ref{parametro_orden_tensorial_004}) o (\ref{parametro_orden_tensorial_005}), se obtiene 
$\text{Tr}\left( \mathbf{Q}^{2}\right) = 3S^{2}/2$, \textit{i. e.}, $Q_{2} = -3 S^{2}/4$; y $Q_{3} = S^{3}/4$.

Otra forma en la que pueden construirse cantidades escalares a partir de un tensor es a trav\'es de operaciones que reducen su 
rango y en matem\'aticas se conocen como \emph{productos internos} o \emph{contracciones}. Uno de los productos internos m\'as conocidos 
es el producto escalar entre dos vectores (tensores de rango uno), el cual toma a dichos vectores, $\mathbf{a}$ y $\mathbf{b}$, y devuelve 
un escalar (tensor de rango cero),~\footnote{En este art\'iculo las 
propiedades tensoriales de las diversas cantidades que estudiaremos no ser\'an analizadas exhaustivamente. Para nuestros prop\'ositos
ser\'a suficiente identificar como tensores, aquellas cantidades escalares, vectoriales o matriciales que son invariantes ante
un cambio de la base utilizada para describirlas. Esto implica, \textit{e. .g.}, que un vector es al mismo tiempo un tensor de rango uno, si al cambiar
de base sigue apuntando en la misma direcci\'on y teniendo la misma magnitud, aunque sus componentes sean distintas.} 
\begin{equation}
c = \mathbf{a} \cdot \mathbf{b} = \sum_{\alpha=1}^{3} a_{\alpha} b_{\alpha},
\label{energia_landau_degennes_001}
\end{equation}

Otra contracci\'on muy conocida es la multiplicaci\'on de una matriz, $\mathbf{A}$, (tensor de rango dos), por un vector, $\mathbf{a}$, (tensor de rango uno), 
que devuelve otro vector cuyas componentes est\'an dadas por
\begin{equation}
c_{\alpha} = \sum_{\beta=1}^{3} A_{\alpha\beta} a_{\beta}.
\label{energia_landau_degennes_002}
\end{equation}

Es com\'un expresar los productos en las ecuaciones (\ref{energia_landau_degennes_001}) y (\ref{energia_landau_degennes_002}) en
t\'erminos de la llamda convenci\'on de suma sobre \'indices repetidos, la cual fue introducida en F\'isica por Albert Einstein.
En esta convenci\'on, un \'indice que aparece dos veces en un t\'ermino implica una sumatoria sobre todos sus valores posibles. 
As\'i, no es necesario escribir los s\'imbolos de sumatoria y  se tienen expresiones m\'as compactas. Por ejemplo,  las ecuaciones 
(\ref{energia_landau_degennes_001}) y (\ref{energia_landau_degennes_002}) se reducen a
\begin{equation}
c = \mathbf{a} \cdot \mathbf{b} = a_{\alpha} b_{\alpha},
\label{energia_landau_degennes_003}
\end{equation}
y
\begin{equation}
c_{\alpha} = A_{\alpha\beta} a_{\beta},
\label{energia_landau_degennes_004}
\end{equation}
respectivamente.
 
El tensor $\mathbf{Q}$ puede dar lugar a un escalar al multiplicarse internamente consigo repetidamente. Por ejemplo, la
llamada doble contracci\'on de $\mathbf{Q}$ consigo mismo da lugar a un escalar relacionado con los invariantes de $\mathbf{Q}$,
\begin{equation}
Q_{\alpha\beta}Q_{\beta\alpha} = \text{Tr}\left( \mathbf{Q}^{2} \right) = \frac{3}{2} S^{2}.
\label{energia_landau_degennes_005}
\end{equation}

Al contraer $\mathbf{Q}$ tres veces consigo mismo y utilizar sus propiedades de simetr\'ia y traza nula tambi\'en se obtiene un
escalar relacionado con los invariantes
\begin{equation}
Q_{\alpha\beta}Q_{\beta\gamma}Q_{\gamma\alpha} = 3\, \text{det}\left(\mathbf{Q}\right) = \frac{3}{4} S^{3}.
\label{energia_landau_degennes_006}
\end{equation}

% ====================================================================================================================
\section{Energ\'{\i}as \label{energias_seccion}}
% ====================================================================================================================

% --------------------------------------------------------------------------------------------------------------------
\subsection{Energ\'{\i}a volum\'etrica \label{energia_formacion_fase_seccion}}
% --------------------------------------------------------------------------------------------------------------------

Los CL pueden realizar diversas transiciones de fase cuando cambia su temperatura o, en el caso de los CL liotr\'opicos, la 
concentraci\'on de sus componentes. Estas transiciones pueden describirse mediante
dos m\'etodos te\'oricos alternativos. En el primero, se modelan las propiedades y energ\'ias moleculares 
y se intenta calcular el rango de los par\'ametros, como la temperatura, para el cual es estable una fase espec\'ifica. Este 
enfoque est\'a en la l\'inea de la \textit{mec\'anica estad\'istica}, la rama de la f\'isica que infiere las propiedades 
macrosc\'opicas de la materia a partir de promediar la din\'amica a nivel molecular. El segundo m\'etodo tiene un enfoque 
fenomenol\'ogico, siendo su prop\'osito, el describir cuantitativamente la transici\'on en t\'erminos del tensor 
$Q_{\alpha\beta}$, que es la variable macrosc\'opica que contiene la informaci\'on de la simetr\'ia y el orden de la fase.

%Debido a que la mayor\'ia de los experimentos se llevan a cabo a temperatura y volumen constantes, el potencial termodin\'amico
%apropiado es la energ\'ia libre $\mathcal{F}$ y el equilibrio termodin\'amico corresponde al m\'inimo de $\mathcal{F}$.

Este enfoque fenomenol\'ogico se basa en la teor\'ia de las transiciones de fase de segundo orden de Lev D. Landau~\cite{landau_1965}
que fue generalizado por primera vez por de Gennes para el caso de fases l\'iquido-cristalinas~\cite{de_gennes_mol_cryst_liq_cryst_1971}.
Debido a que la mayor\'ia de los experimentos se llevan a cabo a temperatura y volumen constantes, el potencial termodin\'amico
apropiado es la energ\'ia libre $F$. La fase que corresponde a las condiciones impuestas sobre el sistema es aquella en donde 
$F$ adquiere un m\'inimo. En muchos contextos resulta muy conveniente llevar a cabo la descripci\'on en t\'erminos de
la energ\'ia por unidad de volumen, $f$. En particular, la contribuci\'on a esta densidad de energ\'ia %por unidad de volumen 
que describe la transici\'on de fase se conoce como densidad de energ\'ia volum\'etrica o \textit{bulk energy density}, en ingl\'es, 
%de Landau--de Gennes 
y aqu\'i ser\'a representada mediante el s\'imbolo $f_{\text{bulk}}$.

En el modelo de Landau--de Gennes, la energ\'ia libre de la fase se considera una funci\'on anal\'itica 
del par\'ametro de orden. Adem\'as, se toma en cuenta que el orden es peque\~no cerca de la transici\'on de fase, por lo que la energ\'ia 
se expande en una serie de potencias de esta cantidad. Al extender la expansi\'on hasta la cuarta potencia resulta~\cite{andrienko_j_mol_liq_2018}
\begin{equation}
f_{\text{bulk}} = f_{0} + \frac{A}{2} S_{\alpha\beta}S_{\beta\alpha} %\nonumber \\
%F & = & F_{0} + \frac{A}{2} \sum_{\alpha=1}^{3}\sum_{\beta=1}^{3} Q_{\alpha\beta}Q_{\beta\alpha} \nonumber \\
 -\frac{B}{3} S_{\alpha\beta}S_{\beta\gamma}S_{\gamma\beta} %\nonumber \\
%  &   & +\frac{B}{3} \sum_{\alpha = 1}^{3} \sum_{\beta = 1}^{3} \sum_{\gamma = 1}^{3} Q_{\alpha\beta}Q_{\beta\gamma}Q_{\gamma\beta} \nonumber \\
 +\frac{C}{4} \left( S_{\alpha\beta}S_{\beta\alpha}  \right)^{2},
%  &   & +\frac{C}{4} \left( \sum_{\alpha=1}^{3}\sum_{\beta=1}^{3} Q_{\alpha\beta}Q_{\beta\alpha}  \right)^{2}
\label{landau}
\end{equation}
donde $f_{0}$ es una constante y el tensor $S_{\alpha\beta}$ est\'a emparentado con 
$Q_{\alpha\beta}$ mediante un factor de proporcionalidad de
$2/3$, \textit{i. e.}, $S_{\alpha\beta} = 2 Q_{\alpha\beta}/3$. Los coeficientes $A$, $B$ y $C$, que se denominan par\'ametros fenomenol\'ogicos, 
son funciones de la temperatura y la presi\'on. Sin embargo, es usual suponer que $B$ y $C$ son constantes y considerar \'unicamente la dependencia 
de $A$ con $T$, a trav\'es de  la igualdad $A = A^{\prime} \left(T-T^{*}\right)$, siendo $A^{\prime}$ y $T^{*}$ otras constantes. 
Algunos valores que se consideran t\'ipicos para los coeficientes fenomenol\'ogicos en modelos te\'oricos son~\cite{musevic_liquid_crystal_colloids_2017}
$A^{\prime} = 10^5\,\text{J}/\left(\text{m}^{3}\text{K}\right)$, $B = 10^{6}\, \text{J}/\text{m}^{3}$ y $C = 10^{6}\,\text{J}/\text{m}^{3}$.
Para el 5CB los valores estimados son $A^{\prime} = 0.044 \times 10^6\,\text{J}/\left(\text{m}^{3}\text{K}\right)$, 
$B = 0.816 \times 10^{6}\, \text{J}/\text{m}^{3}$ y $C = 0.45 \times 10^{6}\,\text{J}/\text{m}^{3}$.~\cite{andrienko_j_mol_liq_2018} Para el MBBA, las 
estimaciones experimentales resultan en  $A^{\prime} = 0.42 \times 10^3\,\text{J}/\left(\text{m}^{3}\text{K}\right)$, 
$B = 0.64 \times 10^{4}\, \text{J}/\text{m}^{3}$ y $C = 0.35 \times 10^{4}\,\text{J}/\text{m}^{3}$.~\cite{mkaddem_phys_rev_e_2000,majumdar_eur_j_appl_math_2010}

Todos los t\'erminos en la ecuaci\'on~(\ref{landau}) involucran a los invariantes de $\mathbf{S}$ (o equivalentemente a los de $\mathbf{Q}$), lo que garantiza que la energ\'ia
ser\'a la misma, independientemente del sistema de referencia en el que se calcule este tensor. Al utilizar la relaci\'on entre $\mathbf{S}$ y $\mathbf{Q}$ y las 
ecuaciones~(\ref{energia_landau_degennes_005}) y (\ref{energia_landau_degennes_006}) en la ecuaci\'on~(\ref{landau}), $f_{\text{bulk}}$ puede escribirse como un polinomio de cuarto grado de $S$,
\begin{equation}
f_{\text{bulk}} = f_{0} + \frac{1}{3} A'\left(T - T^{*}\right) S^{2} - \frac{2}{27} B S^{3} + \frac{1}{9}C S^{4} ,
\label{landau_002}
\end{equation}
la cual muestra que $F_{0}$ es la energ\'{\i}a correspondiente al caso $S = 0$, \textit{i. e.}, la energ\'ia de la fase isotr\'opica.

La ecuaci\'on~(\ref{landau_002}) permite describir muy bien, a nivel fenomenol\'ogico, la transici\'on entre las fases isotr\'opica y nem\'atica. 
Para ello, conviene introducir el exceso de energ\'ia, $\Delta f_{\text{bulk}}\left( S\right) = f_{\text{bulk}} - f_{0}$, o sea
\begin{equation}
\Delta f_{\text{bulk}}\left( S\right) = \frac{1}{3} A'\left(T - T^{*}\right) S^{2} - \frac{2}{27} B S^{3} + \frac{1}{9}C S^{4}.
\label{landau_003}
\end{equation} 

\begin{figure}[h]
\includegraphics[scale=0.4]{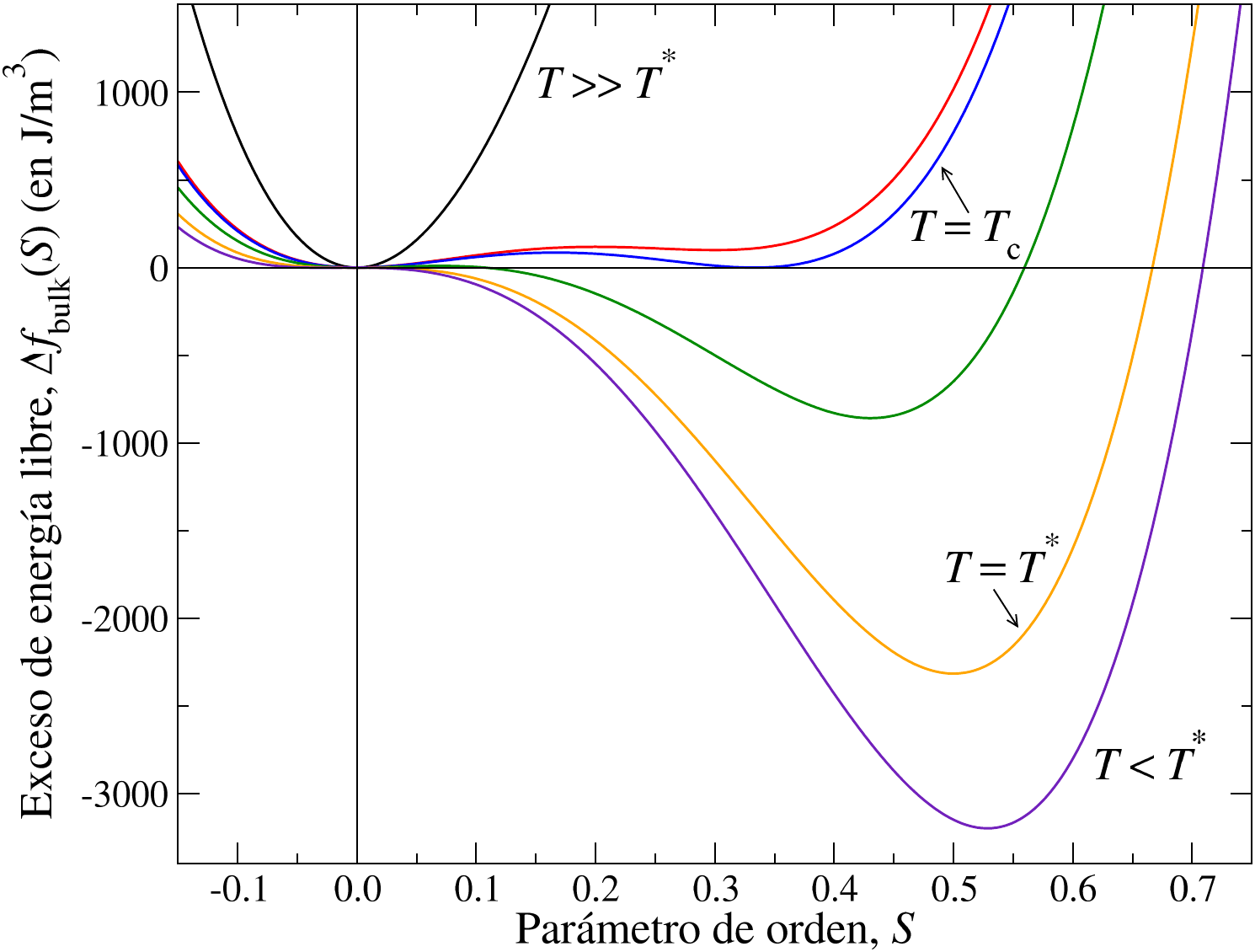}
\caption{Energ\'ia libre volum\'etrica por unidad de volumen de un CLN en el modelo de Landau--de Gennes dado por la 
ecuaci\'on~(\ref{landau_003}).}
\label{figura_010}
\end{figure}

El comportamiento de esta funci\'on para diferentes valores de $T$ se muestra en la figura~\ref{figura_010}. 

Cuando $T \gg T^{*}$, el t\'ermino cuadr\'atico en la ecuaci\'on~(\ref{landau_003}) domina y $\Delta f_{\text{bulk}}$ es similar a una par\'abola 
con v\'ertice en el origen $S = 0$. Esto indica que a temperaturas altas la \'unica fase estable puede ser la fase isotr\'opica. 

Conforme $T$ disminuye, la relevancia del t\'ermino cuadr\'atico disminuye tambi\'en y empiezan a ser notorios los efectos de los t\'erminos 
c\'ubico y cu\'artico, lo que da lugar a la existencia de un m\'inimo local para $S > 0$, aunque $S = 0$ sigue siendo el m\'inimo global de 
$\Delta f_{\text{bulk}}(S)$. Esto implica que la fase isotr\'opica ($S=0$) sigue siendo la fase estable. A esa temperatura, la fase nem\'atica se considera 
\emph{metaestable}. Esto significa que, al corresponder a un m\'inimo local de la energ\'ia, el estado nem\'atico podr\'ia subsistir por 
alg\'un periodo de tiempo, pero la agitaci\'on t\'ermica o una perturbaci\'on externa podr\'ian hacer que el sistema brinque f\'acilmente 
la barrera de energ\'ia que separa a los dos m\'inimos, oblig\'andolo a adoptar la configuraci\'on estable.

Al disminuir a\'un m\'as la temperatura se encontrar\'a un valor $T_{\text{c}}$, referido como la \emph{temperatura cr\'itica}, a la cual son 
igualmente estables la fase isotr\'opica y la fase ordenada.
En otras palabras, a $T = T_{\text{c}}$ los dos m\'inimos de la energ\'ia adquieren el mismo valor $\Delta f_{\text{bulk}} = 0$. Si la temperatura disminuye a\'un m\'as, 
$T < T_{\text{c}}$, el m\'inimo en $S = 0$ se convertir\'a en local y el que ocurre en $S>0$ ser\'a el m\'inimo global. Entonces, $T_{\text{c}}$ puede
considerarse la temperatura a partir de la cual la fase nem\'atica es la fase estable. Durante alg\'un intervalo de temperaturas menores, la fase isotr\'opica ser\'a
metaestable, pero a $T = T^{*}$, el t\'ermino cuadr\'atico en la ecuaci\'on~(\ref{landau_003}) se vuelve negativo y $S = 0$ ya no correponde con un m\'{\i}nimo
de la energ\'{\i}a. Entonces, a temperaturas $T < T^{*}$ la fase isotr\'opica es inestable y la nem\'atica ser\'a la \'unica que se observar\'a. 

La ocurrencia de las fases estables y los valores concretos de equilibrio de $S$ pueden identificarse al obtener los m\'inimos de la funci\'on $\Delta f_{\text{bulk}}(S)$.
Para ello pueden usarse los m\'etodos conocidos del c\'alculo y el \'algebra, tal como se detalla en el ap\'endice~\ref{apendice_003}. Esto conduce a
la soluci\'on
\begin{equation}
S = \frac{B}{4C}\left(1 + \sqrt{1-24 \beta }\right), 
\label{landau_004}
\end{equation}
para el par\'ametro de orden de la fase nem\'atica, donde $\beta = A^{\prime} \left( T- T^{*}\right)C/B^{2}$. Tambi\'en pueden obtenerse
el valor de la temperatura cr\'itica, 
\begin{equation}
T_{\text{c}} = T^{*} + \frac{B^{2}}{27A^{\prime}C},
\label{landau_005}
\end{equation}
y la cantidad de orden a la temperatura cr\'itica,
\begin{equation}
S_{\text{c}} = \frac{1}{3} \frac{B}{C}.
\label{landau_006}
\end{equation}

\begin{figure}[h]
\includegraphics[scale=0.4]{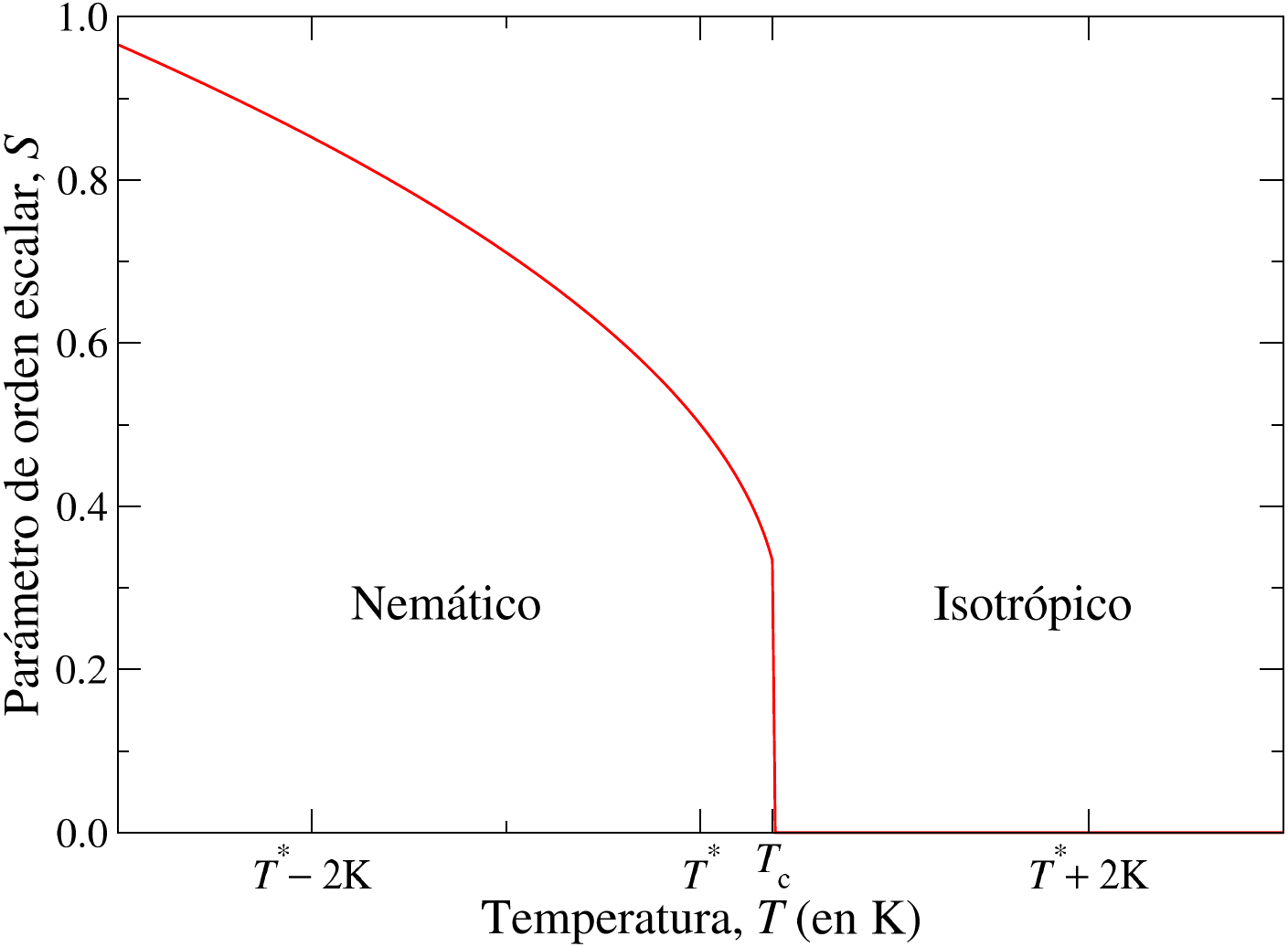}
\caption{La transici\'on entre las fases isotr\'opica y nem\'atica es evidenciada por la aparici\'on de orden. $T_{\text{c}}$, dada por la 
ecuaci\'on~(\ref{landau_005}), representa la temperatura a la cual cambia el car\'acter estable de dichas fases.}
\label{figura_011}
\end{figure}

La transici\'on de fase isotr\'opica--nem\'atica se puede ilustrar tambi\'en mediante una gr\'afica de los valores estables de $S$ contra $T$.
Esa gr\'afica se muestra en la figura~\ref{figura_011} para el modelo de Landau--de Gennes y los valores t\'ipicos de los par\'ametros 
fenomenol\'ogicos mencionados previamente. Para $T > T_{\text{c}}$, la fase estable es la 
isotr\'opica y $S = 0$. En $T=T_{\text{c}}$ hay una discontinuidad pues $S$ aumenta hasta $S_{\text{c}}$, su valor de equilibrio en la
fase nem\'atica. A partir de all\'i, $S$ aumenta de manera mon\'otona tomando los valores descritos por la ecuaci\'on~(\ref{landau_004}).
Este comportamiento discontinuo es la caracter\'istica principal de las llamadas transiciones de fase de primer orden.~\cite{binder_phase_transformations_2001}

A partir de este an\'alisis puede darse un interpretaci\'on retrospectiva de los coeficientes fenomenol\'ogicos en la ecuaci\'on~(\ref{landau_003}).
El coeficiente de $S^{2}$ es el responsable de conducir la transici\'on debido a su dependencia con la 
temperatura. El t\'ermino c\'ubico, permitido por la simetr\'ia, es el responsable de que la transici\'on sea discontinua.   
El coeficiente $C$ permite estabilizar la energ\'ia al impidir que decrezca indefinidamente cuando el t\'ermino cuadr\'atico cambia de signo. 

% --------------------------------------------------------------------------------------------------------------------
\subsection{Energ\'{\i}a el\'astica \label{energia_elastica_seccion}}
% --------------------------------------------------------------------------------------------------------------------

% ....................................................................................................................
\subsubsection{Deformaciones del alineamiento y su descripci\'on matem\'atica \label{deformaciones_seccion}}
% ....................................................................................................................

En un CLN, los ejes moleculares tienden a alinearse alrededor de 
$\hat{\mathbf{n}}$. A\'un en el caso est\'atico, $\hat{\mathbf{n}}$ puede ser obligado a cambiar de un punto a otro  por la acci\'on 
de fuerzas externas o condiciones de frontera, volvi\'endose un campo dependiente de la posici\'on, $\hat{\mathbf{n}}\left(\mathbf{r}\right)$. 
Los cambios de $\hat{\mathbf{n}}$ %en la orientaci\'on promedio 
entre regiones cercanas 
reciben el nombre de \textit{deformaciones} y representan desviaciones con respecto al estado ideal de alineamiento uniforme
de una muestra de CLN. 

Para describir matem\'aticamente a las deformaciones, se toma en cuenta que en la mayor\'ia de las situaciones las distancias para las cuales
son apreciables los cambios del campo director son muy grandes en comparaci\'on con las dimensiones
moleculares. Entonces, puede asumirse que $\hat{\mathbf{n}}\left(\mathbf{r}\right)$ es un campo continuo que cambia de manera suave.
As\'i, es v\'alido medir las variaciones de $\hat{\mathbf{n}}\left(\mathbf{r}\right)$ en t\'erminos de sus derivadas espaciales.
Dado que $\hat{\mathbf{n}}$ tiene tres componentes ($n_{1}$, $n_{2}$ y $n_{3}$) que dependen de las tres direcciones espaciales 
($x_{1}$, $x_{2}$ y $x_{3}$), se pueden formar nueve derivadas del director, que representaremos mediante
\begin{equation}
\frac{\partial n_{\beta}}{\partial x_{\alpha}} = \partial_{\alpha} n_{\beta},
\label{energia_elastica_001}
\end{equation}
para $\alpha,\beta=1,2,3$.
%se consideran cantidades peque\~nas. 

%Antes de ver c\'omo se utilizan estas derivadas en el c\'alculo de la energ\'ia de un CL, convendr\'a introducir algunas ideas y
%expresiones matem\'aticas relacionadas con ellas que ser\'an de mucha utilidad. 
El prop\'osito de esta secci\'on es analizar c\'omo pueden utilizarse estas derivadas determinar las diferentes deformaciones que se pueden producir en un CLN.
Posteriormente, en la secci\'on~\ref{energia_oseen_frank_seccion} estudiaremos c\'omo se relacionan las deformaciones con la energ\'ia de estas mesofases.

Al igual que hacemos en otros campos de la F\'isica para inferir la geometr\'ia de los campos, podemos utilizar operadores diferenciales 
para caracterizar las propiedades espaciales del director.~\footnote{En electromagnetimo, \textit{e. g.}, la ley de Gauss nos dice que los campos
electrost\'aticos emanan de los puntos con densidad de carga el\'etrica diferente de cero, $\boldsymbol{\nabla}\cdot\mathbf{E}=\rho_{\text{e}}/\epsilon_{0}$;
y la ley de Amp\`{e}re nos dice que un campo magnetost\'atico circula alrededor de una densidad de corriente, 
$\boldsymbol{\nabla} \times \mathbf{B} = \mu_{0} \mathbf{J}$.} 
Por ejemplo, podemos saber si $\hat{\mathbf{n}}$ tiene una ``fuente'' local al calcular su divergencia
\begin{equation}
\boldsymbol{\nabla}\cdot\hat{\mathbf{n}} = \partial_{1} n_{1} + \partial_{2} n_{2} + \partial_{3} n_{3} , %= \partial_{\alpha} n_{\alpha},
\label{energia_elastica_002}
\end{equation}
o saber si tiene una circulaci\'on alrededor de un punto al calcular su rotacional,
\begin{eqnarray}
\boldsymbol{\nabla}\times\hat{\mathbf{n}} & = & \left( \partial_{2} n_{3} - \partial_{3} n_{2} \right) \hat{\mathbf{e}}_{1} 
                                               -\left( \partial_{1} n_{3} - \partial_{3} n_{1} \right) \hat{\mathbf{e}}_{2} \nonumber \\
                                          &   &+\left( \partial_{1} n_{2} - \partial_{2} n_{1} \right) \hat{\mathbf{e}}_{3} .           
\label{energia_elastica_003}
\end{eqnarray}

La notaci\'on de \'indices junto con la convenci\'on de suma de Einstein, son muy \'utiles en el an\'alisis de la deformaciones. Con ellas se puede escribir 
de manera compacta la ecuaci\'on~(\ref{energia_elastica_002}) como 
\begin{equation}
\boldsymbol{\nabla}\cdot\hat{\mathbf{n}} = \partial_{\alpha} n_{\alpha}.
\label{energia_elastica_003a}
\end{equation}

En \'algebra de tensores, el \'indice que indica la sumatoria, \textit{e. g.} $\alpha$ en la ecuaci\'on~(\ref{energia_elastica_003a}), se llama un \'indice ``mudo''.
Esto se debe a que el nombre que utilicemos para \'el es irrelevante porque, de todas formas, corre sobre todos sus valores posibles. Dicho de otra manera,
tambi\'en pudimos haber escrito, \textit{e. g.}, $\boldsymbol{\nabla}\cdot\hat{\mathbf{n}} = \partial_{\beta} n_{\beta}$, sin alterar la suma impl\'icita.
Una regla importante al utilizar \'indices repetidos es que nunca debemos usar el mismo \'indice mudo m\'as dos veces en un t\'ermino de ninguna ecuaci\'on,
debido a que esto generar\'ia confusi\'on acerca de c\'omo llevar a cabo la sumatoria. Consideremos, \textit{e. g.}, el cuadrado de la divergencia de $\hat{\mathbf{n}}$,
que tendr\'a un significado f\'isico concreto como veremos m\'as adelante en la secci\'on~\ref{energia_oseen_frank_seccion}. Al tomar en cuenta la 
ecuaci\'on~(\ref{energia_elastica_003a}), podr\'iamos pensar en primera instancia en escribir esta cantidad como: 
$\left( \boldsymbol{\nabla}\cdot\hat{\mathbf{n}}\right)^{2}= \partial_{\alpha} n_{\alpha} \, \partial_{\alpha} n_{\alpha}$. Sin embargo, esto no es lo m\'as
conveniente pues los \'indices repetidos podr\'ian indicar m\'as de una sumatoria incorrecta, \textit{e. g.}, podr\'iamos pensar en expandir primero una sumatoria sobre 
los \'indices de las derivadas parciales y luego expandir la sumatoria sobre el \'indice del director:
\begin{eqnarray}
\left(\boldsymbol{\nabla}\cdot\hat{\mathbf{n}}\right)^{2} & = & \partial_{\alpha} n_{\alpha} \, \partial_{\alpha} n_{\alpha} \nonumber \\ 
                                                          & = & \partial_{1} n_{\alpha} \, \partial_{1} n_{\alpha} 
                                                              + \partial_{2} n_{\alpha} \, \partial_{2} n_{\alpha} 
                                                              + \partial_{3} n_{\alpha} \, \partial_{3} n_{\alpha} \nonumber \\
                                                          & = & \partial_{1} n_{1} \, \partial_{1} n_{1} 
                                                              + \partial_{1} n_{2} \, \partial_{1} n_{2} 
                                                              + \partial_{1} n_{3} \, \partial_{1} n_{3} \nonumber \\ 
                                                          &   & + \partial_{2} n_{1} \, \partial_{2} n_{1} + \dots . \leftarrow \textbf{!`Incorrecto!} 
\nonumber
\end{eqnarray}

Este error se evita si para cada factor de $\left( \boldsymbol{\nabla}\cdot\hat{\mathbf{n}}\right)^{2}$ utilizamos un \'indice mudo diferente, \textit{i. e.},
\begin{equation}
\left(\boldsymbol{\nabla}\cdot\hat{\mathbf{n}}\right)^{2} = \partial_{\alpha} n_{\alpha} \, \partial_{\beta} n_{\beta} ,
\label{energia_elastica_003b}
\end{equation}
en la cual es claro c\'omo deben expandirse las sumatorias:
\begin{eqnarray}
\partial_{\alpha} n_{\alpha} \, \partial_{\beta} n_{\beta} & = & \left(\partial_{1} n_{1} + \partial_{2} n_{2}+ \partial_{3} n_{3} \right) \partial_{\beta} n_{\beta} \nonumber \\
							   & = & \left(\partial_{1} n_{1} + \partial_{2} n_{2}+ \partial_{3} n_{3} \right)^{2} .
\nonumber
\end{eqnarray}

\begin{figure}[h]
\includegraphics[width=\linewidth]{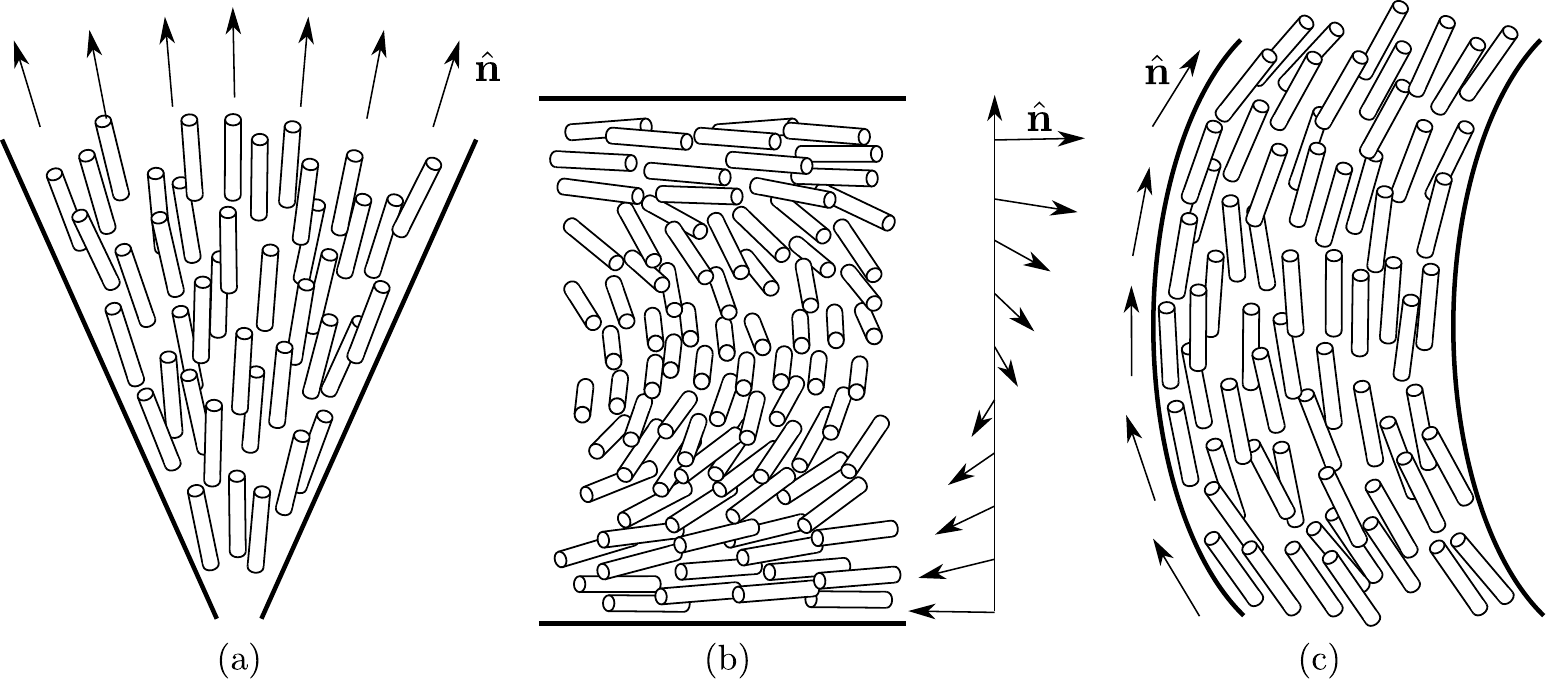}
\caption{Tres deformaciones en un CLN: (a)~deformaci\'on de arista (\textit{splay}), (b)~deformaci\'on torcida (\textit{twist}); (c)~deformaci\'on doblada (\textit{bend}).}
\label{figura_012}
\end{figure}

Como mencionamos, la divergencia del director mide una caracter\'istica geom\'etrica de las deformaciones. Un campo director que diverge tiene
un comportamiento espacial como el que se ilustra en la figura~\ref{figura_012}~(a). Cuando, $\boldsymbol{\nabla}\cdot \hat{\mathbf{n}} \neq 0$, se
dice que el CL tiene una deformaci\'on de tipo \textit{splay}. No es muy com\'un nombrar las diferentes deformaciones de los CL en espa\~nol, pero 
si tuvi\'eramos que proponer una traducci\'on para las deformaciones \textit{splay}, esta ser\'ia ``deformaciones de arista'', por el hecho de que se 
producir\'ian al confinar al CL entre dos placas planas que se van aproximando una a la otra y eventualmente se intersectar\'ian sobre una l\'inea, 
tal como se ilustra en la figura~\ref{figura_012}~(a).

Una relaci\'on \'util, que se demuestra en detalle en el ap\'endice~\ref{apendice_004}, conecta las segundas derivadas espaciales de $\hat{\mathbf{n}}$ 
con $\left( \boldsymbol{\nabla}\cdot\hat{\mathbf{n}}\right)^{2}$ de la siguiente manera
\begin{equation}
\partial_{\alpha} n_{\beta} \, \partial_{\beta}n_{\alpha} = \left( \boldsymbol{\nabla}\cdot\hat{\mathbf{n}}\right)^{2} + \boldsymbol{\nabla}\cdot\mathbf{m},
%                                                          + \partial_{\alpha} \left( n_{\beta} \partial_{\beta} n_{\alpha}
%                                                                                    -n_{\alpha}\partial_{\beta} n_{\beta}\right) ,
\label{energia_elastica_003c}
\end{equation}
donde el vector $\mathbf{m}$ se define por sus componentes como
\begin{equation}
m_{\alpha} = n_{\beta} \partial_{\beta} n_{\alpha} -n_{\alpha}\partial_{\beta} n_{\beta}.
\label{energia_elastica_003cc}
\end{equation}

Otra relaci\'on resulta del hecho de que el director es un vector unitario, $n_{\alpha} n_{\alpha} = 1$. Entonces es claro que cualquier derivada de $n_{\alpha}n_{\alpha}$ 
siempre se anula, lo que conduce a 
$\partial_{\beta}\left( n_{\alpha} n_{\alpha} \right) = 2 n_{\alpha} \partial_{\beta} n_{\alpha} = 0$ y, por lo tanto,
\begin{equation}
n_{\alpha} \partial_{\beta} n_{\alpha}  = 0.
\label{energia_elastica_003d}
\end{equation}

Por otra parte,  la escritura de las derivadas ``cruzadas'' que conforman las componentes del rotacional de $\hat{\mathbf{n}}$, tambi\'en puede simplificarse 
en la notaci\'on de \'indices como sigue,
%
%Como sabemos y es claro de la ecuaci\'on~(\ref{energia_elastica_003}), en cada direcci\'on, $\boldsymbol{\nabla}\times\hat{\mathbf{n}}$ combinan 
%derividas 
%``cruzadas'' de las componentes de $\hat{\mathbf{n}}$. Estas combinaciones pueden escribirse de manera compacta al utilizar la notaci\'on
%tensorial como sigue:
\begin{equation}
\left[\boldsymbol{\nabla}\times\hat{\mathbf{n}}\right]_{\alpha} = \varepsilon_{\alpha\beta\gamma} \partial_{\beta} n_{\gamma},
\label{energia_elastica_004}
\end{equation}
donde $\varepsilon_{\alpha\beta\gamma}$ es el \textit{s\'imbolo de Levi--Civita}, definido por la regla
%\end{multicols}
\begin{equation}
\varepsilon_{\alpha\beta\gamma} = 
   \begin{cases}
      +1, & \text{si } \left\{\alpha,\beta,\gamma\right\} = \left\{1,2,3\right\}, \,\left\{2,3,1\right\} \text{ \'o } \left\{3,1,2\right\},\\
      -1, & \text{si } \left\{\alpha,\beta,\gamma\right\} = \left\{1,3,2\right\}, \,\left\{2,1,3\right\} \text{ \'o } \left\{3,2,3\right\},\\
       0, & \text{si } \alpha=\beta,\, \alpha=\gamma \text{ \'o } \beta=\gamma.
   \end{cases}
\label{energia_elastica_005}
\end{equation}
%\medline
%\begin{multicols}{2}
%\noindent

En otras palabras, $\varepsilon_{\alpha\beta\gamma}$ vale $1$, si sus \'indices forman una permutaci\'on c\'iclica de $\left\{1,2,3\right\}$; vale $-1$, si 
sus \'indices no est\'an no est\'an repetidos y no forman una permutaci\'on c\'iclica de $\left\{1,2,3\right\}$; y se anula si cualquiera de sus \'indices 
se repite. El s\'imbolo de Levi-Civita es muy \'util para expresar determinantes y productos vectoriales en notaci\'on de \'indices, de all\'i que se use 
tambi\'en en la f\'ormula del rotacional. En particular, dados dos vectores, $\mathbf{a}$ y $\mathbf{b}$, el producto vectorial entre ellos es otro vector 
$\mathbf{c} = \mathbf{a} \times \mathbf{b}$ con componentes 
\begin{equation}
c_{\alpha} = \varepsilon_{\alpha\beta\gamma} a_{\beta} b_{\gamma}. 
\label{energia_elastica_005a}
\end{equation} 

Por definici\'on, las componentes del s\'imbolo de Levi-Civita cuyos \'indices est\'an permutados c\'iclicamente son id\'enticas, \textit{i. e.},
$\varepsilon_{\alpha\beta\gamma} = \varepsilon_{\beta\gamma\alpha} = \varepsilon_{\gamma\alpha\beta}$; mientras que dos componentes
relacionadas por una permutaci\'on no c\'iclica de \'indices tienen signos opuestos, \textit{i. e.}, $\varepsilon_{\alpha\beta\gamma} = -\varepsilon_{\beta\alpha\gamma}$.
Otras dos propiedades importantes de $\varepsilon_{\alpha\beta\gamma}$ son la contracci\'on consigo 
mismo sobre uno de sus \'indices, %que su producto externo puede calcularse como un determinante
\begin{equation}
\varepsilon_{\alpha\beta\gamma}\varepsilon_{\alpha\mu\nu} = \delta_{\beta\mu} \delta_{\gamma\nu} - \delta_{\beta\nu} \delta_{\gamma\mu},
\label{energia_elastica_006}
\end{equation}
y su producto externo consigo mismo
\begin{eqnarray}
\varepsilon_{\alpha\beta\gamma}\varepsilon_{\lambda\mu\nu} & = & 
   \begin{vmatrix} 
      \delta_{\alpha\lambda}  & \delta_{\alpha\mu} & \delta_{\alpha\nu} \\
      \delta_{\beta\lambda}   & \delta_{\beta\mu}  & \delta_{\beta\nu} \\
      \delta_{\gamma\lambda}  & \delta_{\gamma\mu} & \delta_{\gamma\nu} \\
   \end{vmatrix} \nonumber \\  
& = & \delta_{\alpha\lambda} \left( \delta_{\beta\mu} \delta_{\gamma\nu} - \delta_{\beta\nu} \delta_{\gamma\mu}  \right) \nonumber \\
&   &-\delta_{\alpha\mu} \left( \delta_{\beta\lambda} \delta_{\gamma\nu} - \delta_{\beta\nu} \delta_{\gamma\lambda} \right) \nonumber \\
&   &+\delta_{\alpha\nu} \left( \delta_{\beta\lambda} \delta_{\gamma\mu} - \delta_{\beta\mu} \delta_{\gamma\lambda} \right) .
\label{energia_elastica_010}
\end{eqnarray}

El rotacional de %$\boldsymbol{\nabla}\times\hat{\mathbf{n}}$ es un vector, \'este puede multiplicarse escalar o vectorialmente por cualquier otro. En
$\hat{\mathbf{n}}$ es un vector que puede multiplicarse escalar o vectorialmente por cualquier otro. En
particular, las multiplicaciones escalar y vectorial entre $\hat{\mathbf{n}}$ y $\boldsymbol{\nabla}\times\hat{\mathbf{n}}$, son muy importantes
por su significado geom\'etrico. Un campo de orientaciones con $\hat{\mathbf{n}} \cdot \boldsymbol{\nabla}\times\hat{\mathbf{n}} \neq 0$, tiene
una estructura distorsionada como la que se ilustra en la figura~\ref{figura_012}~(b) a la que se le llama una deformaci\'on \textit{twist}, en ingl\'es,
y que aqu\'i referiremos como una ``deformaci\'on torcida'', por el hecho de que puede producirse al confinar al CL entre dos placas paralelas 
giradas una con respecto a la otra, como se ilustra en la propia figura~\ref{figura_012}~(b).

Matem\'aticamente, el producto $\hat{\mathbf{n}} \cdot \boldsymbol{\nabla}\times\hat{\mathbf{n}}$ resulta en un escalar que, de acuerdo con la 
ecuaci\'on~(\ref{energia_elastica_004}), podemos escribir como
\begin{equation}
\hat{\mathbf{n}} \cdot \boldsymbol{\nabla}\times\hat{\mathbf{n}} 
= n_{\alpha} \left[ \boldsymbol{\nabla}\times\hat{\mathbf{n}} \right]_{\alpha} 
= n_{\alpha} \varepsilon_{\alpha\beta\gamma} \partial_{\beta} n_{\gamma}.
\label{energia_elastica_010a}
\end{equation}

La multiplicaci\'on vectorial $\hat{\mathbf{n}} \times \left( \boldsymbol{\nabla}\times\hat{\mathbf{n}}\right)$ tambi\'en esta asociada con una deformaci\'on 
caracter\'istica con la geometr\'ia ilustrada en la figura~\ref{figura_012}~(c), la cual se conoce en ingl\'es como deformaci\'on \textit{bend}. Esta deformaci\'on,
que podr\'iamos llamar ``doblada'' en espa\~nol, se produce al confinar a un CLN entre dos placas paralelas y doblarlas posteriormente para darles la forma ilustrada en
la figura~\ref{figura_012}~(c).

La componentes del producto $\hat{\mathbf{n}} \times \left( \boldsymbol{\nabla}\times\hat{\mathbf{n}}\right)$ las podemos encontrar a partir de las 
ecuaciones~(\ref{energia_elastica_004}) y (\ref{energia_elastica_005a}),
\begin{eqnarray}
\left[\hat{\mathbf{n}} \times \left( \boldsymbol{\nabla}\times\hat{\mathbf{n}} \right)\right]_{\alpha} 
& = & \varepsilon_{\alpha\beta\gamma} n_{\beta} \left[ \boldsymbol{\nabla}\times\hat{\mathbf{n}} \right]_{\gamma} \nonumber \\
& = & \varepsilon_{\alpha\beta\gamma} n_{\beta} \varepsilon_{\gamma\lambda\mu} \partial_{\lambda} n_{\mu}. \label{energia_elastica_011}
\end{eqnarray}

En el ap\'endice se demuestra que los productos en la ecuaci\'on~(\ref{energia_elastica_011}) se reducen a
\begin{equation}
\left[\hat{\mathbf{n}} \times \left( \boldsymbol{\nabla}\times\hat{\mathbf{n}} \right)\right]_{\alpha}
= - n_{\beta} \partial_{\beta} n_{\alpha} 
. \label{energia_elastica_012}
\end{equation}

Para finalizar esta secci\'on, escribiremos un par de resultados matem\'aticos que ser\'an muy valiosos por razones que se aclarar\'an m\'as adelante en la 
secci\'on~\ref{energia_oseen_frank_seccion}. Estos consisten en expresar la intensidad cuadr\'atica de las deformaciones torcida y doblada. Comencemos 
directamente con el segundo caso. De acuerdo con la ecuaci\'on~(\ref{energia_elastica_010a}), la norma al cuadrado del vector 
$\hat{\mathbf{n}} \times \left( \boldsymbol{\nabla}\times\hat{\mathbf{n}}\right)$, es
\begin{eqnarray}
\left( \hat{\mathbf{n}} \times \left( \boldsymbol{\nabla}\times\hat{\mathbf{n}}\right) \right)^{2} 
& = & \left[\hat{\mathbf{n}} \times \left( \boldsymbol{\nabla}\times\hat{\mathbf{n}} \right)\right]_{\alpha} 
  \left[\hat{\mathbf{n}} \times \left( \boldsymbol{\nabla}\times\hat{\mathbf{n}} \right)\right]_{\alpha} \nonumber \\
& = & n_{\beta} n_{\gamma} \, \partial_{\beta} n_{\alpha} \, \partial_{\gamma} n_{\alpha}
. \label{energia_elastica_013}
\end{eqnarray}

Por otra parte, el cuadrado del producto que define la deformaci\'on torcida es, de acuerdo con la ecuaci\'on~(\ref{energia_elastica_010a}),
\begin{eqnarray}
\left( \hat{\mathbf{n}} \cdot \boldsymbol{\nabla}\times\hat{\mathbf{n}} \right)^{2}
& = & n_{\alpha} n_{\lambda} \varepsilon_{\alpha\beta\gamma} \varepsilon_{\lambda\mu\nu} \, \partial_{\beta} n_{\gamma} \, \partial_{\mu} n_{\nu} \label{energia_elastica_014} \\
& = & \partial_{\alpha} n_{\beta} \, \partial_{\alpha} n_{\beta} - \partial_{\alpha} n_{\beta} \, \partial_{\beta} n_{\alpha} \nonumber \\
&   & -n_{\beta} n_{\gamma} \, \partial_{\beta} n_{\alpha} \, \partial_{\gamma} n_{\alpha} \label{energia_elastica_014a} \\ 
& = & \partial_{\alpha} n_{\beta}  \partial_{\alpha} n_{\beta} 
     -\boldsymbol{\nabla}\cdot \mathbf{m} - \left(\boldsymbol{\nabla}\cdot \hat{\mathbf{n}} \right)^{2} \nonumber \\
    %- \partial_{\alpha} 
    %  \left( n_{\alpha} \partial_{\beta} n_{\beta} - n_{\beta} \partial_{\beta} n_{\alpha} \right) \nonumber \\
&   & - \left( \hat{\mathbf{n}} \times \left( \boldsymbol{\nabla}\times\hat{\mathbf{n}}\right) \right)^{2},
\label{energia_elastica_015}
\end{eqnarray}
en donde el vector $\mathbf{m}$ se define por componenetes como
\begin{equation}
m_{\alpha} = n_{\alpha} \partial_{\beta} n_{\beta} - n_{\beta} \partial_{\beta} n_{\alpha}.
\label{vector_m}
\end{equation}

Para obtener las igualdades (\ref{energia_elastica_015}) y (\ref{vector_m}) utilizamos diversas ecuaciones presentadas en esta secci\'on,  
%~(\ref{energia_elastica_003b}), (\ref{energia_elastica_003d}), (\ref{energia_elastica_010}) y (\ref{energia_elastica_013}), 
tal como se explica en el ap\'endice~\ref{apendice_004}.

% ............................................................................................................................
\subsubsection{Energ\'ia de Oseen-Frank \label{energia_oseen_frank_seccion}}
% ............................................................................................................................

En respuesta a las deformaciones, el cristal l\'iquido genera torcas y almacena una forma de energ\'ia potencial que se 
libera cuando desaparecen las fuerzas externas sobre \'el. Para describir esta energ\'ia, se parte del modelo m\'as simple posible y
se supone que las torcas son directamente proporcionales a las deformaciones. Esto equivale a la ley del resorte de Hooke que 
%aprendemos en los cursos de mec\'anica cl\'asica, en la cual la fuerza y la energ\'ia est\'an dadas, respectivamente, por 
aprendemos en los cursos de mec\'anica cl\'asica, en la cual la fuerza del resorte es proporcional al estiramiento, $\Delta x$, y la 
energ\'ia potencial est\'a dada por 
%\begin{equation}
%F_{\text{elas}} = -k \Delta x ,
%\label{hooke_001}
%\end{equation}
\begin{equation}
U_{\text{elas}} = \frac{1}{2} k \left( \Delta x\right)^{2},
\label{hooke_002}
\end{equation}
donde $k$ es la elasticidad del resorte. 

Como se discuti\'o brevemente en la secci\'on~\ref{decaimiento_retorno_seccion}, el problema de calcular la energ\'ia libre de las 
deformaciones de un CL fue abordado primero por Oseen y Zocher~\cite{oseen_1929,zocher_discussions_1933} y posteriormente por 
Frank~\cite{frank_discussions_1958}, quien lo formul\'o en t\'erminos de la simetr\'ia de las fases y deriv\'o la expresi\'on que
se usa hasta el d\'ia de hoy con mayor frecuencia. A la densidad de energ\'ia libre el\'astica, $f_{\text{elas}}$, se le conoce en 
la literatura como \textit{energ\'ia el\'astica de Oseen-Frank}, aunque frecuentemente se le dice s\'olo \textit{energ\'ia de Frank}. 

La energ\'ia el\'astica del CL se anula en ausencia de deformaciones y se incrementa cuando ocurren variaciones espaciales del director.
As\'i, puede considerarse una expansi\'on de $f_{\text{elas}}$ en t\'erminos de las derivadas $\partial_{\alpha} n_{\beta}$, 
$\partial_{\alpha} \partial_{\beta} n_{\beta}$, etc. Al considerar que $\hat{\mathbf{n}}$ cambia de manera suave en el espacio,
dichas derivadas pueden considerarse peque\~nas y la expansi\'on de $f_{\text{elas}}$ puede truncarse hasta los t\'erminos de segundo
orden ($\partial_{\alpha} n_{\beta} \partial_{\gamma} n_{\lambda} $ y $\partial_{\alpha} \partial_{\beta} n_{\gamma}$), lo que resulta en
\begin{equation}
f_{\text{elas}} = \tilde{k}_{1} \mathcal{P}_{\alpha\beta} \partial_{\alpha} n_{\beta} 
		+ \tilde{k}_{2} \mathcal{Q}_{\alpha\beta\gamma\lambda} \partial_{\alpha} n_{\beta} \partial_{\gamma} n_{\lambda}
		+ \tilde{k}_{3} \mathcal{R}_{\alpha\beta\gamma} \partial_{\alpha} \partial_{\beta} n_{\gamma}, 
\label{energia_elastica_016}
\end{equation}
donde $\tilde{k}_{1}$, $\tilde{k}_{2}$ y $\tilde{k}_{3}$ son constantes, mientras que los tensores $\mathcal{P}_{\alpha\beta}$,
$\mathcal{Q}_{\alpha\beta\gamma\lambda}$ y $\mathcal{R}_{\alpha\beta\gamma}$ tienen el papel de agrupar los cambios espaciales
del director. Observa que los \'indices de estos tensores se contraen con todos los \'indices de las derivadas para garantizar que 
$f_{\text{elas}}$ sea un escalar. La idea que permite encontrar la forma de los tensores $\mathcal{P}_{\alpha\beta}$, 
$\mathcal{Q}_{\alpha\beta\gamma\lambda}$ y $\mathcal{R}_{\alpha\beta\gamma}$ es que estos pueden expresarse en t\'erminos de los 
vectores y tensores fundamentales $n_{\alpha}$, $\delta_{\alpha\beta}$ y $\varepsilon_{\alpha\beta\gamma}$.

Al escribir las primeras y segundas derivadas de $\hat{\mathbf{n}}$ con \'indices que pueden correr sobre todos sus valores, 
estamos permitiendo de entrada que todas las combinaciones de ellas puedan contribuir a la energ\'ia. Sin embargo, en la expansi\'on 
resultante s\'olo deben ser permitidos aquellos t\'erminos que cumplan con las dos simetr\'ias de la fase 
nem\'atica discutidas en la secci\'on~\ref{director_seccion}: $\hat{\mathbf{n}}\rightarrow -\hat{\mathbf{n}}$ y $\mathbf{r}\rightarrow -\mathbf{r}$.
El objetivo del siguiente an\'alisis %, que es el m\'as laborioso en este art\'iculo, 
es mostrar que $f_{\text{elas}}$ s\'olo tiene tres
contribuciones permitidas, la primera proporcional a  $\left(\boldsymbol{\nabla}\cdot\hat{\mathbf{n}}\right)^{2}$, la segunda a
$\left( \hat{\mathbf{n}} \cdot \boldsymbol{\nabla}\times \hat{\mathbf{n}}\right)^{2}$ y la \'ultima a
$\left( \hat{\mathbf{n}} \cdot \boldsymbol{\nabla}\times \hat{\mathbf{n}}\right)^{2}$. 

Antes de comenzar con dicho an\'alisis conviene comentar que las simetr\'ias $\hat{\mathbf{n}}\rightarrow -\hat{\mathbf{n}}$ y $\mathbf{r}\rightarrow -\mathbf{r}$
permiten una contribuci\'on adicional proporcional a la divergencia del vector $\mathbf{m}$ definido por la ecuaci\'on~(\ref{vector_m}).
Ahora bien, $f_{\text{elas}}$, al igual que $f_{\text{bulk}}$ en la secci\'on~\ref{energia_formacion_fase_seccion}, es una densidad de energ\'ia,
\textit{i. e.}, energ\'ia por unidad de volumen. $f_{\text{elas}}$ representa la densidad de energ\'ia causada por las distorsiones, almacenada 
en un elemento de volumen infinitesimal, $dV$, centrado en una posici\'on $\mathbf{r}$. La energ\'ia el\'astica total de una muestra de CL se 
obtendr\'a integrando $f_{\text{elas}}\left( \mathbf{r}\right)\, dV$ sobre todo el volumen ocupado por dicha muestra,
\begin{equation}
F_{\text{elas}} = \iiint_{V} f_{\text{elas}}\left( \mathbf{r}\right) \, dV,
\nonumber
\end{equation}
en donde hemos enfatizado el car\'acter tridimensional de la operaci\'on involucrada al escribir una integral triple. 

Para cualquier t\'ermino en la ecuaci\'on~(\ref{energia_elastica_016}) que sea una divergencia, \textit{e. g.} $\boldsymbol{\nabla}\cdot\mathbf{m}$, 
podemos utilizar el teorema de la dievergencia de Gauss y escribir su contribuci\'on a $F_{\text{elas}}$ de la siguiente manera
\begin{equation}
\iiint_{V} \boldsymbol{\nabla}\cdot\mathbf{m} \, dV = \varoiint_{\mathcal{S}} \mathbf{m} \cdot \hat{\mathbf{s}}\, d\mathcal{S} ,
\nonumber
\end{equation}
en donde $\mathcal{S}$ es la superficie cerrada que delimita a $V$, $d\mathcal{S}$ es el elemento de superficie, $\hat{\mathbf{s}}$ es el vector
unitario que apunta en la direcci\'on normal
a $\mathcal{S}$ desde el interior hacia el exterior y cuya magnitud es igual al \'area del elemento infinitesimal de superficie y la doble integral cerrada doble
enfatiza el car\'acter bidimensional de la operaci\'on. En otras palabras, la contribuci\'on de $\boldsymbol{\nabla}\cdot\mathbf{m}$ a 
la energ\'ia total se reduce al flujo de $\mathbf{m}$ sobre la superficie de la muestra. Esta contribuci\'on se interpreta entonces como
una energ\'ia superficial y no se considera cuando s\'olo se estudia la energ\'ia volum\'etrica del CL o bien, cuando se sabe que la contribuci\'on
superficial ser\'a peque\~na. Esta es la situaci\'on que consideraremos a partir de ahora.

Dicho todo lo anterior, consideremos el primer t\'ermino de la expansi\'on de $f_{\text{elas}}$ en la ecuaci\'on~(\ref{energia_elastica_016}). 
El tensor de rango dos, $\mathcal{P}_{\alpha\beta}$, puede tener tres contribuciones formadas a partir del director, la delta de Kronecker y el
s\'imbolo de Levi-Civita: $n_{\alpha}n_{\beta}$, $\delta_{\alpha\beta}$ y $\varepsilon_{\alpha\beta\gamma}n_{\gamma}$. Sin embargo, la primera
conduce a un resultado nulo al multiplicarse por $\partial_{\alpha} n_{\beta}$,
\begin{equation}
%\mathcal{P}_{\alpha\beta}^{\prime\prime} \partial_{\alpha} n_{\beta} = n_{\alpha}n_{\beta} \partial_{\alpha} n_{\beta} = 0,
n_{\alpha}n_{\beta} \partial_{\alpha} n_{\beta} = 0,
\label{energia_elastica_018}
\end{equation}
debido a la ecuaci\'on~(\ref{energia_elastica_003d}). Mientras que las \'ultimas dos contribuciones 
%
%Dos tensores de rango dos formados a partir de $n_{\alpha}$, $\delta_{\alpha\beta}$ y $\varepsilon_{\alpha\beta\gamma}$, que conducen
%a resultados distintos de cero son: $\mathcal{P}_{\alpha\beta}^{\prime} = \delta_{\alpha\beta}$,
%y $\mathcal{P}_{\alpha\beta}^{\prime\prime} =\varepsilon_{\alpha\beta\gamma}n_{\gamma}$. Al contraer estos tensores con
%$\partial_{\alpha} n_{\beta}$, se 
producen los siguientes resultados
\begin{equation}
%\mathcal{P}_{\alpha\beta}^{\prime} \partial_{\alpha} n_{\beta} = \delta_{\alpha\beta} \partial_{\alpha} n_{\beta} = \partial_{\alpha} n_{\alpha},
\delta_{\alpha\beta} \partial_{\alpha} n_{\beta} = \partial_{\alpha} n_{\alpha},
\label{energia_elastica_017}
\end{equation}
y
\begin{equation}
%\mathcal{P}_{\alpha\beta}^{\prime\prime\prime} \partial_{\alpha} n_{\beta} = \varepsilon_{\alpha\beta\gamma}n_{\gamma} \partial_{\alpha} n_{\beta}.
\varepsilon_{\alpha\beta\gamma}n_{\gamma} \partial_{\alpha} n_{\beta}.
\label{energia_elastica_019}
\end{equation}

Aunque los resultados de las ecuaciones~(\ref{energia_elastica_017}) y (\ref{energia_elastica_019}) no se cancelan, podemos ver que 
tales contribuciones no est\'an permitidas por violar una o ambas de las simetr\'ias $\hat{\mathbf{n}}\rightarrow -\hat{\mathbf{n}}$ y
$\mathbf{r}\rightarrow -\mathbf{r}$. En otras palabras, estos t\'erminos cambian de signo al invertir el director o el vector de posici\'on. 
En particular, observa que al invertir las coordenadas se tiene: $\partial n_{\alpha} /\partial (-x_{\beta}) = -\partial n_{\alpha} /\partial x_{\beta}$. 

Lo anterior se resume en la tabla~\ref{tabla_001}, en donde se consideran los productos de las contribuciones de 
$\mathcal{P}_{\alpha\beta} \partial_{\alpha} n_{\beta}$ que no se anulan y se indica con \checkmark o $\times$, respectivamente, si los resultados 
satisfacen o no las simetr\'ias y si son permitidos o no.
De este an\'alisis podemos concluir que no pueden existir contribuciones lineales en las derivadas de $\hat{\mathbf{n}}$ a $f_{\text{elas}}$.

Para el segundo t\'ermino en el lado derecho de la ecuaci\'on~(\ref{energia_elastica_016}), podemos ver que existen diversas combinaciones $n_{\alpha}$, 
$\delta_{\alpha\beta}$ y $\varepsilon_{\alpha\beta\gamma}$ que forman tensores de cuarto rango. Al utilizar s\'olo el director, se puede formar la combinaci\'on
$n_{\alpha}n_{\beta}n_{\gamma}n_{\lambda}$. Al combinar el director y la delta de Kronecker, se pueden formar seis t\'erminos: 
$\delta_{\alpha\beta}n_{\gamma}n_{\lambda}$, $\delta_{\alpha\gamma}n_{\beta}n_{\lambda}$, $\delta_{\alpha\lambda}n_{\gamma}n_{\beta}$,
$\delta_{\beta\gamma}n_{\alpha}n_{\lambda}$, $\delta_{\beta\lambda}n_{\alpha}n_{\gamma}$ y $\delta_{\gamma\lambda}n_{\alpha}n_{\beta}$. Al utilizar
el director y el s\'imbolo de Levi--Civita se forman las siguientes cuatro combinaciones: $\varepsilon_{\alpha\beta\gamma}n_{\lambda}$, 
$\varepsilon_{\alpha\gamma\lambda}n_{\beta}$, $\varepsilon_{\alpha\beta\lambda}n_{\gamma}$ y $\varepsilon_{\beta\gamma\lambda}n_{\alpha}$.
Por \'ultimo, al utilizar \'unicamente la delta de Kronecker, resultan tres t\'erminos: $\delta_{\alpha\beta}\delta_{\gamma\lambda}$, 
$\delta_{\alpha\gamma}\delta_{\beta\lambda}$ y $\delta_{\alpha\lambda}\delta_{\beta\gamma}$. La enorme mayor\'ia de las combinaciones anteriores
%tensor de rango cuatro $\mathcal{Q}_{\alpha\beta\gamma\lambda}$ 
se anulan al multiplicarse por 
$\partial_{\alpha}n_{\beta}\partial_{\gamma}n_{\lambda}$, debido a la ecuaci\'on~({\ref{energia_elastica_003d}}). Concretamente,
\begin{eqnarray}
  n_{\alpha}n_{\beta}n_{\gamma}n_{\lambda} \partial_{\alpha}n_{\beta}\,\partial_{\gamma}n_{\lambda} & = & 0, \nonumber \\
 \delta_{\alpha\beta}n_{\gamma}n_{\lambda} \partial_{\alpha}n_{\beta}\,\partial_{\gamma}n_{\lambda} & = & 0, \nonumber \\
 \delta_{\alpha\gamma}n_{\beta}n_{\lambda} \partial_{\alpha}n_{\beta}\,\partial_{\gamma}n_{\lambda} & = & 0, \nonumber \\
 \delta_{\alpha\lambda}n_{\gamma}n_{\beta} \partial_{\alpha}n_{\beta}\,\partial_{\gamma}n_{\lambda} & = & 0, \nonumber \\
 \delta_{\beta\gamma}n_{\alpha}n_{\lambda} \partial_{\alpha}n_{\beta}\,\partial_{\gamma}n_{\lambda} & = & 0, \nonumber \\
 \delta_{\gamma\lambda}n_{\alpha}n_{\beta} \partial_{\alpha}n_{\beta}\,\partial_{\gamma}n_{\lambda} & = & 0, \nonumber \\
\varepsilon_{\alpha\beta\gamma}n_{\lambda} \partial_{\alpha}n_{\beta}\,\partial_{\gamma}n_{\lambda} & = & 0, \nonumber \\
\varepsilon_{\alpha\gamma\lambda}n_{\beta} \partial_{\alpha}n_{\beta}\,\partial_{\gamma}n_{\lambda} & = & 0. \nonumber
\end{eqnarray}

Las contribuciones a $\mathcal{Q}_{\alpha\beta\gamma\lambda}$ que no se anulan %son
%$\delta_{\alpha\beta}\delta_{\gamma\lambda}$,
%$\delta_{\alpha\gamma}\delta_{\beta\lambda}$,
%$\delta_{\alpha\lambda}\delta_{\beta\gamma}$,
%$n_{\alpha}n_{\gamma}\delta_{\beta\lambda}$,
%$\varepsilon_{\alpha\beta\lambda}n_{\gamma}$ y
%$\varepsilon_{\beta\gamma\lambda}n_{\alpha}$. Al
cuando se multiplican por $\partial_{\alpha}n_{\beta}\partial_{\gamma}n_{\lambda}$ conducen a los
resultados mostrados en la tabla~\ref{tabla_002}, en donde hemos ignorado los t\'erminos que tienen la forma de una divergencia. 
La demostraci\'on de estos resultados pueden consultarse en el ap\'endice~\ref{apendice_004}.

%e ignorar los t\'erminos que tienen la forma de una divergencia, 
%se obtienen los 
%resultados mostrados en la tabla~\ref{tabla_002}. La demostraci\'on de estos resultados pueden consultarse en el ap\'endice~\ref{apendice_004}.

Por otra parte, el tensor $\mathcal{R}_{\alpha\beta\gamma}$ en la ecuaci\'on~(\ref{energia_elastica_016}), puede tener las contribuciones 
$n_{\alpha}n_{\beta}n_{\gamma}$, $\varepsilon_{\alpha\beta\gamma}$, $\delta_{\alpha\beta}n_{\gamma}$, $\delta_{\alpha\gamma}n_{\beta}$ y
$\delta_{\beta\gamma}n_{\alpha}$, que dan lugar a los resultados que se muestran en la tabla~\ref{tabla_003} y se obtienen en el 
ap\'endice~\ref{apendice_004}, despu\'es de ignorar aquellos t\'erminos con la forma de una divergencia. 
%En la misma tabla~\ref{tabla_003}, las componentes del vector $\mathbf{m}^{\prime}$ est\'an definidas como
%\begin{equation}
%m_{\alpha}^{\prime} = n_{\alpha}\, \partial_{\beta} n_{\beta}.
%\nonumber
%\end{equation}

%\end{multicols}

\begin{table}[h]
 \centering
 \caption{Posibles contribuciones de $\partial_{\alpha} n_{\beta}$ a la energ\'ia el\'astica.}
 \begin{tabular}{|c|c| >{\centering\arraybackslash}m{1.6 cm}| >{\centering\arraybackslash}m{1.6 cm}|c|}
  \hline
  Producto & Resultado & Simetr\'ia $\hat{\mathbf{n}}\rightarrow -\hat{\mathbf{n}}$ & Simetr\'ia $\mathbf{r}\rightarrow -\mathbf{r}$ & ?`Permitido? \\
  \hline
  \hline
  $\delta_{\alpha\beta} \partial_{\alpha} n_{\beta}$ & $\partial_{\alpha}n_{\alpha}$ & $\times$   & $\times$   & $\times$ \\
 %$n_{\alpha}n_{\beta} \partial_{\alpha} n_{\beta}$  & $0$                           & \checkmark & \checkmark & es nulo  \\
  $\varepsilon_{\alpha\beta\gamma}n_{\gamma}\partial_{\alpha}n_{\beta}$ & $\varepsilon_{\alpha\beta\gamma}n_{\gamma}\partial_{\alpha}n_{\beta}$ & \checkmark &$\times$&$\times$\\
  \hline
 \end{tabular}\label{tabla_001}
\end{table}

\begin{table}[h]
 \centering
 \caption{Posibles contribuciones de $\partial_{\alpha} n_{\beta} \partial_{\gamma} n_{\lambda}$ a la energ\'ia el\'astica. La columna de resultados ignora
  aquellos t\'erminos que tienen la forma de una divergencia.}
  \begin{tabular}{|c|c| >{\centering\arraybackslash}m{1.6 cm}| >{\centering\arraybackslash}m{1.6 cm}|c|}
  \hline
  Producto & Resultado & Simetr\'ia $\hat{\mathbf{n}}\rightarrow -\hat{\mathbf{n}}$ & Simetr\'ia $\mathbf{r}\rightarrow -\mathbf{r}$ & ?`Permitido? \\
  \hline
  \hline
  $\delta_{\alpha\beta}\delta_{\gamma\lambda}\partial_{\alpha}n_{\beta}\partial_{\gamma}n_{\lambda}$ 
      & $\left(\boldsymbol{\nabla}\cdot\hat{\mathbf{n}}\right)^{2}$ 
         & \checkmark & \checkmark & \checkmark \\
  $\delta_{\alpha\gamma}\delta_{\beta\lambda}\partial_{\alpha}n_{\beta}\partial_{\gamma}n_{\lambda}$ 
      & $\left(\boldsymbol{\nabla}\cdot\hat{\mathbf{n}}\right)^{2} 
	+ \left(\hat{\mathbf{n}}  \cdot \boldsymbol{\nabla} \times \hat{\mathbf{n}}\right)^{2} 
	+ \left(\hat{\mathbf{n}}  \times \boldsymbol{\nabla} \times \hat{\mathbf{n}}\right)^{2}
        $%+ \boldsymbol{\nabla}\cdot\mathbf{m}$ %+ \partial_{\alpha} \left( n_{\alpha} \partial_{\beta}n_{\beta} - n_{\beta} \partial_{\beta}n_{\alpha} \right)$
         & \checkmark & \checkmark & \checkmark  \\
  $\delta_{\alpha\lambda}\delta_{\beta\gamma}\partial_{\alpha}n_{\beta}\partial_{\gamma}n_{\lambda}$ & $\left(\boldsymbol{\nabla}\cdot\hat{\mathbf{n}}\right)^{2}$ & \checkmark & \checkmark & \checkmark \\
  $\delta_{\beta\lambda}n_{\alpha}n_{\gamma}\partial_{\alpha}n_{\beta}\partial_{\gamma}n_{\lambda}$ 
      & $\left(\hat{\mathbf{n}} \times \boldsymbol{\nabla} \times \hat{\mathbf{n}}\right)^{2}$ & \checkmark & \checkmark& \checkmark \\
  $\varepsilon_{\alpha\beta\lambda}n_{\gamma}\partial_{\alpha}n_{\beta}\partial_{\gamma}n_{\lambda}$ 
      & $\varepsilon_{\alpha\beta\lambda}n_{\gamma}\partial_{\alpha}n_{\beta}\partial_{\gamma}n_{\lambda}$    
      & $\times$ & \checkmark& $\times$ \\
  $\varepsilon_{\beta\gamma\lambda}n_{\alpha}\partial_{\alpha}n_{\beta}\partial_{\gamma}n_{\lambda}$ 
      & $\varepsilon_{\beta\gamma\lambda}n_{\alpha}\partial_{\alpha}n_{\beta}\partial_{\gamma}n_{\lambda}$    
      & $\times$ & \checkmark& $\times$ \\
 \hline
 \end{tabular}\label{tabla_002}
\end{table}

\begin{table}[h]
 \centering
 \caption{Posibles contribuciones de $\partial_{\alpha} \partial_{\beta} n_{\gamma}$ a la energ\'ia el\'astica. La columna de resultados ignora
  aquellos t\'erminos que tienen la forma de una divergencia.}
  \begin{tabular}{|c|c| >{\centering\arraybackslash}m{1.6 cm}| >{\centering\arraybackslash}m{1.6 cm}|c|}
  \hline
  Producto & Resultado & Simetr\'ia $\hat{\mathbf{n}}\rightarrow -\hat{\mathbf{n}}$ & Simetr\'ia $\mathbf{r}\rightarrow -\mathbf{r}$ & ?`Permitido? \\
  \hline
  \hline
  $n_{\alpha}n_{\beta}n_{\gamma}\partial_{\alpha}\partial_{\beta}n_{\gamma}$ 
      & $-\left(\hat{\mathbf{n}}  \times \left( \boldsymbol{\nabla} \times \hat{\mathbf{n}}\right) \right)^{2}$ 
         & \checkmark & \checkmark & \checkmark \\
  $\varepsilon_{\alpha\beta\gamma}\partial_{\alpha}\partial_{\beta}n_{\gamma}$ 
      & $\varepsilon_{\alpha\beta\gamma}\partial_{\alpha}\partial_{\beta}n_{\gamma}$ %+ \partial_{\alpha} \left( n_{\alpha} \partial_{\beta}n_{\beta} - n_{\beta} \partial_{\beta}n_{\alpha} \right)$
         & $\times$ & \checkmark & $\times$  \\
  $\delta_{\alpha\beta} n_{\gamma}\partial_{\alpha}\partial_{\beta}n_{\gamma}$ 
      & $-\left(\boldsymbol{\nabla}\cdot\hat{\mathbf{n}}\right)^{2} 
        - \left(\hat{\mathbf{n}}  \cdot \boldsymbol{\nabla} \times \hat{\mathbf{n}}\right)^{2} 
        - \left(\hat{\mathbf{n}}  \times \left( \boldsymbol{\nabla} \times \hat{\mathbf{n}}\right) \right)^{2}
        $ %- \boldsymbol{\nabla}\cdot\mathbf{m}$
         & \checkmark & \checkmark & \checkmark \\
  $\delta_{\alpha\gamma}n_{\beta} \partial_{\alpha}\partial_{\beta}n_{\gamma}$ 
      %& $\boldsymbol{\nabla}\cdot\mathbf{m}^{\prime}-\left(\boldsymbol{\nabla}\cdot\hat{\mathbf{n}}\right)^{2}$ & \checkmark & \checkmark & \checkmark \\
      & $-\left(\boldsymbol{\nabla}\cdot\hat{\mathbf{n}}\right)^{2}$ & \checkmark & \checkmark & \checkmark \\
  $\delta_{\beta\gamma} n_{\alpha}\partial_{\alpha}\partial_{\beta}n_{\gamma}$ 
      %& $\boldsymbol{\nabla}\cdot\mathbf{m}^{\prime}-\left(\boldsymbol{\nabla}\cdot\hat{\mathbf{n}}\right)^{2}$ & \checkmark & \checkmark & \checkmark \\
      & $-\left(\boldsymbol{\nabla}\cdot\hat{\mathbf{n}}\right)^{2}$ & \checkmark & \checkmark & \checkmark \\
  \hline
 \end{tabular}\label{tabla_003}
\end{table}

%\medline
%\begin{multicols}{2}

Los resultados de las tablas~\ref{tabla_001} a \ref{tabla_003} demuestran, como hab\'iamos anticipado, que todas las contribuciones permitidas a $f_{\text{elas}}$ pueden
agruparse en las intensidades cuadr\'aticas de las deformaciones de arista, $\left(\boldsymbol{\nabla}\cdot\hat{\mathbf{n}}\right)^{2}$, 
torcida, $\left(\hat{\mathbf{n}}  \cdot \boldsymbol{\nabla} \times \hat{\mathbf{n}}\right)^{2}$, y doblada, 
$\left(\hat{\mathbf{n}}  \times \left( \boldsymbol{\nabla} \times \hat{\mathbf{n}}\right)\right)^{2}$. %, as\'i como en la divergencia de un vector. 
La forma est\'andar de agrupar estas contribuciones es 
\begin{eqnarray}
f_{\text{elas}} & = & \frac{1}{2} K_{1} \left(\boldsymbol{\nabla}\cdot\hat{\mathbf{n}}\right)^{2}
                 +\frac{1}{2} K_{2} \left(\hat{\mathbf{n}}  \cdot \boldsymbol{\nabla} \times \hat{\mathbf{n}}\right)^{2} \nonumber \\
                 &   &  +\frac{1}{2} K_{3} \left(\hat{\mathbf{n}}  \times \boldsymbol{\nabla} \times \hat{\mathbf{n}}\right)^{2}.
              % + \boldsymbol{\nabla}\cdot\mathbf{p}^{\prime}.
\label{energia_elastica_021}
\end{eqnarray}
%\end{multicols}
%\begin{equation}
%f_{\text{elas}} = \frac{1}{2} K_{1} \left(\boldsymbol{\nabla}\cdot\hat{\mathbf{n}}\right)^{2}
%                 +\frac{1}{2} K_{2} \left(\hat{\mathbf{n}}  \cdot \boldsymbol{\nabla} \times \hat{\mathbf{n}}\right)^{2}
%		 +\frac{1}{2} K_{3} \left(\hat{\mathbf{n}}  \times \left(\boldsymbol{\nabla} \times \hat{\mathbf{n}}\right) \right)^{2}.
%\label{energia_elastica_021}
%\end{equation}
%\medline
%\begin{multicols}{2}

La ecuaci\'on~(\ref{energia_elastica_021}) resalta la similitud entre la elasticidad de un CL con la elasticidad de un resorte, ecuaci\'on~(\ref{hooke_002}). 
En efecto, para un resorte de Hooke,
en el que la fuerza es proporcional al estiramiento, la energ\'ia crece cuadr\'aticamente con la distorsi\'on. Para un CL la energ\'ia el\'astica es
tambi\'en es cuadr\'atica, s\'olo que en las tres deformaciones independientes ilustradas en las figuras~\ref{figura_012}~(a) a (c). Esto es una indicaci\'on
de que el modelo de Oseen-Frank tambi\'en est\'a basado en una aproximaci\'on lineal, en el que las torcas sobre el director son proporcionales a 
las deformaciones, aunque, por brevedad, no abundaremos m\'as sobre la descripci\'on matem\'atica de dicha relaci\'on. En la 
ecuaci\'on~(\ref{energia_elastica_021}), cada contribuci\'on tiene su propio coeficiente de elasticidad. A $K_{1}$, $K_{2}$ y $K_{3}$ se les conoce
como las constantes de Frank o constantes el\'asticas \textit{splay}, \textit{twist} y \textit{bend}, respectivamente. Ya que $f_{\text{elas}}$ se mide en 
unidades de energ\'ia sobre volumen y $\hat{\mathbf{n}}$ es adimensional, las unidades de las constantes de Frank son las mismas que las de la fuerza.
T\'ipicamente sus valores son del orden de los $\text{pN}$ ($10^{-12}~\text{N}$) y tienen una dependencia fuerte con la temperatura. Mediciones 
reportan para el 5CB a $24^{\circ}\text{C}$, $K_{1} = 6.5~\text{pN}$, $K_{2} = 3.5~\text{pN}$ y 
$K_{3} = 9.8~\text{pN}$~\cite{iglesias_acs_appl_mater_interfaces_2012}. Para la misma substancia a los $30^{\circ}\text{C}$ se reportan 
$K_{1} = 5.3~\text{pN}$, $K_{2} = 3.2~\text{pN}$ y $K_{3} = 5.8~\text{pN}$ y cerca de los $26^{\circ}\text{C}$ otras mediciones dan resultados
similares: $K_{1} = 6.2~\text{pN}$, $K_{2} = 3.9~\text{pN}$ y $K_{3} = 8.2~\text{pN}$~\cite{stewart_2019}. Para el CLN MBBA cerca de los 
$25^{\circ}\text{C}$, los valores reportados son~\cite{haller_prog_solid_state_ch_1975,stewart_2019}
$K_{1} = 6.0~\text{pN}$, $K_{2} = 3.8~\text{pN}$ y $K_{3} = 7.5~\text{pN}$. Mientras que para el PAA a los $125^{\circ}\text{C}$ se han reportado~\cite{stewart_2019}
$K_{1} = 4.5~\text{pN}$, $K_{2} = 2.9~\text{pN}$ y $K_{3} = 9.5~\text{pN}$.

Se puede observar que la constante asociada al doblamiento, $K_{3}$, es significativamente mayor que las otras dos. Por su parte, la constante de torsi\'on, 
$K_{2}$, es peque\~na.

En muchas situaciones de inter\'es, la energ\'ia descrita por la ecuaci\'on~(\ref{energia_elastica_021}) es demasiado compleja para tratarse anal\'itica o num\'ericamente.
Cuando esto ocurre, suele utilizarse una simplificaci\'on que consiste en suponer que todas las constantes de Frank tienen el mismo valor, \textit{i. e.}, 
$K_{1} = K_{2} = K_{3} = K$. A esta simplificaci\'on se le llama la \textit{aproximaci\'on de constantes el\'asticas iguales}. Bajo ella, $f_{\text{elas}}$ adquiere
la siguiente forma, que es sencilla de recordar pues s\'olo involucra a la divergencia y rotacional del director,
\begin{equation}
f_{\text{elas}} = \frac{1}{2} K 
                  \left[ \left(\boldsymbol{\nabla} \cdot  \hat{\mathbf{n}}\right)^{2}
		        +\left(\boldsymbol{\nabla} \times \hat{\mathbf{n}}\right)^{2} 
                  \right].
\label{energia_elastica_021a}
\end{equation}

El uso de la notaci\'on de \'indices facilita mucho el poder reducir la ecuaci\'on~(\ref{energia_elastica_021}) a la ecuaci\'on~(\ref{energia_elastica_021a}),
como se demuestra en el ap\'endice~\ref{apendice_005}.

Los valores experimentales de las constantes $K_{1}$, $K_{2}$ y $K_{3}$ que presentamos previamente para el 5CB, el MBBA y el PAA, claramente prohiben
que podamos considerar exactas las predicciones basadas en la ecuaci\'on~(\ref{energia_elastica_021a}). A pesar de esto, la aproximaci\'on de constantes el\'asticas
iguales es muy \'util para comprender de manera cualitativa y describir aproximadamente las distorsiones del campo director. 

% .....................................................................................................................
\subsubsection{Representaci\'on tensorial de la energ\'ia el\'astica \label{representacion_tensorial_energia_elastica_seccion}}
% .....................................................................................................................

Adem\'as de las consideraciones que hemos hecho en la seccion~\ref{energia_oseen_frank_seccion}, una limitaci\'on importante de la ecuaci\'on~(\ref{energia_elastica_021}) 
que podemos se\~nalar es que, al deducirla, no hemos tomado en cuenta que adem\'as del director,
el par\'ametro de orden escalar tambi\'en puede sufrir variaciones espaciales. Esto es particularmente importante en el caso de los llamados CL coloidales, que se forman
al introducir part\'iculas de tama\~no micro o nanom\'etrico en un CL~\cite{stark_phys_rep_2001,musevic_liquid_crystal_colloids_2017}. En estos casos se forman singularidades 
alrededor de las part\'iculas intrusas, conocidos como defectos topol\'ogicos, 
que pueden tomar la forma de puntos o curvas con un n\'ucleo con un valor muy bajo de $S$, como si all\'i el CL se hubiera fundido.
Esto genera variaciones muy grandes en la energ\'ia el\'astica que la ecuaci\'on~(\ref{energia_elastica_021}) no puede recuperar. 
M\'as a\'un, en el n\'ucleo de los defectos el director queda indefinido y la ecuaci\'on~(\ref{energia_elastica_021}) pierde
sentido. En estos casos es necesario representar la energ\'ia el\'astica en t\'erminos de los cambios espaciales del tensor $Q_{\alpha\beta}$, 
los cuales cuantifican de manera m\'as acertada las deformaciones del CL.

De manera similar al proceso que llevamos a cabo para obtener la ecuaci\'on~(\ref{energia_elastica_021}), las derivadas espaciales
$\partial_{\alpha} Q_{\beta\gamma}$, $\partial_{\alpha}\partial_{\beta}Q_{\gamma\lambda}$, etc., pueden considerarse peque\~nas y $f_{\text{elas}}$ puede
expandirse en t\'erminos de ellas. Sin embargo, debido a que $Q_{\alpha\beta}$ tiene un rango tensorial mayor que $\hat{\mathbf{n}}$, llevar
cabo esta expasi\'on es un procedimiento matem\'atico m\'as complejo. Aqu\'i nos limitaremos a presentar la forma est\'andar de la expansi\'on resultante y
se\~nalar su conexi\'on con la ecuaci\'on~(\ref{energia_elastica_021}). Espec\'ificamente, $f_{\text{elas}}$ en la representaci\'on tensorial suele
escribirse como
\begin{eqnarray}
f_{\text{elas}} & = & \frac{1}{2} L_{1} \partial_{\gamma} Q_{\alpha\beta}\, \partial_{\gamma} Q_{\alpha\beta}
                + \frac{1}{2} L_{2} \partial_{\beta} Q_{\alpha\beta} \, \partial_{\gamma} Q_{\alpha\gamma} \nonumber \\
                &   & + \frac{1}{2} L_{3} Q_{\alpha\beta} \, \partial_{\alpha} Q_{\gamma\lambda} \, \partial_{\beta} Q_{\gamma\lambda}.
\label{energia_elastica_022}
\end{eqnarray}
%\end{multicols}
%\begin{equation}
%f_{\text{elas}} = \frac{1}{2} L_{1} \partial_{\gamma} Q_{\alpha\beta}\, \partial_{\gamma} Q_{\alpha\beta}
%                + \frac{1}{2} L_{2} \partial_{\beta} Q_{\alpha\beta} \, \partial_{\gamma} Q_{\alpha\gamma}
%		+ \frac{1}{2} L_{3} Q_{\alpha\beta} \, \partial_{\alpha} Q_{\gamma\lambda} \, \partial_{\beta} Q_{\gamma\lambda}.
%\label{energia_elastica_022}
%\end{equation}
%\medline
%\begin{multicols}{2}

Observa que en la ecuaci\'on~(\ref{energia_elastica_022}) hay sumatorias impl\'icitas sobre todos los \'indices involucrados, lo que implica que $f_{\text{elas}}$ es,
como deber\'ia, un escalar. Cabe resaltar aqu\'i la utilidad de la notaci\'on de \'indices que permite condensar en una f\'ormula relativamente compacta una expansi\'on
con una forma expl\'icita muy larga. Espec\'ificamente, si se expandieran las sumatorias sobre \'indices repetidos en la ecuaci\'on~(\ref{energia_elastica_022}), se
observar\'ia que el primer, segundo y tercer t\'erminos en su lado derecho consisten a su vez de $27$, $27$ y $81$ t\'erminos, respectivamente. Por ello, se recomienda 
prestar mucho atenci\'on a dicha expansi\'on al acercarse por primera vez a su uso. 

%puede resultar confuso lidiar con dichas sumatorias, por lo que que aqu\'i se ejemplifica c\'omo
%desarrollarlas, con el caso del primer t\'ermino del lado derecho de la ecuaci\'on~(\ref{energia_elastica_022}). Al desarrollar primero la suma sobre el \'indice $\alpha$
%y posteriormente sumar sobre $\beta$ se obtiene
%\begin{eqnarray}
%\partial_{\gamma} Q_{\alpha\beta}\, \partial_{\gamma} Q_{\alpha\beta} & = & \partial_{\gamma} Q_{\alpha\beta}\, \partial_{\gamma} Q_{\alpha\beta} \nonumber \\
%\end{eqnarray}

Los coeficientes de la expansi\'on en la ecuaci\'on~(\ref{energia_elastica_022}), $L_{1}$, $L_{2}$ y $L_{3}$, son las constantes el\'asticas correspondientes a la representaci\'on
tensorial de tres deformaciones fundamentales y tienen la caracter\'istica de ser indepenedientes de $S$. La relaci\'on entre las constantes $K_{i}$ y $L_{i}$,
para $i = 1,2,3$, puede establecerse al sustituir la ecuaci\'on~(\ref{parametro_orden_tensorial_004}) en la ecuaci\'on~(\ref{energia_elastica_022}), desarrollar
las derivadas involucradas suponiendo que $S$ es uniforme para as\'i comparar miembro a miebro el resultado con la ecuaci\'on~(\ref{energia_elastica_021}). 
Esto conduce a
\begin{equation}
L_{1} = 2 \frac{3 K_{2} - K_{1} + K_{3}}{27 S^{2}},
\label{energia_elastica_023}
\end{equation}
\begin{equation}
L_{2} = 4\frac{K_{1} - K_{2}}{9 S^{2}},
\label{energia_elastica_024}
\end{equation}
y
\begin{equation}
L_{3} = 4 \frac{K_{3} - K_{1}}{27 S^{3}},
\label{energia_elastica_025}
\end{equation}
tal como se demuestra en detalle en el ap\'endice~\ref{apendice_006}.

En la pr\'actica, los valores de $L_{1}$, $L_{2}$ y $L_{3}$ suelen asignarse utilizando las ecuaciones~(\ref{energia_elastica_023}) a (\ref{energia_elastica_025})
y los valores experimentales para las constantes de Frank y el par\'ametro de orden escalar~\cite{blanc_phys_rev_lett_2005,mandal_phys_rev_e_2019,mandal_eur_phys_j_e_2021}.
En la aproximaci\'on de constantes iguales, las ecuaciones~(\ref{energia_elastica_023}) a (\ref{energia_elastica_025}) resultan en
\begin{equation}
L_{1} = L = \frac{2}{9}\frac{K}{S^{2}}
\label{energia_elastica_026}
\end{equation}
y $L_{2} = L_{3} = 0$, lo que reduce la ecuaci\'on~(\ref{energia_elastica_022}) a
\begin{equation}
f_{\text{elas}} = \frac{1}{2} L \partial_{\gamma} Q_{\alpha\beta}\, \partial_{\gamma} Q_{\alpha\beta}.
\label{energia_elastica_027}
\end{equation}

La ecuaciones~(\ref{energia_elastica_021}) y (\ref{energia_elastica_022}) son las f\'ormulas fundamentales de la teor\'ia de los CL descritos
como materia continua y son ampliamente utilizadas en investigaciones actuales, donde se consideran las contribuciones a la energ\'ia estudiadas aqu\'i,
$f_{\text{LdG}}+f_{\text{elas}}$, y otras que representan la interacci\'on del CLN con superficies y campos externos, para analizar las distorsiones
resultantes en el alineamiento molecular~\cite{hashemi_nature_2017,musevic_materials_2017,smalyukh_rep_prog_phys_2020}.

%\begin{equation}
%\boldsymbol{\nabla} \hat{\mathbf{n}} = \left(
%                                       \begin{matrix} 
%                                       \frac{\partial n_{x}}{\partial x} & \frac{\partial n_{y}}{\partial x} & \frac{\partial n_{z}}{\partial x} \\
%                                       \frac{\partial n_{x}}{\partial y} & \frac{\partial n_{y}}{\partial y} & \frac{\partial n_{z}}{\partial y} \\
%                                       \frac{\partial n_{x}}{\partial z} & \frac{\partial n_{y}}{\partial z} & \frac{\partial n_{z}}{\partial z} \\
%                                       \end{matrix} 
%                                       \right) .
%\label{energia_elastica_001}
%\end{equation}

% ====================================================================================================================================================
\section{Reflexiones finales \label{reflexiones_finales_seccion}}
% ====================================================================================================================================================

En este art\'iculo hemos presentado algunas de las propiedades b\'asicas de las fases intermedias de la materia conocidas como CL. Despu\'es de discutir 
sobre su historia y tipos principales, hemos presentado una descripci\'on matem\'atica detallada de las caracter\'isticas que defienden la estructura, 
simetr\'ia y energ\'ia de la fase de CL m\'as simple, \textit{i. e.}, la llamada fase nem\'atica. Al entender que la lectura de este trabajo pudiera 
representar un primer acercamiento al tema de los CL, hemos excluido la exposici\'on de situaciones en donde los CLN est\'an fluyendo o se ven afectados 
por campos externos, superficies o part\'iculas intrusas. Todas estas situaciones son de enorme inter\'es fundamental y pr\'actico pero, dada su complejidad 
y extensi\'on, consideramos que es mejor discutirlas en entregas subsecuentes. En una segunda parte de este curso se presentar\'an, precisamente, los efectos
de campos electrost\'aticos y magnetost\'aticos sobre el campo director y el anclaje que causan superficies s\'olidas en contacto con CLN y se
analizar\'an en detalle diversos patrones de orientaci\'on que surgen como consecuencia de estos efectos.

Como hemos podido apreciar, a\'un en este caso, el tratamiento matem\'atico de un CL no es sencillo. El entendimiento de sus conceptos y propiedades 
b\'asicas requiere del uso de herramientas matem\'aticas que abarcan, entre otras, la probabilidad, el c\'alculo vectorial, el \'algebra lineal y el 
\'algebra de tensores. En este art\'iculo, hemos procurado explicar c\'omo deben utilizarse estas herramientas y en algunos casos hemos, incluso, presentado 
las definiciones de las operaciones utilizadas. Lo anterior vuelve a este primer curso de CL autocontenido y, en nuestra opini\'on, brinda dos ventajas 
adicionales. Primero, se aprecia la relevancia de todas estas ramas de las matem\'aticas en la f\'isica, con \'enfasis en el estudio de la materia suave. 
Segundo, las t\'ecnicas aprendidas aqu\'i pueden utilizarse para el estudio de otros sistemas, \textit{e. g.} el manejo de tensores para el estudio de la 
mec\'anica de fluidos o el medio continuo. 

Asimismo, hemos discutido sobre la conexi\'on de los CL con ramas de la f\'isica y la qu\'imica como el electromagnetismo, la termodin\'amica, la \'optica 
y la qu\'imica org\'anica. En la investigaci\'on actual, los CL son de inter\'es en una diversidad mucho m\'as grande de campos. Proveen de ejemplos tangibles
y casos de estudio a la topolog\'ia y la teor\'ia de nudos~\cite{smalyukh_rep_prog_phys_2020}, a los m\'etodos num\'ericos de soluci\'on de ecuaciones 
diferenciales parciales~\cite{han_siam_j_appl_math_2020,han_nonlinearity_2021} y de la f\'isica estad\'istica 
computacional~\cite{andrienko_j_mol_liq_2018,hijar_fluct_noise_lett_2019,reyes_physica_a_2020,gonzalez_rev_mex_fis_2022} donde muchas de las f\'ormulas
que hemos ense\~nado aqu\'i son ampliamente utilizadas. Tambi\'en son enormemente
estudiados por su relevancia en potenciales aplicaciones biom\'edicas~\cite{stasiek_opt_laser_technol_2006,coles_nat_photonics_2010,bisoyi_chem_rev_2021}.

Todo lo anterior demuestra que el estudio de los CL es multidisciplinario y complejo y que su aprendizaje deber\'ia considerarse fundamental para todos los 
interesados en desarrollar investigaci\'on de frontera en el campo de la materia condensada suave. 

\section*{Agradecimientos}

H. H. agradece el apoyo econ\'onmico del CONAHCyT a trav\'es del proyecto CB 2017-2018 A1-S-46608 ``Simulaci\'on multi-escala de 
cristales l\'iquidos nem\'aticos'' y al Dr. V\'ictor Duarte Alaniz por comentarios muy valiosos que han servido principalmente
para escribir de manera correcta diversos t\'erminos en la secci\'on~\ref{fases_liquidocristalinas_seccion}.

\appendix

% ==============================================================================================================
\section{Valores l\'imite del par\'ametro de orden\label{apendice_001}}
% ==============================================================================================================

\begin{figure}[h]
\centering
\includegraphics[scale = 0.60]{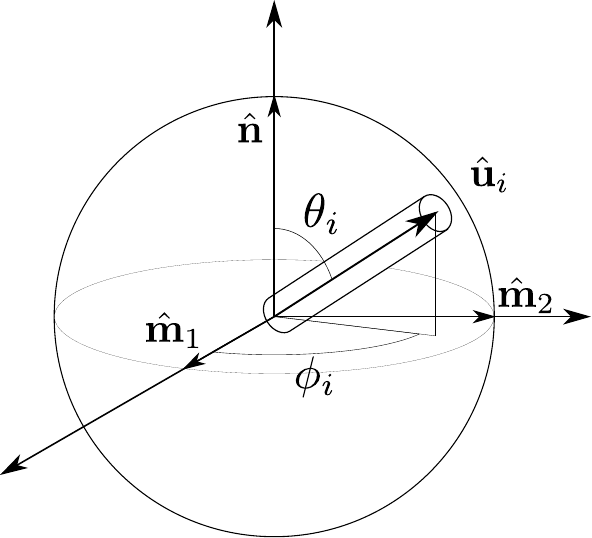}
\caption{Sistema de referencia utilizado para calcular el par\'ametro de orden escalar.}
\label{figura_013}
\end{figure}

Para calcular $S$, puede considerarse un marco de referencia en donde $\hat{\mathbf{n}}$ coincide con el eje polar de un sistema
cartesiano, tal como se muestra en la figura~\ref{figura_013}. El eje molecular, $\hat{\mathbf{u}}_{i}$, queda determinado por
los \'angulos polar, $\theta_{i}$, y azimutal, $\phi_{i}$. Dada la simetr\'ia uniaxial, \'este \'ultimo puede medirse con respecto 
a cualquier eje sobre el plano perpendicular a $\hat{\mathbf{n}}$. 

De acuerdo con la ecuaci\'on~(\ref{descripcion_matematica_001}), $S$ puede calcularse de la forma
\begin{equation}
S = \frac{1}{2} \iint\limits_{\text{esfera unitaria}} d\Omega_{i} P\left(\phi_{i}, \theta_{i}\right) \left(3\cos^{2}\theta_{i} - 1\right) ,
\label{apendice_001_001}
\end{equation}
donde 
\begin{equation}
d\Omega_{i} = \text{sen}\theta_{i}\,d\phi_{i}\,d\theta_{i},
\nonumber
\end{equation}
es el elemento de \'angulo s\'olido de las coordenadas esf\'ericas y $P\left(\phi_{i}, \theta_{i}\right)$ la densidad de probabilidad
de que $\hat{\mathbf{u}}_{i}$ est\'e orientado a lo largo de los \'angulos $\phi_{i}$ y $\theta_{i}$.

Cuando un CLN est\'a completamente desordenado, $P\left(\phi_{i}, \theta_{i}\right)$ se distribuye uniformemente sobre la esfera 
unitaria. La probabilidad de encontrar $\hat{\mathbf{u}}_{i}$ dentro del elemento de \'angulo s\'olido $d\Omega_{i}$ es 
%proporcional a dicho elemento,
%concretamente,
%$d\Omega_{i} = \text{sen}\theta_{i}\,d\phi_{i}\,d\theta_{i}$, en donde $\hat{\mathbf{n}}$ se considera el eje
%polar y $\phi_{i}$ un \'angulo azimutal que, dada la simetr\'{\i}a uniaxial, puede medirse con respecto a cualquier eje sobre el plano
%perpendicular a $\hat{\mathbf{n}}$. La distribuci\'on uniforme sobre la esfera es
\begin{equation}
P(\phi_{i},\theta_{i}) d\Omega_{i} = \frac{1}{4 \pi}  \text{sen}\theta_{i}\,d\phi_{i}\,d\theta_{i}.
\label{apendice_001_002}
\end{equation}

Al sustituir la ecuaci\'on~(\ref{apendice_001_002}) en (\ref{apendice_001_001}) se obtiene
\begin{eqnarray}
S & = & \frac{1}{8\pi} \int_{0}^{2\pi} d\phi_{i}\int_{0}^{\pi} d\theta_{i} \, \text{sen}\theta_{i} \left(3\cos^{2}\theta_{i} - 1\right) \nonumber \\
  & = & \frac{1}{4   } 
        \left[
        3 \int_{0}^{\pi} d\theta_{i} \, \text{sen}\theta_{i} \cos^{2}\theta_{i}
       -  \int_{0}^{\pi} d\theta_{i} \, \text{sen}\theta_{i}
        \right] \nonumber \\
  & = & \frac{1}{4   } 
	\left[ 3 \frac{2}{3} - 2
        \right] \nonumber \\
  & = & 0. \nonumber
%\label{descripcion_matematica_003}
\end{eqnarray}

Por otra parte, cuando todas las mol\'eculas est\'an perfectamente alineadas con el director tenemos $\hat{\mathbf{u}}_{i} = \hat{\mathbf{n}}$ y al
sustituir esta igualdad en la ecuaci\'on~(\ref{descripcion_matematica_001}) nos queda
\begin{equation}
S = \frac{1}{2} \langle 3 \left(\hat{\mathbf{n}} \cdot \hat{\mathbf{n}} \right)^2 - 1 \rangle = \frac{1}{2} \langle 3 - 1 \rangle = 1.
\nonumber
%\label{descripcion_matematica_004}
\end{equation}

% ==============================================================================================================
\section{Valores y vectores propios del tensor de par\'ametro de orden en el caso uniaxial\label{apendice_002}}
% ==============================================================================================================

Para llevar a cabo los c\'alculos en este ap\'endice utilizaremos notaci\'on matricial, en la cual el tensor de par\'ametro de orden tiene 
la forma %dada por la ecuaci\'on~(\ref{parametro_orden_tensorial_005}).
\begin{equation}
\mathbf{Q} = \frac{1}{2} \langle 3 \hat{\mathbf{u}} \hat{\mathbf{u}} - \mathbf{I} \rangle .
\label{apendice_002_001}
\end{equation}

En el caso uniaxial, el \'unico eje distitivo es paralelo a $\hat{\mathbf{n}}$. Las direcciones perpendiculares a
$\hat{\mathbf{n}}$ son todas equivalentes. Entonces, por simetr\'ia, puede anticiparse que $\hat{\mathbf{n}}$ ser\'a un vector propio de
$\mathbf{Q}$. Consecuentemente, $\mathbf{Q} \cdot \hat{\mathbf{n}} = \lambda \hat{\mathbf{n}}$, donde $\lambda$ es el valor propio correspondiente.
Entonces, al multiplicar por $\hat{\mathbf{n}}^{\text{T}}$ por la izquierda podremos escribir
\begin{equation}
\hat{\mathbf{n}}^{\text{T}}\cdot \mathbf{Q} \cdot \hat{\mathbf{n}} = \hat{\mathbf{n}}^{\text{T}} \cdot \lambda \hat{\mathbf{n}} = \lambda.
\label{apendice_002_004}
\end{equation}

Ahora es posible demostrar que, efectivamente, $\lambda = S$ . Para ello, sustituyamos la 
ecuaci\'on~(\ref{apendice_002_001}) en (\ref{apendice_002_004}),
\begin{equation}
\lambda =   \frac{1}{2} \hat{\mathbf{n}}^{\text{T}}\cdot 
         \langle 3 \hat{\mathbf{u}} \hat{\mathbf{u}} - \mathbf{I} \rangle 
         \cdot \hat{\mathbf{n}} .
\nonumber
\end{equation}

Dado que $\hat{\mathbf{n}}$ es constante, puede entrar en los \textit{brakets} que indican el promedio. As\'i se obtiene
\begin{eqnarray}
\lambda & = & \frac{1}{2} 
              \left( 3 \hat{\mathbf{n}}^{\text{T}}\cdot \left\langle \hat{\mathbf{u}} \hat{\mathbf{u}} \right\rangle \cdot \hat{\mathbf{n}} 
                     - \hat{\mathbf{n}}^{\text{T}}\cdot \mathbf{I} \cdot \hat{\mathbf{n}}
              \right) \nonumber \\
        & = & \frac{1}{2} 
              \left( 3\left\langle  \hat{\mathbf{n}}^{\text{T}}\cdot \hat{\mathbf{u}} \hat{\mathbf{u}} \cdot \hat{\mathbf{n}} \right\rangle 
                     - \hat{\mathbf{n}}^{\text{T}} \cdot \hat{\mathbf{n}}
              \right) \nonumber \\
        & = & \frac{1}{2} 
	      \left[ 3\left\langle \left( \hat{\mathbf{u}} \cdot \hat{\mathbf{n}} \right)^{2} \right\rangle 
                     - 1
	      \right] \nonumber \\
        & = & \frac{1}{2} 
	      \left\langle 3 \left( \hat{\mathbf{u}} \cdot \hat{\mathbf{n}} \right)^{2} 
                     - 1
	      \right\rangle = S . \nonumber %\\
\end{eqnarray}

%en donde en la tercera igualdad hemos utilizado las ecuaciones~(\ref{apendice_002_002}) y (\ref{apendice_002_003}). 
De la \'ultima igualdad tambi\'en se obtiene el siguiente resultado que ser\'a de utilidad
\begin{equation}
\left\langle \left( \hat{\mathbf{u}} \cdot \hat{\mathbf{n}} \right)^{2} \right\rangle = \frac{2 S+1}{3}.
\label{apendice_002_005}
\end{equation}

Tambi\'en por simetr\'ia, puede anticiparse que $\mathbf{Q}$ tendr\'a un valor propio degenerado,
$\lambda^{\prime}$, asociado con la libertad de que cualquier vector en el plano perpendicular a $\hat{\mathbf{n}}$ debe ser un vector propio. Sin p\'erdida 
de generalidad, consideremos el sistema ortonormal de vectores $\{\hat{\mathbf{m}}_{1},\hat{\mathbf{m}}_{2},\hat{\mathbf{n}}\}$ mostrado en la 
figura~\ref{figura_013}. Entonces, de manera an\'aloga a la ecuaci\'on~(\ref{apendice_002_004}), se cumple
\begin{equation}
\hat{\mathbf{m}}_{1}^{\text{T}}  \cdot \mathbf{Q} \cdot \hat{\mathbf{m}}_{1} = \lambda^{\prime}
\label{apendice_002_006}
\end{equation}
y al sustituir la ecuaci\'on~(\ref{apendice_002_001}) y desarrollar los productos tambi\'en se obtiene
\begin{equation}
\lambda^{\prime} = \frac{1}{2} 
	      \left\langle 3 \left( \hat{\mathbf{u}} \cdot \hat{\mathbf{m}}_{1} \right)^{2} 
                     - 1
              \right\rangle .
\label{apendice_002_007}
\end{equation}

Las orientaciones moleculares se pueden expresar en t\'erminos de la base $\{\hat{\mathbf{m}}_{1},\hat{\mathbf{m}}_{2},\hat{\mathbf{n}}\}$, de la siguiente forma
\begin{equation}
\hat{\mathbf{u}}_{i} = \left( \hat{\mathbf{u}}_{i} \cdot \hat{\mathbf{m}}_{1} \right) \hat{\mathbf{m}}_{1} 
                     + \left( \hat{\mathbf{u}}_{i} \cdot \hat{\mathbf{m}}_{2} \right) \hat{\mathbf{m}}_{2} 
                     + \left( \hat{\mathbf{u}}_{i} \cdot \hat{\mathbf{n}} \right) \hat{\mathbf{n}}.
\label{nonumber}
\end{equation}

Dado que $\hat{\mathbf{m}}_{1}$, $\hat{\mathbf{m}}_{2}$ y $\hat{\mathbf{n}}$ son unitarios y mutuamente perpendiculares, tenemos 
$u_{i}^{2} = 1 = \left( \hat{\mathbf{u}}_{i} \cdot \hat{\mathbf{m}}_{1} \right)^{2} 
+ \left( \hat{\mathbf{u}}_{i} \cdot \hat{\mathbf{m}}_{2} \right)^{2} 
+ \left( \hat{\mathbf{u}}_{i} \cdot \hat{\mathbf{n}} \right)^{2}$ y, al promediar,
\begin{equation}
\langle \left( \hat{\mathbf{u}} \cdot \hat{\mathbf{m}}_{1} \right)^{2} \rangle 
+ \langle \left( \hat{\mathbf{u}} \cdot \hat{\mathbf{m}}_{2} \right)^{2} \rangle 
+ \langle \left( \hat{\mathbf{u}} \cdot \hat{\mathbf{n}} \right)^{2} \rangle = 1.
\nonumber
\end{equation}

Sin embargo, dado que las direcciones $\hat{\mathbf{m}}_{1}$ y $\hat{\mathbf{m}}_{2}$ son equivalentes, podemos esperar
que los ejes moleculares se proyecten igualmente sobre ellas. Consecuentemente, 
$ \langle \left( \hat{\mathbf{u}} \cdot \hat{\mathbf{m}}_{1} \right)^{2} \rangle 
= \langle \left( \hat{\mathbf{u}} \cdot \hat{\mathbf{m}}_{2} \right)^{2} \rangle $, 
lo que conduce a
\begin{equation}
\langle \left( \hat{\mathbf{u}}\cdot \hat{\mathbf{m}}_{1} \right)^{2} \rangle =\frac{1-\langle \left( \hat{\mathbf{u}} \cdot \hat{\mathbf{n}} \right)^{2} \rangle}{2}
= \frac{1}{3} \left(1-S\right),
\label{apendice_002_008}
\end{equation}
en donde, para obtener la \'ultima igualdad se ha utilizado la ecuaci\'on~(\ref{apendice_002_005}). Finalmente, al sustituir la ecuaci\'on~(\ref{apendice_002_008}) en
la ecuaci\'on~(\ref{apendice_002_007}) se obtiene el menor de los eigenvalores de $\mathbf{Q}$,
\begin{equation}
\lambda^{\prime} = -\frac{S}{2}.
\label{apendice_002_009}
\end{equation}

%Es posible demostrar, primero, que $\hat{\mathbf{n}}$ es, efectivamente un vector propio de $\mathbf{Q}$, \textit{i. e.}, que satisface la 
%igualdad,

% ==============================================================================================================================
\section{$S$ y $T_{\text{c}}$ en el modelo de Landau--de Gennes \label{apendice_003}}
% ==============================================================================================================================

La cantidad de orden en la fase nem\'atica puede encontrarse al calcular el valor en el que $\Delta f_{\text{bulk}}$ adquiere
un m\'inimo. Al derivar la expresi\'on~(\ref{landau_003}) con respecto a $S$. Al igualar el resultado a cero y considerar el caso 
en el que $S \ne 0$, se obtiene
\begin{equation}
\frac{2A}{3} -\frac{2 B}{9} S + \frac{4C}{9}S^{2} = 0,
\label{apendice_003_000}
\end{equation}
la cual, despu\'es de un poco de \'algebra, se reduce a
\begin{equation}
\frac{3A}{2C} -\frac{B}{2 C} S + S^{2} = 0.
\label{apendice_003_001}
\end{equation}

Al resolver esta ecuaci\'on cuadr\'atica para $S$ se obtiene, precisamente, la ecuaci\'on~(\ref{landau_004}). La soluci\'on
conjugada no se considera pues corresponde al m\'aximo de la funci\'on $\Delta f_{\text{bulk}}$, que separa a las 
fases isotr\'opica y nem\'atica.

La cantidad de orden a la temperatura cr\'itica puede obtnerse de imponer la condici\'on de que adem\'as de tener un m\'inimo en $S_{\text{c}}$, 
$\Delta f_{\text{bulk}}$ tambi\'en se anula en ese valor, \textit{i. e.}, $\Delta f_{\text{bulk}}(S_{\text{c}}) = 0$, lo que conduce a
\begin{equation}
\frac{A}{3} S_{\text{c}}^{2} -\frac{2 B}{27} S_{\text{c}}^{3} +\frac{C}{9} S_{\text{c}}^{4} = 0,
\label{apendice_003_003}
\end{equation}
o bien, al considerar que $S_{\text{c}}\ne 0$ y multiplicar por $2$,
\begin{equation}
\frac{2 A}{3} -\frac{4 B}{27} S_{\text{c}} +\frac{2 C}{9} S_{\text{c}}^{2} = 0.
\label{apendice_003_004}
\end{equation}

Entonces, al evaluar la ecuaci\'on~(\ref{apendice_003_000}) en $S = S_{\text{c}}$, y restar el resultado t\'ermino a t\'ermino con la
ecuaci\'on~(\ref{apendice_003_004}), resulta
\begin{equation}
\left(-\frac{2}{9}+\frac{4}{27}\right)B S_{\text{c}} + \left( \frac{4}{9}-\frac{2}{9} \right) C S_{\text{c}}^{2} = 0,
\end{equation}
cuya soluci\'on no trivial es justamente la ecuaci\'on~(\ref{landau_006}).

Finalmente, al sustituir $S_{\text{c}}$ dado por la ecuaci\'on~(\ref{landau_006}) en la ecuaci\'on~(\ref{apendice_003_001}) se obtiene
\begin{equation}
\frac{3A}{2C} - \frac{B^{2}}{6 C^{2}} + \frac{B^{2}}{9 C^{2}} = 0,
\label{apendice_003_006}
\end{equation}
de la cual, al sustituir $A = A^{\prime}(T_{\textit{c}}-T^{*})$ y despejar $T_{\text{c}}$ se obtiene la expresi\'on~(\ref{landau_005}).

% ==============================================================================================================
\section{Demostraciones adicionales relacionadas con las derivadas de $\hat{\mathbf{n}}$\label{apendice_004}}
% ==============================================================================================================

La ecuaci\'on~(\ref{energia_elastica_003c}) puede demostrarse al calcular primero 
\begin{eqnarray}
\partial_{\alpha} \left( n_{\alpha} \partial_{\beta} n_{\beta}\right) 
& = & n_{\alpha} \partial_{\alpha} \partial_{\beta} n_{\beta}
+ \partial_{\alpha} n_{\alpha} \, \partial_{\beta} n_{\beta}, \nonumber \\
& = & n_{\alpha} \partial_{\alpha} \partial_{\beta} n_{\beta}
+ \left( \boldsymbol{\nabla} \cdot \hat{\mathbf{n}}\right)^{2},
\label{apendice_004_001}
\end{eqnarray}
en donde hemos utilizado la ecuaci\'on~(\ref{energia_elastica_003b}), y
\begin{eqnarray}
\partial_{\alpha} \left( n_{\beta} \partial_{\beta} n_{\alpha}\right)
& = & n_{\beta} \partial_{\alpha} \partial_{\beta} n_{\beta}
+ \partial_{\alpha} n_{\beta} \, \partial_{\beta} n_{\alpha}, \nonumber \\
& = & n_{\alpha} \partial_{\alpha} \partial_{\beta} n_{\beta}
+ \partial_{\alpha} n_{\beta} \, \partial_{\beta} n_{\alpha},
\label{apendice_004_002}
\end{eqnarray}
en donde, para obtener la segunda igualdad intercambiamos el order de las derivadas en el primer t\'ermino y despu\'es renombramos los
\'indices $\alpha$ como $\beta$ y viceversa, aprovechando que son mudos. Al restar miembro a miembro las ecuaciones~(\ref{apendice_004_001}) y
(\ref{apendice_004_002}) y reorganizar los t\'erminos, se obtiene la ecuaci\'on~(\ref{energia_elastica_003c}).

Para demostrar la validez de la ecuaci\'on~(\ref{energia_elastica_012}), podemos partir de la ecuaci\'on~(\ref{energia_elastica_011}) reescrita
con un reordenamiento de los factores y una permutaci\'on c\'iclica de \'indices
\begin{equation}
\left[\hat{\mathbf{n}} \times \left( \boldsymbol{\nabla}\times\hat{\mathbf{n}} \right)\right]_{\alpha} 
=  \varepsilon_{\gamma\alpha\beta} \varepsilon_{\gamma\lambda\mu} n_{\beta} \partial_{\lambda} n_{\mu} .
\nonumber
\end{equation}

Al utilizar la ecuaci\'on~(\ref{energia_elastica_006}) y expandir los productos involucrados obtenemos el resultado deseado,
\begin{eqnarray}
\left[\hat{\mathbf{n}} \times \left( \boldsymbol{\nabla}\times\hat{\mathbf{n}} \right)\right]_{\alpha}
& = & \left( \delta_{\alpha \lambda} \delta_{\beta \mu} - \delta_{\alpha \mu} \delta_{\beta \lambda} \right) n_{\beta} \partial_{\lambda} n_{\mu} \nonumber \\
& = & n_{\beta} \partial_{\alpha} n_{\beta} - n_{\beta} \partial_{\beta} n_{\alpha} \nonumber \\
& = & - n_{\beta} \partial_{\beta} n_{\alpha} . \nonumber
\end{eqnarray}
en donde en el \'ultimo paso hemos utilizado la ecuaci\'on~(\ref{energia_elastica_003d}).

Con el fin de demostrar la ecuaci\'on~(\ref{energia_elastica_015}), podemos sustituir el producto $\varepsilon_{\alpha\beta\gamma}\varepsilon_{\lambda\mu\nu}$
dado por la ecuaci\'on~(\ref{energia_elastica_010}) en la igualdad (\ref{energia_elastica_014}). Al desarrollar los productos y contraer sobre las deltas de 
Kronecker obtenemos
\begin{eqnarray}
\left( \hat{\mathbf{n}} \cdot \boldsymbol{\nabla}\times\hat{\mathbf{n}} \right)^{2}
& = & n_{\alpha} n_{\lambda} \partial_{\beta} n_{\gamma} \, \partial_{\mu} n_{\nu} \nonumber \\
&   & \times \left[
             \delta_{\alpha\lambda} \left( \delta_{\beta\mu} \delta_{\gamma\nu} - \delta_{\beta\nu} \delta_{\gamma\mu}  \right) \right. \nonumber \\
&   &-\delta_{\alpha\mu} \left( \delta_{\beta\lambda} \delta_{\gamma\nu} - \delta_{\beta\nu} \delta_{\gamma\lambda} \right) \nonumber \\
&   &\left. +\delta_{\alpha\nu} \left( \delta_{\beta\lambda} \delta_{\gamma\mu} - \delta_{\beta\mu} \delta_{\gamma\lambda} \right)
	     \right] \nonumber \\
& = & \partial_{\beta} n_{\gamma} \, \partial_{\mu} n_{\nu} \left( \delta_{\beta\mu} \delta_{\gamma\nu} - \delta_{\beta\nu} \delta_{\gamma\mu}  \right) \nonumber \\
&   & - n_{\alpha} n_{\lambda} \partial_{\beta} n_{\gamma} \, \partial_{\alpha} n_{\nu} \left( \delta_{\beta\lambda} \delta_{\gamma\nu} - \delta_{\beta\nu} \delta_{\gamma\lambda}  \right) \nonumber \\
&   & - \cancelto{0}{n_{\alpha} n_{\lambda} \partial_{\beta} n_{\gamma} \, \partial_{\mu} n_{\alpha}} 
        \left( \delta_{\beta\lambda} \delta_{\gamma\mu} - \delta_{\beta\mu} \delta_{\gamma\lambda}  \right) \nonumber \\
& = & \partial_{\beta} n_{\gamma} \, \partial_{\beta} n_{\gamma} - \partial_{\beta} n_{\gamma} \, \partial_{\gamma} n_{\beta} \nonumber \\
&   & -n_{\alpha} n_{\beta} \, \partial_{\beta} n_{\gamma} \, \partial_{\alpha} n_{\gamma} \nonumber \\ 
&   & -\cancelto{0}{n_{\alpha} n_{\beta} \, \partial_{\gamma} n_{\gamma}\, \partial_{\alpha} n_{\beta}} \nonumber\\ 
& = & \partial_{\alpha} n_{\beta} \, \partial_{\alpha} n_{\beta} - \partial_{\alpha} n_{\beta} \, \partial_{\beta} n_{\alpha} \nonumber \\
&   & -n_{\beta} n_{\gamma} \, \partial_{\beta} n_{\alpha} \, \partial_{\gamma} n_{\alpha} , \label{apendice_004_003}
\end{eqnarray}
en donde hemos podido cancelar varios t\'erminos id\'enticamente debido que contienen productos de la forma dada por la ecuaci\'on~(\ref{energia_elastica_003d})
y en el \'ultimo paso simplemente hemos renombrado los \'indices mudos para facilitar la comparaci\'on subsecuente del resultado.
El segundo t\'ermino del lado derecho de la ecuaci\'on~(\ref{apendice_004_003}), est\'a dado por la ecuaci\'on~(\ref{energia_elastica_003c}), mientras
que el tercero es el negativo de $\left( \hat{\mathbf{n}} \times \left(\boldsymbol{\nabla}\times \hat{\mathbf{n}} \right)\right)^{2}$, tal como se
exhibe en la ecuaci\'on~(\ref{energia_elastica_013}). Al hacer estas identificaciones y sustituir, se demuestra que la ecuaci\'on~(\ref{energia_elastica_015}) es correcta.

Los productos que aparecen en la segunda columna de la tabla~\ref{tabla_002} pueden obtenerse de la siguiente manera. Para el primer rengl\'on, la contracci\'on
directa de los \'indices $\beta$ y $\lambda$ produce
\begin{equation}
\delta_{\alpha\beta}\delta_{\gamma\lambda}\partial_{\alpha}n_{\beta} \, \partial_{\gamma}n_{\lambda} 
= \partial_{\alpha}n_{\alpha} \, \partial_{\gamma}n_{\gamma} 
= \left(\boldsymbol{\nabla}\cdot\hat{\mathbf{n}}\right)^{2}.
\nonumber
\end{equation}

Para el tercer rengl\'on, al contraer $\gamma$ y $\lambda$ resulta
\begin{eqnarray}
\delta_{\alpha\gamma}\delta_{\beta\lambda} \partial_{\alpha}n_{\beta} \, \partial_{\gamma}n_{\lambda}
& = & \partial_{\alpha}n_{\beta}\partial_{\alpha}n_{\beta} \nonumber \\
& = & \left(\boldsymbol{\nabla}\cdot\hat{\mathbf{n}}\right)^{2} 
    + \left(\hat{\mathbf{n}}  \cdot \boldsymbol{\nabla} \times \hat{\mathbf{n}}\right)^{2} \nonumber \\
&   & + \left(\hat{\mathbf{n}}  \times \boldsymbol{\nabla} \times \hat{\mathbf{n}}\right)^{2} + \boldsymbol{\nabla}\cdot\mathbf{m},
\nonumber
\end{eqnarray}
donde se ha utilizado la ecuaci\'on~(\ref{energia_elastica_015}). Al ignorar el t\'ermino $\boldsymbol{\nabla}\cdot\mathbf{m}$ se obtiene el resultado
deseado.

De manera similar, para el cuarto rengl\'on se tiene, de acuerdo con la ecuaci\'on~(\ref{energia_elastica_003c}),
\begin{equation}
\delta_{\alpha\lambda}\delta_{\beta\gamma} \partial_{\alpha}n_{\beta} \, \partial_{\gamma}n_{\lambda}
= \partial_{\alpha}n_{\beta} \, \partial_{\beta}n_{\alpha}
= \left(\boldsymbol{\nabla}\cdot\hat{\mathbf{n}}\right)^{2} + \boldsymbol{\nabla}\cdot\mathbf{m},
\nonumber
\end{equation}
que tambi\'en produce el resultado deaseado al ignorar la divergencia de $\mathbf{m}$.

El producto del quinto rengl\'on es
\begin{equation}
\delta_{\beta\lambda}n_{\alpha}n_{\gamma}\partial_{\alpha}n_{\beta}\,\partial_{\gamma}n_{\lambda} =
n_{\alpha} n_{\gamma} \partial_{\alpha}n_{\beta} \partial_{\gamma}n_{\beta} = 
\left(\hat{\mathbf{n}} \times \left( \boldsymbol{\nabla} \times \hat{\mathbf{n}}\right)\right)^{2},
\nonumber
\end{equation}
donde la identificaci\'on es directa de acuerdo con la ecuaci\'on~(\ref{energia_elastica_013}).

Los \'ultimos dos renglones de la tabla~\ref{tabla_002} no requieren desarrollo.

Para obtener los resultados de la segunda columna en la tabla~\ref{tabla_003}, lo m\'as conveniente es seguir un paso similar al que se usa en la integraci\'on
por partes, seguido de anular las productos que tienen la forma dada por la ecuaci\'on~(\ref{energia_elastica_003d}). Espec\'ificamente, para el producto del segundo
rengl\'on se tiene
\begin{eqnarray}
n_{\alpha}n_{\beta}n_{\gamma}\partial_{\alpha}\partial_{\beta}n_{\gamma}
& = & n_{\alpha}n_{\beta} \cancelto{0}{\partial_{\alpha}\left(n_{\gamma} \partial_{\beta}n_{\gamma} \right)} 
     -n_{\alpha}n_{\beta}\partial_{\alpha}n_{\gamma} \, \partial_{\beta}n_{\gamma} \nonumber \\
& = & - \left(\hat{\mathbf{n}}  \times \boldsymbol{\nabla} \times \hat{\mathbf{n}}\right)^{2},
\nonumber
\end{eqnarray}
de acuerdo con la ecuaci\'on~(\ref{energia_elastica_013}).

El tercer rengl\'on en la tabla~\ref{tabla_003} no requiere desarrollo, mientras que para el cuarto se tiene
\begin{eqnarray}
\delta_{\alpha\beta} n_{\gamma}\partial_{\alpha}\partial_{\beta}n_{\gamma} 
& = & n_{\gamma}\partial_{\alpha}\partial_{\alpha}n_{\gamma} \nonumber \\
& = & \cancelto{0}{\partial_{\alpha}\left( n_{\gamma} \partial_{\alpha}n_{\gamma} \right)} - \partial_{\alpha} n_{\gamma} \, \partial_{\alpha}n_{\gamma} 
\nonumber
\end{eqnarray}
en donde el t\'ermino que no se cancela puede indentificarse en la ecuaci\'on~(\ref{energia_elastica_015}). Al despejarlo e ignorar la divergencia de
$\mathbf{m}$ se obtiene el resultado deseado.

Los resultados del quinto y sexto rengl\'on en la tabla~\ref{tabla_003} se obtienen de manera id\'entica como sigue
\begin{eqnarray}
\delta_{\alpha\gamma}n_{\beta} \partial_{\alpha}\partial_{\beta}n_{\gamma} & = & n_{\beta} \partial_{\alpha}\partial_{\beta}n_{\alpha} \nonumber \\
& = & n_{\beta} \partial_{\beta}\partial_{\alpha} n_{\alpha}  \nonumber \\
& = & \partial_{\beta} \left( n_{\beta} \partial_{\alpha} n_{\alpha} \right) - \partial_{\beta} n_{\beta} \, \partial_{\alpha} n_{\alpha}   \nonumber \\
\nonumber
\end{eqnarray}
donde se obtiene el resultado buscado al ignorar el primer t\'ermino por ser una divergencia.

% ==============================================================================================================
\section{Aproximaci\'on de constantes el\'asticas iguales \label{apendice_005}}
% ==============================================================================================================

Para obtener la expresi\'on~(\ref{energia_elastica_021a}), puede partirse primero de la igualdad
\begin{eqnarray}
\left(\boldsymbol{\nabla}\cdot\hat{\mathbf{n}}\right)^{2} + \left(\boldsymbol{\nabla}\times\hat{\mathbf{n}}\right)^{2} 
& = & \partial_{\alpha}n_{\alpha} \, \partial_{\beta}n_{\beta} \nonumber \\ 
&   & + \varepsilon_{\alpha\beta\gamma} \partial_{\beta}n_{\gamma} \varepsilon_{\alpha\mu\nu} \partial_{\mu}n_{\nu} \nonumber \\
& = & \partial_{\alpha}n_{\alpha} \, \partial_{\beta}n_{\beta} \nonumber \\ 
&   & + \partial_{\beta}n_{\gamma} \partial_{\mu}n_{\nu} 
      \left( \delta_{\beta\mu} \delta_{\gamma\nu} - \delta_{\beta\nu}\delta_{\gamma\mu} \right) \nonumber \\
& = & \partial_{\alpha}n_{\alpha} \, \partial_{\beta}n_{\beta} + \partial_{\mu}n_{\gamma} \, \partial_{\mu}n_{\gamma} \nonumber \\ 
&   &  - \partial_{\nu}n_{\mu} \partial_{\mu}n_{\nu} \nonumber \\
& = & \partial_{\alpha}n_{\alpha} \, \partial_{\beta}n_{\beta} + \partial_{\alpha}n_{\beta} \, \partial_{\alpha}n_{\beta} \nonumber \\ 
&   & - \partial_{\alpha}n_{\beta} \partial_{\beta}n_{\alpha},
\label{apendice_005_001}
\end{eqnarray}
en donde en la segunda igualdad hemos utilizado la ecuaci\'on~(\ref{energia_elastica_006}) y en la \'ultima hemos renombrado los \'indices mudos.

Por otra parte, al suponer $K_{1} = K_{2} = K_{3}$ y reescribir la ecuaci\'on~(\ref{energia_elastica_021a}) en t\'erminos de las expresiones en notaci\'on de \'indices 
para $\left(\boldsymbol{\nabla}\cdot\hat{\mathbf{n}}\right)^{2}$ (ecuaci\'on~(\ref{energia_elastica_003b})), 
$\left(\hat{\mathbf{n}} \times \left( \boldsymbol{\nabla}\times\hat{\mathbf{n}}\right)\right)^{2}$ (ecuaci\'on~(\ref{energia_elastica_013})), y
$\left(\hat{\mathbf{n}} \cdot \boldsymbol{\nabla}\times\hat{\mathbf{n}}\right)^{2}$ (ecuaci\'on~(\ref{energia_elastica_014a})), obtenemos
\begin{eqnarray}
f_{\text{elas}} 
& = & \frac{1}{2} K \left( \partial_{\alpha}n_{\alpha} \partial_{\beta} n_{\beta} \right. \nonumber \\
&   & + \partial_{\alpha} n_{\beta} \, \partial_{\alpha} n_{\beta} - \partial_{\alpha} n_{\beta} \, \partial_{\beta} n_{\alpha} \nonumber \\
&   & -n_{\beta} n_{\gamma} \, \partial_{\beta} n_{\alpha} \, \partial_{\gamma} n_{\alpha} \nonumber \\
&   & \left. + n_{\beta} n_{\gamma} \, \partial_{\beta} n_{\alpha} \, \partial_{\gamma} n_{\alpha} \right) \nonumber \\
& = & \frac{1}{2} K \left( \partial_{\alpha}n_{\alpha} \partial_{\beta} n_{\beta} \right. \nonumber \\
&   & \left. + \partial_{\alpha} n_{\beta} \, \partial_{\alpha} n_{\beta} - \partial_{\alpha} n_{\beta} \, \partial_{\beta} n_{\alpha} \right). \label{apendice_005_002}
\end{eqnarray}

Al comparar las ecuaciones~(\ref{apendice_005_001}) y (\ref{apendice_005_002}) se observa la validez de la expresi\'on~(\ref{energia_elastica_021a}).

% ==============================================================================================================
\section{Relaci\'on entre las constantes el\'asticas $K_{i}$ y $L_{i}$ \label{apendice_006}}
% ==============================================================================================================

Al utilizar la ecuaci\'on~(\ref{parametro_orden_tensorial_004}) y suponer que $S$ es uniforme, se puede demostrar la siguiente igualdad
\begin{equation}
\partial_{\gamma}Q_{\alpha\beta} = \frac{3}{2} S \left( n_{\alpha} \partial_{\gamma} n_{\beta} + n_{\beta} \partial_{\gamma} n_{\alpha}\right),
\nonumber
\end{equation}
%y
%\begin{equation}
%\partial_{\beta}Q_{\alpha\beta} = \frac{3}{2} S \left( n_{\alpha} \partial_{\beta} n_{\beta} + n_{\beta} \partial_{\beta} n_{\alpha}\right),
%\nonumber
%\end{equation}
de la cual se obtienen las contribuciones a la energ\'ia el\'astica en la representaci\'on tensorial, ecuaci\'on~(\ref{energia_elastica_022}),
\begin{equation}
\partial_{\gamma}Q_{\alpha\beta} \partial_{\gamma}Q_{\alpha\beta} = \frac{9}{2} S^{2} \partial_{\beta} n_{\alpha} \, \partial_{\beta} n_{\alpha},
\label{apendice_006_001}
\end{equation}
\begin{equation}
\partial_{\beta}Q_{\alpha\beta} \partial_{\gamma}Q_{\alpha\gamma} 
 = \frac{9}{4}S^{2} \left(  \partial_{\alpha} n_{\alpha} \, \partial_{\beta} n_{\beta} 
    + n_{\beta}n_{\gamma} \partial_{\beta}n_{\alpha}\, \partial_{\gamma}n_{\alpha}\right),
\label{apendice_006_002}
\end{equation}
\begin{eqnarray}
Q_{\alpha\beta}\partial_{\alpha}Q_{\gamma\lambda} \partial_{\beta}Q_{\gamma\lambda} 
& = & \frac{9}{4}S^{3}
      \left( 3 n_{\alpha}n_{\beta}  \partial_{\alpha} n_{\gamma} \, \partial_{\beta} n_{\gamma} \right. \nonumber \\ 
&   & \left.  - \partial_{\beta} n_{\alpha} \partial_{\beta} n_{\alpha} \right).
\label{apendice_006_003}
\end{eqnarray}

Para obtener las ecuaciones~(\ref{apendice_006_001}) a (\ref{apendice_006_003}) es necesario expandir los productos y utilizar la ecuaci\'on~(\ref{energia_elastica_003d}).

Las cantidades en el lado derecho de las ecuaciones~(\ref{apendice_006_001}) a (\ref{apendice_006_003}) pueden identificarse con las contribuciones cuadr\'aticas de las deformaciones
a trav\'es de las ecuaciones~(\ref{energia_elastica_003b}), (\ref{energia_elastica_013}) y (\ref{energia_elastica_015}). Al utilizar estas ecuaciones, agrupar los t\'erminos 
correspondientes a cada tipo de deformaci\'on, ignorar aquellos que corresponden a una divergencia y comparar el resultado miembro a miembro con la 
ecuaci\'on~(\ref{energia_elastica_021}) se obtienen las igualdades siguientes
\begin{equation}
\frac{9}{2} S^{2} L_{1} + \frac{9}{4} S^{2} L_{2} - \frac{9}{4} S^{3} L_{3} = K_{1} ,
\label{apendice_006_004}
\end{equation}
\begin{equation}
\frac{9}{2} S^{2} L_{1} -\frac{9}{4} S^{3} L_{3} = K_{2}  
\label{apendice_006_005}
\end{equation}
y
\begin{equation}
\frac{9}{2} S^{2} L_{1} + \frac{9}{4} S^{2} L_{2} + \frac{9}{2} S^{3} L_{3} = K_{3} .
\label{apendice_006_006}
\end{equation}

La ecuaci\'on~(\ref{energia_elastica_025}) resulta de restar miembro a miembro las ecuaciones~(\ref{apendice_006_006}) y (\ref{apendice_006_004}) y despejar
$L_{3}$. La ecuaci\'on~(\ref{energia_elastica_024}) se obtiene al sustituir la f\'ormula obtenida para $L_{3}$ en la ecuaci\'on~(\ref{apendice_006_005}) y
despejar $L_{1}$. Finalmente, la ecuaci\'on~(\ref{energia_elastica_023}) se obtiene al sustituir los resultados para $L_{1}$ y $L_{3}$ en la 
ecuaci\'on~(\ref{apendice_006_004}) y despejar $L_{2}$.

Como nota final se menciona que las ecuaciones~(\ref{energia_elastica_023}) a (\ref{energia_elastica_025}) son equivalentes a las reportadas en~\cite{hernandez_j_chem_phys_2011},
excepto porque la representaci\'on de la energ\'ia en esa referencia se da en t\'erminos del tensor que aqu\'i llamamos $S_{\alpha\beta}$ y por un aparente error en la potencia
de $S$ en la f\'ormula para $L_{3}$.

%\end{multicols}

%\medline
%\begin{multicols}{2}
%%%%%%%%%%%%%%%
%
%Using BibTeX
\nocite{*}
%\bibliographystyle{rmf-style}
%\bibliography{liquid_crystals,statistical_mechanics,mpcd,active_particles,lc_topological_defects}
\bibliography{chapter_001}
%
%%%%%%%%
%
%Introducing references manually
%

%\begin{thebibliography}{99}
%\bibitem{ELK} E. Ley-Koo, Recent progress in confined atoms and molecules: Superintegrability and symmetry breakings, Rev. Mex. Fis. 64 (2018) 326, \url{https://doi.org/10.31349/RevMexFis.64.326}
%
%\bibitem{Griffiths} D.J. Griffiths, Introduction to Electrodynamics, 2nd ed. (Prentice Hall, Englewood Cliffs, NJ, 1989), pp. 331–334.
%
%\end{thebibliography}
%\end{multicols}
\end{document}